\documentclass[aps, prb, twocolumn,floatfix, superscriptaddress]{revtex4-2}

\usepackage{graphicx} 
\usepackage{float}
\usepackage{mathptmx}
\usepackage[utf8]{inputenc}
\usepackage[T1]{fontenc}
\usepackage[english]{babel}
\usepackage{amsmath}
\usepackage{amssymb}
\usepackage{xcolor}
\usepackage{braket}
\usepackage{bm}
\usepackage{physics}
\usepackage[colorlinks=true, linkcolor=blue,
citecolor=red, 
urlcolor=blue, hypertexnames=true]{hyperref}

\usepackage{multibib}

\newcommand{\eref}[1]{Eq.~\eqref{#1}}

\newcommand{\fref}[1]{Fig.~\ref{#1}}

\begin{document}

\title{Nonlinear Breathing and Caustic Collapse in Quenched Long-Range Interacting Gases}

\author{Manas Kulkarni}
\email{manas.kulkarni@icts.res.in }
\affiliation{International Centre for Theoretical Sciences, Tata Institute of Fundamental Research, Bangalore 560089, India}

\author{Sanjib Sabhapandit}
\email{sanjib@rri.res.in}
\affiliation{Raman Research Institute, Bangalore 560080, India}

\date{\today}

\begin{abstract}
Exact descriptions of time evolution of out-of-equilibrium interacting many-body systems are generically rare. We show
that for the harmonically confined Riesz gas with pair potential
$\propto|x|^{-k}$, starting from its zero-temperature equilibrium initial configuration, the positions can be exactly described under any time-dependent trap,
for every particle number $N$ and every $k>-2$, in terms of only two effective degrees of freedom: one for the center of mass motion and another for the common dilation scale parameter. These bulk-translation and dilation parameters also  dictate the time evolution of the coarse-grained density profile predicted by the hydrodynamics perfectly. For a sudden quench, the gas undergoes periodic undamped breathing for $k>0$, even though the dynamics is generically non-integrable. For $-2<k<0$, compressing the trap beyond a factor
$\sqrt{2/|k|}$ triggers the breathing to a sharp transition yielding perfect caustics, where all $N$
particles meet at one point at once and re-emerge in reversed order.
\end{abstract}

\maketitle

\textit{Introduction.}--- 
After a sudden change of its parameters, an
isolated interacting many-body system generically scrambles the
information stored in its initial state and undergoes out-of-equilibrium dynamics~\cite{polkovnikov2011,dalessio2016,eisert2015}. Exact descriptions of such far-from-equilibrium evolution are  rare. They exist
almost exclusively for integrable
models~\cite{,calogero1971,moser1975,perelomov1990,ap_review, castroalvaredo2016,bertini2016,kinoshita2006,doyon2020}, or for
harmonically trapped systems with scale-invariant interactions, such as
inverse-square or two-dimensional contact interactions, where a hidden
SO(2,1) symmetry gives rise to a breathing dynamics~\cite{pitaevskii1997}.
For generic non-integrable systems, the fully nonlinear dynamics of a
trapped cloud, at arbitrary particle number and with no small
parameter, is known only in special cases~\cite{castin1996,kagan1996}.

The problem becomes even more challenging for long-range interactions, whose pair
potential varies as $r^{-k}$ with $k\le d$ in $d$
dimensions, as for gravitating and Coulomb systems~\cite{rmp0,rmp1,rmp2}.
Since $\int d^dr\,r^{-k}$  diverges at large distances, the
energy of each particle is dominated by distant particles, the total energy is not
additive, and relaxation toward equilibrium is extremely slow, often
passing through quasi-stationary states whose lifetime grows with the number of particles
$N$~\cite{campa2009, campa2014book,defenu}. The far-from-equilibrium dynamics therefore persists for long times,
which makes exact descriptions of long-range systems particularly
valuable.

\begin{figure}[t]
  \centering
  \includegraphics[width=\columnwidth]{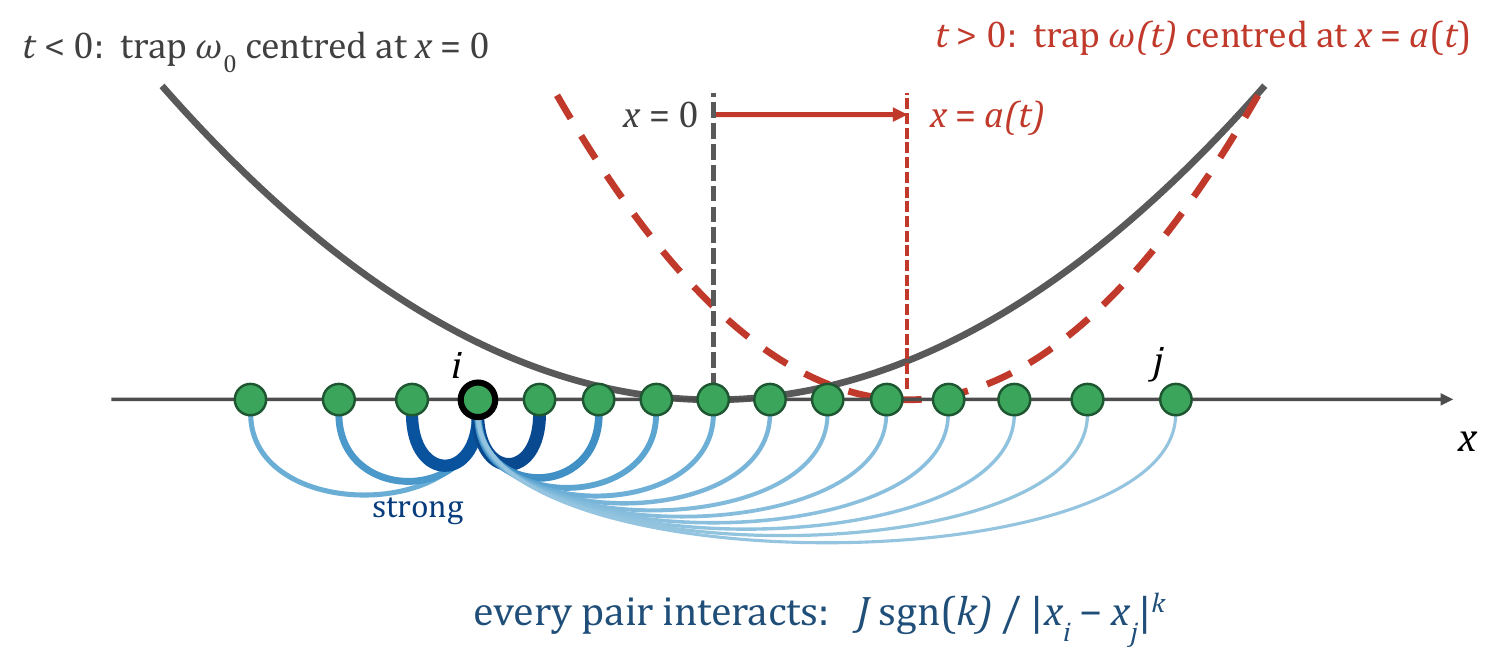}
  \caption{Setup and quench protocol: $N$ identical particles (green)
  on a line are confined by a harmonic trap and interact through the
  long-range Riesz potential [\eref{eq:H}]. For $t<0$ the trap has
  frequency $\omega_0$ and is centred at $x=0$ (grey, solid), and the
  particles rest at the zero-temperature equilibrium positions $x_i^0$
  fixed by the force balance [\eref{eq:equil}]. At $t=0^+$,  the trap
  frequency and center are switched to $\omega(t)$ and $a(t)$ (red,
  dashed). The vertical dashed lines mark the two trap centers. The
  arcs show that a tagged particle $i$ interacts with every other
  particle $j$, with strength $J\,\mathrm{sgn}(k)\,|x_i-x_j|^{-k}$.}
  \label{fig:schematic}
\end{figure}

A natural setting for studying this problem
is the Riesz
gas~\cite{riesz1938,lewin2022,serfaty2024}, where  $N$ particles on a line are
confined in a harmonic trap and they interact through the pair potential
$J\,\mathrm{sgn}(k)\,|x_i-x_j|^{-k}$ with $k>-2$~[Fig.~\ref{fig:schematic}], and $x_{i}$'s denoting the positions of the particles. It is relatively long-range for
$k\le1$ and short-range for $k>1$, despite the presence of all-to-all coupling. The single power-law exponent
$k$ connects well-known models such as the one-dimensional one-component plasma
($k=-1$)~\cite{ocp1, ocp2, ocp3, ocp4, ocp6, dhar2017}, the log-gas of random-matrix theory
($k\to0$), chains of trapped ions ($k=1$)~\cite{james1998}, the
integrable Calogero--Moser system ($k=2$)~\cite{calogero1971, moser1975,ap_review, perelomov1990, Abanov2011}, 
dipolar gases ($k=3$)~\cite{lahaye2009, chomaz2023}, among many others~\cite{brown_CUP_2003}. The equilibrium properties of this family are now
understood in considerable detail~\cite{agarwal2019,lewin2022,kethepalli2021,kethepalli2022}. On the other hand, 
its real-time dynamics is only beginning to be charted, through
fluctuating~\cite{dandekar2023} and dissipative~\cite{mukherjee2026}
hydrodynamics, and through spatiotemporal
spread of perturbations~\cite{bks2021}. 
Even for
the simplest nonequilibrium protocol, namely, a quench of the trap,  analytical
results beyond linear response exist only for special values of $k$,
such as the integrable case $k=2$~\cite{calogero1971, moser1975,ap_review, perelomov1990, Abanov2011} and the
 Euler--Poisson description of the one-component plasma, $k=-1$~\cite{engelberg2001}. Otherwise they are
confined to the linear response regime, where the breathing frequency
$\sqrt{k+2}\,\omega_0$ is universal, independent of both $N$ and
dimension~\cite{henning2008}.

In this Letter we show that for a $N$ particle Riesz gas, the protocol of quenching of the trap frequency is exactly solvable for any $k>-2$, in terms of only two effective independent degrees of freedom, far beyond
linear response. More precisely, starting from the zero-temperature equilibrium crystal, represented by the positions $\{x_1^0, x_2^0, \dotsc, x_N^0 \}$, in a trap of frequency $\omega_0$, we let the trap frequency
$\omega(t)$ and the trap center $a(t)$ vary arbitrarily in time $t>0$~[Fig.~\ref{fig:schematic}]. For
any value of  $N$, each particle  moves as
$x_i(t)=X_{CM}(t)+b(t)\,x_i^0$, i.e., the crystal translates rigidly by $X_{CM}(t)$, as in a
generalized Kohn mode~\cite{kohn1961,dobson1994}, and dilates
uniformly through the common scale factor $b(t)$. Therefore, the $N$-body problem collapses exactly onto two independent second order
ordinary differential equations, one for $X_{CM}(t)$ and another for $b(t)$. The equation for the scale factor
$b(t)$ is fully nonlinear, which reduces to the Ermakov--Pinney equation~\cite{pinney1950}
only in the integrable case $k=2$. The same dynamics emerges as an exact
self-similar solution of the continuum Euler equations, both in the
nonlocal regime $k<1$ and in the local, polytropic regime $k>1$,
generalizing the scaling solutions known for Bose--Einstein
condensates~\cite{castin1996,kagan1996,GPE_reset}. Microscopic and hydrodynamic
descriptions therefore agree exactly
without any adjustable parameters. 

\begin{figure}[t]
  \centering
  \includegraphics[width=\columnwidth]{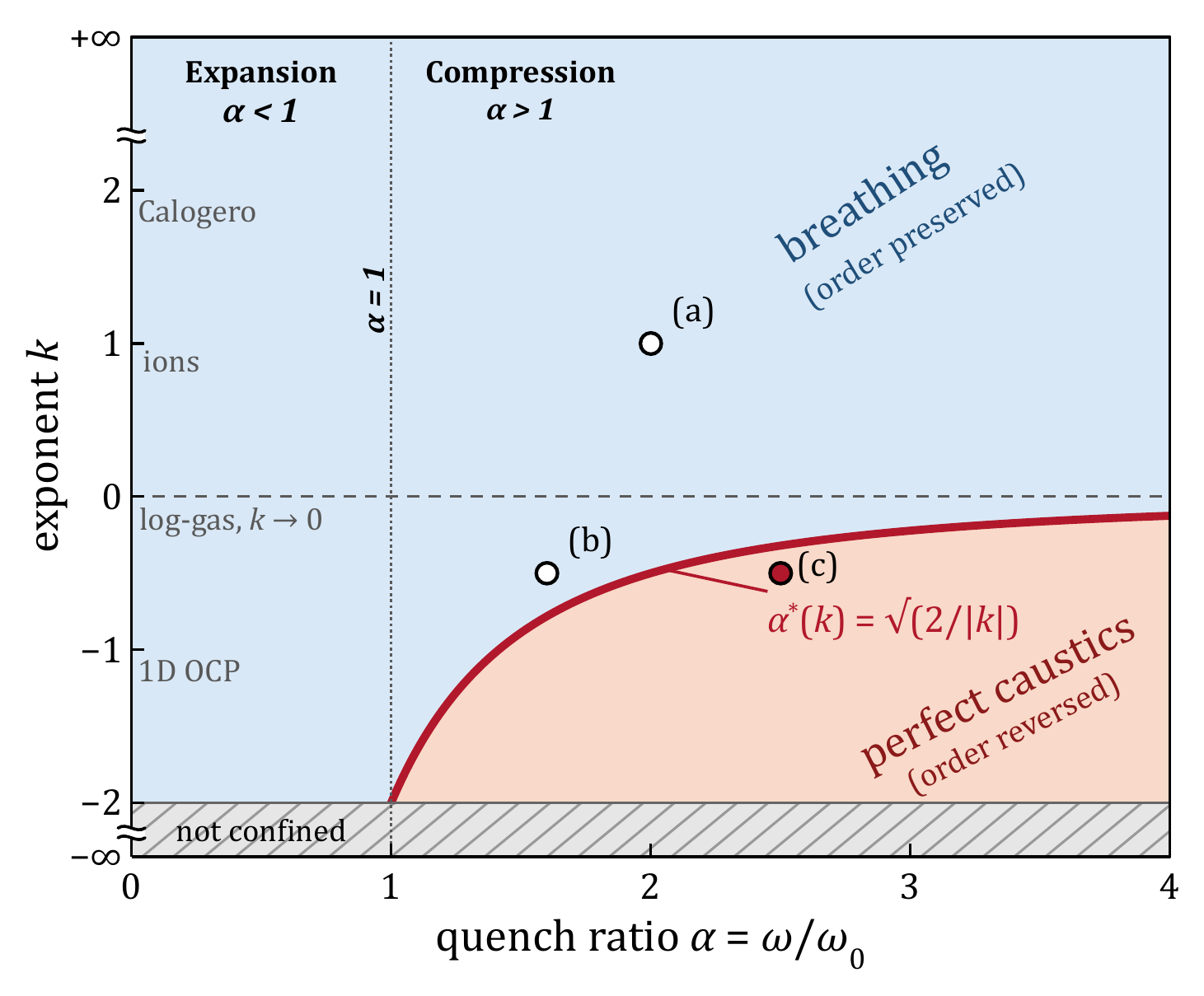}
  \caption{Dynamical phase diagram of the sudden trap quench $\omega_0\to \alpha \omega_0$ in the
  $(\alpha,k)$ plane. The vertical dotted line $\alpha=1$ separates expansion ($\alpha<1$)
  from compression ($\alpha>1$) of the trap. For $-2<k<0$ the effective
  potential $V(b)$ in \eref{eq:Vb} stays finite at $b=0$. The cloud then
  collapses into a perfect caustic and re-emerges with its order
  reversed whenever $\alpha>\alpha^*(k)=\sqrt{2/|k|}$,
  shown by the red region, bounded by the curve $k=-2/\alpha^2$. Everywhere else,
  including all $k>0$ where $V(b)$ diverges at $b=0$, the cloud
  breathes with its order preserved (blue region). On the boundary, $b(t)$
  follows Eq~\eqref{eq:separatrix}. The threshold
  $\alpha^*\to 1$ as $k\to-2^+$ and diverges as $k\to0^-$ (dashed line), so the
  log-gas never forms a caustic. The labels on the left mark the
  Calogero model ($k=2$), trapped ions ($k=1$), the log-gas ($k\to0$),
  and the one-component plasma ($k=-1$).
  The trap does not confine the gas for $k\le-2$ (hatched). Circles
  (a)--(c) mark the quenches shown in the corresponding columns of
  \fref{fig:dynamics}.}
  \label{fig:phase}
\end{figure}

\begin{figure*}[t]
  \centering
  \includegraphics[width=\textwidth]{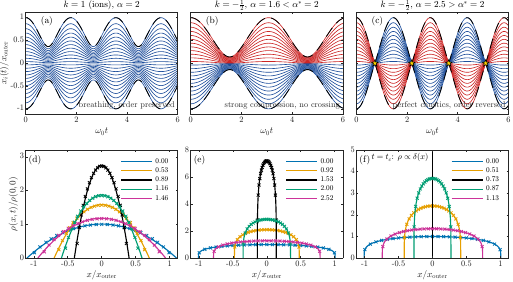}
  \caption{Breathing and caustic collapse after a sudden compression of
  the trap, $\omega_0\to\alpha\,\omega_0$ with $\alpha>1$, at fixed
  centre ($a=0$, so that $X_{CM}=0$). Left column: $k=1$, $\alpha=2$. Middle
  column: $k=-1/2$ below the caustic threshold,
  $\alpha=1.6<\alpha^*=2$. Right column: $k=-1/2$ above it,
  $\alpha=2.5>\alpha^*$, with $\alpha^*$ given by \eref{eq:alphastar}.
  Positions are in units of $x_{\rm outer}\equiv x_N^0$, the initial position of the
  outermost particle, and times in units of $1/\omega_0$.
  (a)--(c)~Trajectories $x_i(t)$ of $N=24$ particles, obtained by
  integrating Newton's equations~\eqref{eq:newton}, starting from the equilibrium
  configuration~\eqref{eq:equil}. They coincide with the exact dilation,
  \eref{eq:ansatz}. The dashed lines are the exact cloud edges
  $\pm|b(t)|\,x_{\rm{outer}}$, with $b(t)$ obtained from \eref{eq:b}. In (c) all
  particles meet at $x=0$ at the caustic times $(2n+1)\,t_c$ (stars),
  with $n=0, 1, 2, \dotsc$ and $t_c$ in \eref{eq:tc}, and re-emerge in reversed order. In (b) and (c) the particles are coloured by the sign of their initial
position, red for $x_i^0>0$ and blue for $x_i^0<0$: the colours keep
their sides while the cloud breathes (b) and swap sides at every
caustic (c).
  (d)--(f)~Density profiles for $N=512$ at the five times given in the
legends. Crosses: particle density obtained from the inverse
interparticle distance,
$\rho(x_{i+1/2},t)=1/\{N[x_{i+1}(t)-x_i(t)]\}$, with the positions
ordered and $x_{i+1/2}=(x_i+x_{i+1})/2$. Solid lines: the hydrodynamic
scaling form~\eqref{eq:selfsim} with $\lambda=b$, where $\rho_0(x)$ is the
continuum equilibrium density given in 
the Supplemental Material~\cite{SM}  with the chemical potential fixed by
normalization. All densities
are divided by the same initial central density $\rho(0,0)$ of the
particles and no fitting parameter is used. In (d) and (e) the times span
one breathing period: the profile narrows, peaks at maximal compression
(black) and widens again, as a rescaled copy of the initial profile.
In (f) the snapshots are taken before, at and after the first caustic:
the cloud collapses to $\rho\propto\delta(x)$ at $t=t_c$ (black
vertical line) and re-forms afterwards.}
\label{fig:dynamics}
\end{figure*}

We find that, the most dramatic consequence occurs for $-2<k<0$, where the pair
potential stays finite at contact and particles can pass through each other. 
For a sudden quench to a constant
frequency, $\omega(t)=\omega=\alpha\,\omega_0$, the exact solution exhibits a sharp
dynamical transition at the critical compression quench ratio $\alpha^*(k)=\sqrt{2/|k|}$ [\fref{fig:phase}]. For
$\alpha<\alpha^*$ the gas breathes periodically without ever collapsing to a single point [Fig.~\ref{fig:dynamics}~(b) and (e)]. On the other hand, 
for $\alpha>\alpha^*$, all $N$ particles arrive at a single point at the
same instant $t_c$ in time, thereby, giving rise to a perfect \emph{caustic} in which the whole cloud is focused to
a single point, and then re-emerges with the spatial order of the particles reversed [Fig.~\ref{fig:dynamics}~(c) and (f)]. Here, we recall that caustics are singularities of families of rays or classical
trajectories, where neighboring trajectories cross and the density
diverges~\cite{nye1999,berry1981, BERRY1980257, Chalker2009,  mumford2017, ODell2012, agarwal2023, kirkby2022}.  Exactly
at the threshold $\alpha=\alpha^*(k)$, the cloud reaches the focus with zero velocity along the
closed-form trajectories $x_i(t)=X_{CM}(t)+b^*(t)\,x_i^0$ with the separatrix given by
$b^*(t)=\left[\cos\!\left(\tfrac{k+2}{2}\,\alpha^* t\right)\right]^{2/(k+2)}$. For $k>0$, by contrast, the interaction energy diverges at contact and
acts as an infinite barrier against collapse. For every quench ratio
$\alpha\in (0,\infty)$ and any $N>1$, the cloud then breathes forever, periodically and
without damping, even though the system is non-integrable for
$k\neq2$ [Fig.~\ref{fig:dynamics}~(a) and (d)]. Starting from the
equilibrium crystal, the information in the initial state is thus never
lost.

\textit{Model.}--- We consider $N$ classical particles, each of mass $m$, on a
line interacting through the Riesz
potential and held by a harmonic trap whose frequency $\omega(t)$ and center
$a(t)$ can be varied arbitrarily in time~[Fig.~\ref{fig:schematic}]. The Hamiltonian is given by
\begin{equation}
H=  \sum_{i=1}^{N}\left[\frac{p_i^2}{2m} 
  +\frac{m\,\omega^2(t)}{2}\big[x_i-a(t)\big]^2\right]
  +\frac{J\,\mathrm{sgn}(k)}{2}\sum_{i\neq j}\frac{1}{|x_i-x_j|^{k}} ,
\label{eq:H}
\end{equation}
with $J>0$. The interaction is thus repulsive for every $k$, and the trap
confines the gas for any $k>-2$. Newton's equations corresponding to Eq.~\eqref{eq:H} read
\begin{equation}
m\ddot x_i=-m\,\omega^2(t)\big[x_i-a(t)\big]+F_i(\{x_j\}) ,
\label{eq:newton}
\end{equation}
with the interaction force
\begin{equation}
F_i(\{x_j\})=J|k|\sum_{j\neq i}
  \frac{\mathrm{sgn}(x_i-x_j)}{|x_i-x_j|^{k+1}} .
\label{eq:force}
\end{equation}
For $k>0$, the pair potential in~\eref{eq:H} diverges at contact and the particles keep
their order. For $-2<k<0$, the pair potential between any two particles $-J|x_i-x_j|^{|k|}$ vanishes at contact
and the pair force is integrable. Hence, 
particles can pass through each other. 

\textit{Quench protocol.}--- For $t<0$ the gas is in its
zero-temperature equilibrium in a trap of frequency $\omega_0$ centered
at the origin, $a(0^{-})=0$. All momenta are zero, and the initial positions $\{x_i^0\}$ thus satisfy
the force balance
\begin{equation}
m\,\omega_0^2\,x_i^0=F_i\big(\{x_j^0\}\big) .
\label{eq:equil}
\end{equation}
Since $\sum_iF_i=0$ from \eref{eq:force},  the crystal is centered, i.e., $\sum_ix_i^0=0$. If the initial positions are considered to be ordered $x_1^0 <x_2^0<\dotsb <x_N^0$, then by
symmetry $x_{N+1-i}^0=-x_i^0$. For large $N$, the equilibrium configuration $\{x_1^0, x_2^0, \dotsc, x_N^0\}$ can be described by a smooth density profile, which is known~\cite{agarwal2019,SM}. At $t=0^+$, the trap frequency and center are changed,
$\omega_0\to\omega(t)=\alpha(t)\omega_0$ and $a(0^-)=0\to a(t)$, with otherwise arbitrary time
dependence. 
For illustration, we consider the sudden quench:
$\omega(t)=\alpha\,\omega_0$ and $a(t)=a$, where both $\alpha$ and $a$ are constants. The cases $\alpha>1$ and $0<\alpha<1$, respectively, correspond to the compression and expansion of the gas [\fref{fig:phase}]. The setup and the protocol are sketched in~\fref{fig:schematic}.

\textit{Exact solution.}--- Two properties of the interaction force make
the problem solvable. It depends only on differences of positions, and
it is homogeneous of degree $-(k+1)$,
\begin{equation}
F_i(\{\Lambda x_j+c\})=\mathrm{sgn}(\Lambda)\,|\Lambda|^{-(k+1)}\,
F_i(\{x_j\}) ,
\label{eq:homog}
\end{equation}
for any real $\Lambda\neq0$ and any arbitrary constant $c$. Inspired by Eq.~\eqref{eq:homog}, we substitute the ansatz
\begin{equation}
x_i(t)=X_{CM}(t)+b(t)\,x_i^0
\label{eq:ansatz}
\end{equation}
in \eref{eq:newton}, and use Eqs.~\eqref{eq:equil} and
\eqref{eq:homog}, which yields~\cite{SM}
$[\ddot X_{CM}+\omega^2(X_{CM}-a)]+x_i^0\,[\ddot b+\omega^2b
-\omega_0^2\,\mathrm{sgn}(b)\,|b|^{-(k+1)}]=0$. Since the initial positions $\{x_i^0\}$ are distinct [\eref{eq:equil}], both brackets vanish separately to yield,
\begin{align}
\ddot X_{CM} &=-\omega^2(t)\,\big[X_{CM}-a(t)\big] ,
\label{eq:X}\\
\ddot b &=-\omega^2(t)\,b+\frac{\omega_0^2\,\mathrm{sgn}(b)}{|b|^{k+1}} ,
\label{eq:b}
\end{align}
with $X_{CM}(0)=\dot X_{CM}(0)=0$, $b(0)=1$ and $\dot b(0)=0$. 
Equations~\eqref{eq:ansatz}--\eqref{eq:b} describe the evolution of the positions $\{x_1(t), x_2(t), \dotsc, x_N(t)\}$ exactly for arbitrary $\omega(t)$
and $a(t)$, for any $N>1$
and every $k>-2$.

The center of mass moves as a single particle in the
trap via \eref{eq:X}. The scale factor obeys a single
nonlinear equation from which $N$, $J$, $m$ and the shape of the
crystal have naturally dropped out, and all microscopic information is only stored in the
initial configuration $\{x_i^0\}$. We remark that for $k=2$,
\eref{eq:b} reduces to the Ermakov--Pinney equation~\cite{pinney1950} and the  $k=1$  has been obtained in the context of ion chains~\cite{ralf_PRL}. 
For general $k>-2$, 
linearizing \eref{eq:b} about its minimum for constant $\omega$ gives~\cite{SM} the
universal breathing frequency $\sqrt{k+2}\,\omega$~\cite{henning2008}.

\textit{Hydrodynamics.}--- For $N\gg1$,  the gas can be described in terms a smooth density field 
$\rho(x,t)$, normalized to unity, and a velocity field $v(x,t)$. These fields 
obey the continuity and Euler equations~\cite{SM}
\begin{gather}
\partial_t\rho+\partial_x(\rho v)=0 ,
\label{eq:cont}\\
\partial_tv+v\,\partial_xv=-\omega^2(t)\big[x-a(t)\big]
  -\frac{1}{m}\,\partial_xV_\mathrm{eff} .
\label{eq:euler}
\end{gather}
Here $V_\mathrm{eff}=N^{-1}\delta\mathcal{E}_\mathrm{int}/\delta\rho$ is
the effective potential for the coarse-grained velocity field. The interaction energy is nonlocal for
$-2<k<1$, whereas for $k>1$ near neighbours dominate and the interaction energy becomes local making the
gas polytrope with the adiabatic index
$k+1$~\cite{agarwal2019,SM}:
\begin{equation}
\mathcal{E}_\mathrm{int}=
\begin{cases}
\displaystyle
 \frac{J\,\mathrm{sgn}(k)}{2}\,N^2 
\iint dx\,dy\,\frac{\rho(x)\,\rho(y)}{|x-y|^{k}} &\text{for}~ -2<k<1\, ,\\
\displaystyle
J\zeta(k)N^{k+1}\!\int dx\,[\rho(x)]^{k+1} &\text{for}~ k>1\, .
\end{cases}
\label{eq:Eint-nonlocal}
\end{equation}
The equilibrium density profile $\rho_0(x)$ corresponding to $\mathcal{E}_\mathrm{int}$ is supported in a finite interval $x \in (-x_0,x_0)$~\cite{agarwal2019}.  In both local and non-local regimes, with $\mathcal{E}_\mathrm{int}$  given in \eref{eq:Eint-nonlocal}, the scaling form
\begin{equation}
\rho(x,t)=\frac{1}{|\lambda|}\,
  \rho_0\!\left(\frac{x-X_{CM}}{|\lambda|}\right),~~
v(x,t)=\dot X_{CM}+\frac{\dot\lambda}{\lambda}\,(x-X_{CM})
\label{eq:selfsim}
\end{equation}
solves Eqs.~\eqref{eq:cont} and \eqref{eq:euler} exactly~\cite{SM}. Here $X_{CM}(t)$
is given by \eref{eq:X}, and $\lambda(t)$ obeys the same equation as
$b(t)$ [\eref{eq:b}]. The derivation uses only two facts~\cite{SM}. First,
$V_\mathrm{eff}$ scales as
$|\lambda|^{-k}V_\mathrm{eff}^0\big((x-X_{CM})/|\lambda|\big)$. Second, $V_\mathrm{eff}^0(x)=\mu-\frac{1}{2}m\omega_0^2x^2$ inside the initial support, which directly follows from Eq.~\eqref{eq:euler}. The explicit form the initial density $\rho_0(x)$ itself is never needed in the calculation. Hence $\lambda(t)=b(t)$,
and the continuum solution is the coarse-grained image of the exact
dilation of the microscopic gas. Indeed, \eref{eq:ansatz} gives the empirical
density at any time as
$\hat\rho(x,t)=|b|^{-1}\hat\rho\big((x-X_{CM})/|b|,0\big)$, for any $N$~\cite{SM}. 

\textit{Sudden quench: breathing and caustic transition.}--- 
For convenience,  we use the rescaled time $\omega_0 t\to t$ and the dimensionless quenched frequency $\alpha(t)=\omega(t)/\omega_0$.
For
constant $\alpha$ and $a$, \eref{eq:X} gives $X_{CM}(t)=a(1-\cos\alpha t)$,
and $b(t)$ in \eref{eq:b}, starting from $b(0)=1$ and $\dot{b}(0)=0$, moves in the effective potential~\cite{SM}
\begin{equation}
V(b)=\frac{\alpha^2b^2}{2}+\frac{1}{k\,|b|^{k}} ,\qquad
\dot b^2=2\big[V(1)-V(b)\big] .
\label{eq:Vb}
\end{equation}

For $k>0$, $V(b)$ diverges at $b=0$. Hence, $b(t)$ oscillates
periodically between $b_1=1$ and a second turning point $b_2$, which is always positive:  $b_2> 1$ for $\alpha<1$ and  $0<b_2<1$ for $\alpha>1$. This phenomena can be analytically illustrated with an explicit solution for $b(t)$ for the case $k=2$~\cite{SM}.

For $-2<k<0$, the potential stays finite at the origin $V(0)=0$, where it has a
cusp $V(b) \sim - |b|^{|k|}$ as $b\to 0$. The gas can therefore reach full collapse, $b(t)=0$ at a finite time $t=t_c$, if
and only if the energy injected by the compression quench exceeds this barrier,
$V(1)=\alpha^2/2-1/|k|>0$, which corresponds to [see \fref{fig:phase}]
\begin{equation}
\alpha>\alpha^*(k)=\sqrt{2/|k|} .
\label{eq:alphastar}
\end{equation}
The threshold $\alpha^*(k)$ decreases from $\infty$ to $1$ as $k$ decreases from
$0^-$ to $-2^+$. 
This phenomena can be analytically demonstrated explicitly for the case $k=-1$~\cite{SM}.

The critical time $t_c$ can be found by integrating \eqref{eq:Vb} as~\cite{SM} 
\begin{equation}
t_c=\int_0^1\frac{db}{\sqrt{2[V(1)-V(b)]}} ,\qquad
|\dot b(t_c)|=\sqrt{\alpha^2-\alpha^{*2}} .
\label{eq:tc}
\end{equation}
All $N$ particles reach the point $X_{CM}(t_c)$ at the same
instant of time $t=t_c$, with finite velocities $\dot b(t_c)\,x_i^0$ relative to the
center of mass, and pass through it. This is a perfect caustic. Because
$V(b)$ in \eref{eq:Vb} is even, $b(t)$ continues to the second turning point $b_2=-1$ and returns with the period $4t_c$. This periodic motion between the two turning points $\pm 1$ continues forever, with caustics occurring at times $t_n=(2n+1)\,t_c$ with $n=0, 1, 2, \dotsc$. Between successive caustics $b(t)$ alternates its sign, thereby, switching the order of particles and the cloud
re-emerges as its own mirror image. Below the threshold $\alpha < \alpha^*(k)$, $b(t)$ breathes between the turning points $b_1=1$ and $b_2>0$ where the gas reaches the maximal compression. As $\alpha\to\alpha^{*-}$, $b_2\to 0$ as   
$b_2\simeq(1-\alpha^2/\alpha^{*2})^{1/|k|}$. Figure~\ref{fig:dynamics} shows both regimes $k>0$ and $-2<k<0$ (the upper and lower rows, respectively, show the particle trajectories and the density profile evolution). For $k=1$ and for $k=-1/2$ (below the threshold), the cloud breathes without collapsing to a single point [\fref{fig:dynamics} (a) and (b) respectively]. Above the threshold, the gas collapses into a perfect caustic and re-forms in reversed order [\fref{fig:dynamics} (c)]. In all cases the particles follow \eref{eq:ansatz}, and the density remains self-similar [\eref{eq:selfsim}].

Exactly at the  threshold $\alpha=\alpha^*(k)$ [see \fref{fig:phase}], \eref{eq:Vb} admits the solution~\cite{SM}
\begin{equation}
b^*(t)=\Big[\cos\Big(\frac{\pi t}{2 t_c}\Big)\Big]^{2/(2+k)}\, ,\quad 0<t<t_c ,
\label{eq:separatrix}
\end{equation}
where $t_c=\pi/[(2+k)\,\alpha^*]$.  Evidently, at $t=t_c$, $b^*(t_c)=0$ and $\dot{b}^*(t_c)=0$. 
Therefore, on the critical line $\alpha=\alpha^*(k)$ [\fref{fig:phase}], the cloud reaches the full
collapse state $b^*(t_c) =0$ in a finite time $t_c$ with zero velocity $\dot{b}^*(t_c)=0$, along the separatrix
\eref{eq:separatrix}. In the vicinity of this point, the force in \eref{eq:b} is proportional to
$\mathrm{sgn}(b^*)\,|b^*|^{|k|-1}$, which is not Lipschitz continuous. Hence, 
Picard's theorem no longer guarantees a unique
solution~\cite{simmons2016,coddington1955,teschl2012}. For $-1\le k<0$, the force at $b=0$ is singular ($-1<k<0$) or
discontinuous ($k=-1$), and the cloud re-emerges at once, either
bouncing back (for $\alpha\to\alpha^{*-}$) or passing through (for $\alpha\to\alpha^{*+}$), with the same
density in both cases. For $-2<k<-1$ the force is continuous and
vanishes at $b=0$, so the collapsed cloud is an exact unstable
equilibrium. Solutions then still exist, by Peano's
theorem~\cite{coddington1955,teschl2012}, but they are not unique.
Besides re-emerging immediately, the cloud may remain collapsed for an
arbitrary time before re-emerging. The continuation beyond $t_c$ is
therefore not fixed by \eref{eq:b} alone, and it is selected by any
short-distance regularization of the interaction or any perturbation of
the initial state.

\textit{Conclusions and outlook.}--- We have shown that the
harmonically confined Riesz gas, prepared in its zero-temperature
equilibrium crystal state at trap frequency $\omega_0$, evolves exactly by a rigid translation and a
uniform dilation under arbitrary time dependence of the trap frequency $\alpha(t) \omega_0$ and
center $a(t)$, for any particle number $N>1$ and every exponent $k>-2$. The
$N$-body problem thus reduces to two independent ordinary differential
equations, one for the center of mass $X_{CM}(t)$ and another for the common scale factor $b(t)$. The continuum hydrodynamics in both its nonlocal ($k<1$) and local
($k>1$) regimes admits a self-similar solution where the time dependent scale factor and the shift obey the same dynamics as the individual particle trajectories. 
For a sudden quench to a constant frequency $\omega_0\to \alpha \omega_0$, the system exhibits rich behavior in the  $(\alpha,k)$ plane
[\fref{fig:phase}]. For $k>0$ the cloud breathes forever without
damping or equilibrating, even though the dynamics is non-integrable for $k\neq2$. For
$-2<k<0$, a compression of the trap beyond $\alpha^*(k)=\sqrt{2/|k|}$ causes all
particle trajectories to focus at the single point $X_{CM}(t_c)$, at the given instant $t_c$, after which the
cloud re-emerges as its own mirror image. This process of focusing and re-emerging continues periodically. The critical
line $\alpha=\alpha^*(k)$ in \fref{fig:phase} itself is a separatrix, on which the continuation through the
focus is not unique.

For any prescribed $b(t)$, inverting \eref{eq:b}
gives the exact trap protocol
$\omega^2(t)=[\omega_0^2\,\mathrm{sgn}(b)\,|b|^{-(k+1)}-\ddot b\,]/b$.
This provides shortcuts to adiabaticity for ion crystals of arbitrary
size, beyond normal-mode
approximations~\cite{palmero2015,guery2019}. Natural next steps are understanding the fate of both the breathing
and the caustic at finite temperature and in the quantum regime~\cite{OlivettiPRL,abraham2014,abraham2014review}. Quenching of trap parameters are routine in
trapped-ion~\cite{pyka2013,ulm2013} and ultracold-atom experiments~\cite{moritz2003,fang2014}, where the cloud size $b(t)$
is directly measurable. The nonlinear breathing can thus be tested in
ion chains ($k=1$) and dipolar gases ($k=3$), while the caustic regime
calls for platforms with $k<0$, such as the one-dimensional
one-component plasma. Our construction in one dimension trivially carries over directly to isotropic traps in
any dimension~\cite{CHAFAI2022126367,BYUN2026130813}. 

\textit{Acknowledgements.---} MK acknowledges the support of the Department of Atomic Energy, Government of India, under project
no. RTI4001. This research was supported in part by the International Centre for Theoretical Sciences (ICTS) for participating in the program - Two and a half decades of the Macroscopic Fluctuation Theory 2026 (code: ICTS/MFTHEORY2026/09).

\bibliography{references}

\begin{thebibliography}{70}%
\makeatletter
\providecommand \@ifxundefined [1]{%
 \@ifx{#1\undefined}
}%
\providecommand \@ifnum [1]{%
 \ifnum #1\expandafter \@firstoftwo
 \else \expandafter \@secondoftwo
 \fi
}%
\providecommand \@ifx [1]{%
 \ifx #1\expandafter \@firstoftwo
 \else \expandafter \@secondoftwo
 \fi
}%
\providecommand \natexlab [1]{#1}%
\providecommand \enquote  [1]{``#1''}%
\providecommand \bibnamefont  [1]{#1}%
\providecommand \bibfnamefont [1]{#1}%
\providecommand \citenamefont [1]{#1}%
\providecommand \href@noop [0]{\@secondoftwo}%
\providecommand \href [0]{\begingroup \@sanitize@url \@href}%
\providecommand \@href[1]{\@@startlink{#1}\@@href}%
\providecommand \@@href[1]{\endgroup#1\@@endlink}%
\providecommand \@sanitize@url [0]{\catcode `\\12\catcode `\$12\catcode
  `\&12\catcode `\#12\catcode `\^12\catcode `\_12\catcode `\%12\relax}%
\providecommand \@@startlink[1]{}%
\providecommand \@@endlink[0]{}%
\providecommand \url  [0]{\begingroup\@sanitize@url \@url }%
\providecommand \@url [1]{\endgroup\@href {#1}{\urlprefix }}%
\providecommand \urlprefix  [0]{URL }%
\providecommand \Eprint [0]{\href }%
\providecommand \doibase [0]{https://doi.org/}%
\providecommand \selectlanguage [0]{\@gobble}%
\providecommand \bibinfo  [0]{\@secondoftwo}%
\providecommand \bibfield  [0]{\@secondoftwo}%
\providecommand \translation [1]{[#1]}%
\providecommand \BibitemOpen [0]{}%
\providecommand \bibitemStop [0]{}%
\providecommand \bibitemNoStop [0]{.\EOS\space}%
\providecommand \EOS [0]{\spacefactor3000\relax}%
\providecommand \BibitemShut  [1]{\csname bibitem#1\endcsname}%
\let\auto@bib@innerbib\@empty
\bibitem [{\citenamefont {Polkovnikov}\ \emph {et~al.}(2011)\citenamefont
  {Polkovnikov}, \citenamefont {Sengupta}, \citenamefont {Silva},\ and\
  \citenamefont {Vengalattore}}]{polkovnikov2011}%
  \BibitemOpen
  \bibfield  {author} {\bibinfo {author} {\bibfnamefont {A.}~\bibnamefont
  {Polkovnikov}}, \bibinfo {author} {\bibfnamefont {K.}~\bibnamefont
  {Sengupta}}, \bibinfo {author} {\bibfnamefont {A.}~\bibnamefont {Silva}},\
  and\ \bibinfo {author} {\bibfnamefont {M.}~\bibnamefont {Vengalattore}},\
  }\bibfield  {title} {\bibinfo {title} {Colloquium: Nonequilibrium dynamics of
  closed interacting quantum systems},\ }\href
  {https://doi.org/10.1103/RevModPhys.83.863} {\bibfield  {journal} {\bibinfo
  {journal} {Rev. Mod. Phys.}\ }\textbf {\bibinfo {volume} {83}},\ \bibinfo
  {pages} {863} (\bibinfo {year} {2011})}\BibitemShut {NoStop}%
\bibitem [{\citenamefont {D'Alessio}\ \emph {et~al.}(2016)\citenamefont
  {D'Alessio}, \citenamefont {Kafri}, \citenamefont {Polkovnikov},\ and\
  \citenamefont {Rigol}}]{dalessio2016}%
  \BibitemOpen
  \bibfield  {author} {\bibinfo {author} {\bibfnamefont {L.}~\bibnamefont
  {D'Alessio}}, \bibinfo {author} {\bibfnamefont {Y.}~\bibnamefont {Kafri}},
  \bibinfo {author} {\bibfnamefont {A.}~\bibnamefont {Polkovnikov}},\ and\
  \bibinfo {author} {\bibfnamefont {M.}~\bibnamefont {Rigol}},\ }\bibfield
  {title} {\bibinfo {title} {From quantum chaos and eigenstate thermalization
  to statistical mechanics and thermodynamics},\ }\href
  {https://doi.org/10.1080/00018732.2016.1198134} {\bibfield  {journal}
  {\bibinfo  {journal} {Adv. Phys.}\ }\textbf {\bibinfo {volume} {65}},\
  \bibinfo {pages} {239} (\bibinfo {year} {2016})}\BibitemShut {NoStop}%
\bibitem [{\citenamefont {Eisert}\ \emph {et~al.}(2015)\citenamefont {Eisert},
  \citenamefont {Friesdorf},\ and\ \citenamefont {Gogolin}}]{eisert2015}%
  \BibitemOpen
  \bibfield  {author} {\bibinfo {author} {\bibfnamefont {J.}~\bibnamefont
  {Eisert}}, \bibinfo {author} {\bibfnamefont {M.}~\bibnamefont {Friesdorf}},\
  and\ \bibinfo {author} {\bibfnamefont {C.}~\bibnamefont {Gogolin}},\
  }\bibfield  {title} {\bibinfo {title} {Quantum many-body systems out of
  equilibrium},\ }\href {https://doi.org/10.1038/nphys3215} {\bibfield
  {journal} {\bibinfo  {journal} {Nat. Phys.}\ }\textbf {\bibinfo {volume}
  {11}},\ \bibinfo {pages} {124} (\bibinfo {year} {2015})}\BibitemShut
  {NoStop}%
\bibitem [{\citenamefont {{Calogero}}(1971)}]{calogero1971}%
  \BibitemOpen
  \bibfield  {author} {\bibinfo {author} {\bibfnamefont {F.}~\bibnamefont
  {{Calogero}}},\ }\bibfield  {title} {\bibinfo {title} {Solution of the
  one-dimensional $n$-body problems with quadratic and/or inversely quadratic
  pair potentials},\ }\href {https://doi.org/10.1063/1.1665604} {\bibfield
  {journal} {\bibinfo  {journal} {J. Math. Phys.}\ }\textbf {\bibinfo {volume}
  {12}},\ \bibinfo {pages} {419} (\bibinfo {year} {1971})}\BibitemShut
  {NoStop}%
\bibitem [{\citenamefont {Moser}(1975)}]{moser1975}%
  \BibitemOpen
  \bibfield  {author} {\bibinfo {author} {\bibfnamefont {J.}~\bibnamefont
  {Moser}},\ }\bibfield  {title} {\bibinfo {title} {Three integrable
  {H}amiltonian systems connected with isospectral deformations},\ }\href
  {https://doi.org/10.1016/0001-8708(75)90151-6} {\bibfield  {journal}
  {\bibinfo  {journal} {Adv. Math.}\ }\textbf {\bibinfo {volume} {16}},\
  \bibinfo {pages} {197} (\bibinfo {year} {1975})}\BibitemShut {NoStop}%
\bibitem [{\citenamefont {Perelomov}(1990)}]{perelomov1990}%
  \BibitemOpen
  \bibfield  {author} {\bibinfo {author} {\bibfnamefont {A.~M.}\ \bibnamefont
  {Perelomov}},\ }\href {https://doi.org/10.1007/978-3-0348-9257-5} {\emph
  {\bibinfo {title} {Integrable Systems of Classical Mechanics and {L}ie
  Algebras}}},\ Vol.~\bibinfo {volume} {I}\ (\bibinfo  {publisher}
  {Birkh{\"a}user},\ \bibinfo {address} {Basel},\ \bibinfo {year}
  {1990})\BibitemShut {NoStop}%
\bibitem [{\citenamefont {Polychronakos}(2006)}]{ap_review}%
  \BibitemOpen
  \bibfield  {author} {\bibinfo {author} {\bibfnamefont {A.~P.}\ \bibnamefont
  {Polychronakos}},\ }\bibfield  {title} {\bibinfo {title} {The physics and
  mathematics of calogero particles},\ }\href
  {https://doi.org/10.1088/0305-4470/39/41/S07} {\bibfield  {journal} {\bibinfo
   {journal} {J. Phys. A: Math. Gen.}\ }\textbf {\bibinfo {volume} {39}},\
  \bibinfo {pages} {12793} (\bibinfo {year} {2006})}\BibitemShut {NoStop}%
\bibitem [{\citenamefont {Castro-Alvaredo}\ \emph {et~al.}(2016)\citenamefont
  {Castro-Alvaredo}, \citenamefont {Doyon},\ and\ \citenamefont
  {Yoshimura}}]{castroalvaredo2016}%
  \BibitemOpen
  \bibfield  {author} {\bibinfo {author} {\bibfnamefont {O.~A.}\ \bibnamefont
  {Castro-Alvaredo}}, \bibinfo {author} {\bibfnamefont {B.}~\bibnamefont
  {Doyon}},\ and\ \bibinfo {author} {\bibfnamefont {T.}~\bibnamefont
  {Yoshimura}},\ }\bibfield  {title} {\bibinfo {title} {Emergent hydrodynamics
  in integrable quantum systems out of equilibrium},\ }\href
  {https://doi.org/10.1103/PhysRevX.6.041065} {\bibfield  {journal} {\bibinfo
  {journal} {Phys. Rev. X}\ }\textbf {\bibinfo {volume} {6}},\ \bibinfo {pages}
  {041065} (\bibinfo {year} {2016})}\BibitemShut {NoStop}%
\bibitem [{\citenamefont {Bertini}\ \emph {et~al.}(2016)\citenamefont
  {Bertini}, \citenamefont {Collura}, \citenamefont {De~Nardis},\ and\
  \citenamefont {Fagotti}}]{bertini2016}%
  \BibitemOpen
  \bibfield  {author} {\bibinfo {author} {\bibfnamefont {B.}~\bibnamefont
  {Bertini}}, \bibinfo {author} {\bibfnamefont {M.}~\bibnamefont {Collura}},
  \bibinfo {author} {\bibfnamefont {J.}~\bibnamefont {De~Nardis}},\ and\
  \bibinfo {author} {\bibfnamefont {M.}~\bibnamefont {Fagotti}},\ }\bibfield
  {title} {\bibinfo {title} {Transport in out-of-equilibrium {XXZ} chains:
  Exact profiles of charges and currents},\ }\href
  {https://doi.org/10.1103/PhysRevLett.117.207201} {\bibfield  {journal}
  {\bibinfo  {journal} {Phys. Rev. Lett.}\ }\textbf {\bibinfo {volume} {117}},\
  \bibinfo {pages} {207201} (\bibinfo {year} {2016})}\BibitemShut {NoStop}%
\bibitem [{\citenamefont {Kinoshita}\ \emph {et~al.}(2006)\citenamefont
  {Kinoshita}, \citenamefont {Wenger},\ and\ \citenamefont
  {Weiss}}]{kinoshita2006}%
  \BibitemOpen
  \bibfield  {author} {\bibinfo {author} {\bibfnamefont {T.}~\bibnamefont
  {Kinoshita}}, \bibinfo {author} {\bibfnamefont {T.}~\bibnamefont {Wenger}},\
  and\ \bibinfo {author} {\bibfnamefont {D.~S.}\ \bibnamefont {Weiss}},\
  }\bibfield  {title} {\bibinfo {title} {A quantum newton's cradle},\ }\href
  {https://doi.org/10.1038/nature04693} {\bibfield  {journal} {\bibinfo
  {journal} {Nature}\ }\textbf {\bibinfo {volume} {440}},\ \bibinfo {pages}
  {900} (\bibinfo {year} {2006})}\BibitemShut {NoStop}%
\bibitem [{\citenamefont {Doyon}(2020)}]{doyon2020}%
  \BibitemOpen
  \bibfield  {author} {\bibinfo {author} {\bibfnamefont {B.}~\bibnamefont
  {Doyon}},\ }\bibfield  {title} {\bibinfo {title} {Lecture notes on
  generalised hydrodynamics},\ }\href
  {https://doi.org/10.21468/SciPostPhysLectNotes.18} {\bibfield  {journal}
  {\bibinfo  {journal} {SciPost Phys. Lect. Notes}\ }\textbf {\bibinfo {volume}
  {18}} (\bibinfo {year} {2020})}\BibitemShut {NoStop}%
\bibitem [{\citenamefont {Pitaevskii}\ and\ \citenamefont
  {Rosch}(1997)}]{pitaevskii1997}%
  \BibitemOpen
  \bibfield  {author} {\bibinfo {author} {\bibfnamefont {L.~P.}\ \bibnamefont
  {Pitaevskii}}\ and\ \bibinfo {author} {\bibfnamefont {A.}~\bibnamefont
  {Rosch}},\ }\bibfield  {title} {\bibinfo {title} {Breathing modes and hidden
  symmetry of trapped atoms in two dimensions},\ }\href
  {https://doi.org/10.1103/PhysRevA.55.R853} {\bibfield  {journal} {\bibinfo
  {journal} {Phys. Rev. A}\ }\textbf {\bibinfo {volume} {55}},\ \bibinfo
  {pages} {R853} (\bibinfo {year} {1997})}\BibitemShut {NoStop}%
\bibitem [{\citenamefont {Castin}\ and\ \citenamefont
  {Dum}(1996)}]{castin1996}%
  \BibitemOpen
  \bibfield  {author} {\bibinfo {author} {\bibfnamefont {Y.}~\bibnamefont
  {Castin}}\ and\ \bibinfo {author} {\bibfnamefont {R.}~\bibnamefont {Dum}},\
  }\bibfield  {title} {\bibinfo {title} {{B}ose--{E}instein condensates in time
  dependent traps},\ }\href {https://doi.org/10.1103/PhysRevLett.77.5315}
  {\bibfield  {journal} {\bibinfo  {journal} {Phys. Rev. Lett.}\ }\textbf
  {\bibinfo {volume} {77}},\ \bibinfo {pages} {5315} (\bibinfo {year}
  {1996})}\BibitemShut {NoStop}%
\bibitem [{\citenamefont {Kagan}\ \emph {et~al.}(1996)\citenamefont {Kagan},
  \citenamefont {Surkov},\ and\ \citenamefont {Shlyapnikov}}]{kagan1996}%
  \BibitemOpen
  \bibfield  {author} {\bibinfo {author} {\bibfnamefont {Y.}~\bibnamefont
  {Kagan}}, \bibinfo {author} {\bibfnamefont {E.~L.}\ \bibnamefont {Surkov}},\
  and\ \bibinfo {author} {\bibfnamefont {G.~V.}\ \bibnamefont {Shlyapnikov}},\
  }\bibfield  {title} {\bibinfo {title} {Evolution of a {B}ose-condensed gas
  under variations of the confining potential},\ }\href
  {https://doi.org/10.1103/PhysRevA.54.R1753} {\bibfield  {journal} {\bibinfo
  {journal} {Phys. Rev. A}\ }\textbf {\bibinfo {volume} {54}},\ \bibinfo
  {pages} {R1753} (\bibinfo {year} {1996})}\BibitemShut {NoStop}%
\bibitem [{\citenamefont {Tomza}\ \emph {et~al.}(2019)\citenamefont {Tomza},
  \citenamefont {Jachymski}, \citenamefont {Gerritsma}, \citenamefont
  {Negretti}, \citenamefont {Calarco}, \citenamefont {Idziaszek},\ and\
  \citenamefont {Julienne}}]{rmp0}%
  \BibitemOpen
  \bibfield  {author} {\bibinfo {author} {\bibfnamefont {M.}~\bibnamefont
  {Tomza}}, \bibinfo {author} {\bibfnamefont {K.}~\bibnamefont {Jachymski}},
  \bibinfo {author} {\bibfnamefont {R.}~\bibnamefont {Gerritsma}}, \bibinfo
  {author} {\bibfnamefont {A.}~\bibnamefont {Negretti}}, \bibinfo {author}
  {\bibfnamefont {T.}~\bibnamefont {Calarco}}, \bibinfo {author} {\bibfnamefont
  {Z.}~\bibnamefont {Idziaszek}},\ and\ \bibinfo {author} {\bibfnamefont
  {P.~S.}\ \bibnamefont {Julienne}},\ }\bibfield  {title} {\bibinfo {title}
  {Cold hybrid ion-atom systems},\ }\href
  {https://doi.org/10.1103/RevModPhys.91.035001} {\bibfield  {journal}
  {\bibinfo  {journal} {Rev. Mod. Phys.}\ }\textbf {\bibinfo {volume} {91}},\
  \bibinfo {pages} {035001} (\bibinfo {year} {2019})}\BibitemShut {NoStop}%
\bibitem [{\citenamefont {Monroe}\ \emph {et~al.}(2021)\citenamefont {Monroe},
  \citenamefont {Campbell}, \citenamefont {Duan}, \citenamefont {Gong},
  \citenamefont {Gorshkov}, \citenamefont {Hess}, \citenamefont {Islam},
  \citenamefont {Kim}, \citenamefont {Linke}, \citenamefont {Pagano},
  \citenamefont {Richerme}, \citenamefont {Senko},\ and\ \citenamefont
  {Yao}}]{rmp1}%
  \BibitemOpen
  \bibfield  {author} {\bibinfo {author} {\bibfnamefont {C.}~\bibnamefont
  {Monroe}}, \bibinfo {author} {\bibfnamefont {W.~C.}\ \bibnamefont
  {Campbell}}, \bibinfo {author} {\bibfnamefont {L.-M.}\ \bibnamefont {Duan}},
  \bibinfo {author} {\bibfnamefont {Z.-X.}\ \bibnamefont {Gong}}, \bibinfo
  {author} {\bibfnamefont {A.~V.}\ \bibnamefont {Gorshkov}}, \bibinfo {author}
  {\bibfnamefont {P.~W.}\ \bibnamefont {Hess}}, \bibinfo {author}
  {\bibfnamefont {R.}~\bibnamefont {Islam}}, \bibinfo {author} {\bibfnamefont
  {K.}~\bibnamefont {Kim}}, \bibinfo {author} {\bibfnamefont {N.~M.}\
  \bibnamefont {Linke}}, \bibinfo {author} {\bibfnamefont {G.}~\bibnamefont
  {Pagano}}, \bibinfo {author} {\bibfnamefont {P.}~\bibnamefont {Richerme}},
  \bibinfo {author} {\bibfnamefont {C.}~\bibnamefont {Senko}},\ and\ \bibinfo
  {author} {\bibfnamefont {N.~Y.}\ \bibnamefont {Yao}},\ }\bibfield  {title}
  {\bibinfo {title} {Programmable quantum simulations of spin systems with
  trapped ions},\ }\href {https://doi.org/10.1103/RevModPhys.93.025001}
  {\bibfield  {journal} {\bibinfo  {journal} {Rev. Mod. Phys.}\ }\textbf
  {\bibinfo {volume} {93}},\ \bibinfo {pages} {025001} (\bibinfo {year}
  {2021})}\BibitemShut {NoStop}%
\bibitem [{\citenamefont {Morigi}\ \emph {et~al.}(2026)\citenamefont {Morigi},
  \citenamefont {Bollinger}, \citenamefont {Drewsen}, \citenamefont
  {Podolsky},\ and\ \citenamefont {Shimshoni}}]{rmp2}%
  \BibitemOpen
  \bibfield  {author} {\bibinfo {author} {\bibfnamefont {G.}~\bibnamefont
  {Morigi}}, \bibinfo {author} {\bibfnamefont {J.}~\bibnamefont {Bollinger}},
  \bibinfo {author} {\bibfnamefont {M.}~\bibnamefont {Drewsen}}, \bibinfo
  {author} {\bibfnamefont {D.}~\bibnamefont {Podolsky}},\ and\ \bibinfo
  {author} {\bibfnamefont {E.}~\bibnamefont {Shimshoni}},\ }\bibfield  {title}
  {\bibinfo {title} {Ion coulomb crystals: An exotic form of condensed
  matter},\ }\href {https://doi.org/10.1103/tn8k-r4w8} {\bibfield  {journal}
  {\bibinfo  {journal} {Rev. Mod. Phys.}\ }\textbf {\bibinfo {volume} {98}},\
  \bibinfo {pages} {035001} (\bibinfo {year} {2026})}\BibitemShut {NoStop}%
\bibitem [{\citenamefont {Campa}\ \emph {et~al.}(2009)\citenamefont {Campa},
  \citenamefont {Dauxois},\ and\ \citenamefont {Ruffo}}]{campa2009}%
  \BibitemOpen
  \bibfield  {author} {\bibinfo {author} {\bibfnamefont {A.}~\bibnamefont
  {Campa}}, \bibinfo {author} {\bibfnamefont {T.}~\bibnamefont {Dauxois}},\
  and\ \bibinfo {author} {\bibfnamefont {S.}~\bibnamefont {Ruffo}},\ }\bibfield
   {title} {\bibinfo {title} {Statistical mechanics and dynamics of solvable
  models with long-range interactions},\ }\href
  {https://doi.org/10.1016/j.physrep.2009.07.001} {\bibfield  {journal}
  {\bibinfo  {journal} {Phys. Rep.}\ }\textbf {\bibinfo {volume} {480}},\
  \bibinfo {pages} {57} (\bibinfo {year} {2009})}\BibitemShut {NoStop}%
\bibitem [{\citenamefont {Campa}\ \emph {et~al.}(2014)\citenamefont {Campa},
  \citenamefont {Dauxois}, \citenamefont {Fanelli},\ and\ \citenamefont
  {Ruffo}}]{campa2014book}%
  \BibitemOpen
  \bibfield  {author} {\bibinfo {author} {\bibfnamefont {A.}~\bibnamefont
  {Campa}}, \bibinfo {author} {\bibfnamefont {T.}~\bibnamefont {Dauxois}},
  \bibinfo {author} {\bibfnamefont {D.}~\bibnamefont {Fanelli}},\ and\ \bibinfo
  {author} {\bibfnamefont {S.}~\bibnamefont {Ruffo}},\ }\href@noop {} {\emph
  {\bibinfo {title} {Physics of Long-Range Interacting Systems}}}\ (\bibinfo
  {publisher} {Oxford University Press},\ \bibinfo {address} {Oxford},\
  \bibinfo {year} {2014})\BibitemShut {NoStop}%
\bibitem [{\citenamefont {Defenu}\ \emph {et~al.}(2023)\citenamefont {Defenu},
  \citenamefont {Donner}, \citenamefont {Macr{\`{\i}}}, \citenamefont {Pagano},
  \citenamefont {Ruffo},\ and\ \citenamefont {Trombettoni}}]{defenu}%
  \BibitemOpen
  \bibfield  {author} {\bibinfo {author} {\bibfnamefont {N.}~\bibnamefont
  {Defenu}}, \bibinfo {author} {\bibfnamefont {T.}~\bibnamefont {Donner}},
  \bibinfo {author} {\bibfnamefont {T.}~\bibnamefont {Macr{\`{\i}}}}, \bibinfo
  {author} {\bibfnamefont {G.}~\bibnamefont {Pagano}}, \bibinfo {author}
  {\bibfnamefont {S.}~\bibnamefont {Ruffo}},\ and\ \bibinfo {author}
  {\bibfnamefont {A.}~\bibnamefont {Trombettoni}},\ }\bibfield  {title}
  {\bibinfo {title} {Long-range interacting quantum systems},\ }\href
  {https://doi.org/10.1103/RevModPhys.95.035002} {\bibfield  {journal}
  {\bibinfo  {journal} {Rev. Mod. Phys.}\ }\textbf {\bibinfo {volume} {95}},\
  \bibinfo {pages} {035002} (\bibinfo {year} {2023})}\BibitemShut {NoStop}%
\bibitem [{\citenamefont {Riesz}(1938)}]{riesz1938}%
  \BibitemOpen
  \bibfield  {author} {\bibinfo {author} {\bibfnamefont {M.}~\bibnamefont
  {Riesz}},\ }\bibfield  {title} {\bibinfo {title} {Int{\'e}grales de
  {R}iemann--{L}iouville et potentiels},\ }\href@noop {} {\bibfield  {journal}
  {\bibinfo  {journal} {Acta Sci. Math. (Szeged)}\ }\textbf {\bibinfo {volume}
  {9}},\ \bibinfo {pages} {1} (\bibinfo {year} {1938})}\BibitemShut {NoStop}%
\bibitem [{\citenamefont {Lewin}(2022)}]{lewin2022}%
  \BibitemOpen
  \bibfield  {author} {\bibinfo {author} {\bibfnamefont {M.}~\bibnamefont
  {Lewin}},\ }\bibfield  {title} {\bibinfo {title} {Coulomb and {R}iesz gases:
  The known and the unknown},\ }\href {https://doi.org/10.1063/5.0086835}
  {\bibfield  {journal} {\bibinfo  {journal} {J. Math. Phys.}\ }\textbf
  {\bibinfo {volume} {63}},\ \bibinfo {pages} {061101} (\bibinfo {year}
  {2022})}\BibitemShut {NoStop}%
\bibitem [{\citenamefont {Serfaty}(2024)}]{serfaty2024}%
  \BibitemOpen
  \bibfield  {author} {\bibinfo {author} {\bibfnamefont {S.}~\bibnamefont
  {Serfaty}},\ }\href@noop {} {\bibinfo {title} {Lectures on {C}oulomb and
  {R}iesz gases}} (\bibinfo {year} {2024}),\ \Eprint
  {https://arxiv.org/abs/2407.21194} {arXiv:2407.21194 [math-ph]} \BibitemShut
  {NoStop}%
\bibitem [{\citenamefont {Choquard}\ \emph {et~al.}(1981)\citenamefont
  {Choquard}, \citenamefont {Kunz}, \citenamefont {Martin},\ and\ \citenamefont
  {Navet}}]{ocp1}%
  \BibitemOpen
  \bibfield  {author} {\bibinfo {author} {\bibfnamefont {P.}~\bibnamefont
  {Choquard}}, \bibinfo {author} {\bibfnamefont {H.}~\bibnamefont {Kunz}},
  \bibinfo {author} {\bibfnamefont {P.~A.}\ \bibnamefont {Martin}},\ and\
  \bibinfo {author} {\bibfnamefont {M.}~\bibnamefont {Navet}},\ }\bibfield
  {title} {\bibinfo {title} {One-dimensional {C}oulomb systems},\ }in\
  \href@noop {} {\emph {\bibinfo {booktitle} {Physics in One Dimension}}},\
  \bibinfo {series} {Springer Series in Solid-State Sciences}, Vol.~\bibinfo
  {volume} {23},\ \bibinfo {editor} {edited by\ \bibinfo {editor}
  {\bibfnamefont {J.}~\bibnamefont {Bernasconi}}\ and\ \bibinfo {editor}
  {\bibfnamefont {T.}~\bibnamefont {Schneider}}}\ (\bibinfo  {publisher}
  {Springer},\ \bibinfo {address} {Berlin, Heidelberg},\ \bibinfo {year}
  {1981})\ pp.\ \bibinfo {pages} {335--350}\BibitemShut {NoStop}%
\bibitem [{\citenamefont {Lenard}(1961)}]{ocp2}%
  \BibitemOpen
  \bibfield  {author} {\bibinfo {author} {\bibfnamefont {A.}~\bibnamefont
  {Lenard}},\ }\bibfield  {title} {\bibinfo {title} {Exact statistical
  mechanics of a one-dimensional system with {C}oulomb forces},\ }\href
  {https://doi.org/10.1063/1.1703757} {\bibfield  {journal} {\bibinfo
  {journal} {J. Math. Phys.}\ }\textbf {\bibinfo {volume} {2}},\ \bibinfo
  {pages} {682} (\bibinfo {year} {1961})}\BibitemShut {NoStop}%
\bibitem [{\citenamefont {Prager}(1962)}]{ocp3}%
  \BibitemOpen
  \bibfield  {author} {\bibinfo {author} {\bibfnamefont {S.}~\bibnamefont
  {Prager}},\ }\bibfield  {title} {\bibinfo {title} {The one-dimensional
  plasma},\ }in\ \href {https://doi.org/10.1002/9780470143506.ch5} {\emph
  {\bibinfo {booktitle} {Advances in Chemical Physics}}},\ Vol.~\bibinfo
  {volume} {4},\ \bibinfo {editor} {edited by\ \bibinfo {editor} {\bibfnamefont
  {I.}~\bibnamefont {Prigogine}}}\ (\bibinfo  {publisher} {John Wiley \&
  Sons},\ \bibinfo {address} {Hoboken, NJ},\ \bibinfo {year} {1962})\ p.\
  \bibinfo {pages} {201}\BibitemShut {NoStop}%
\bibitem [{\citenamefont {Baxter}(1963)}]{ocp4}%
  \BibitemOpen
  \bibfield  {author} {\bibinfo {author} {\bibfnamefont {R.~J.}\ \bibnamefont
  {Baxter}},\ }\bibfield  {title} {\bibinfo {title} {Statistical mechanics of a
  one-dimensional {C}oulomb system with a uniform charge background},\ }\href
  {https://doi.org/10.1017/S0305004100003790} {\bibfield  {journal} {\bibinfo
  {journal} {Proc. Cambridge Philos. Soc.}\ }\textbf {\bibinfo {volume} {59}},\
  \bibinfo {pages} {779} (\bibinfo {year} {1963})}\BibitemShut {NoStop}%
\bibitem [{\citenamefont {T{\'e}llez}\ and\ \citenamefont
  {Trizac}(2015)}]{ocp6}%
  \BibitemOpen
  \bibfield  {author} {\bibinfo {author} {\bibfnamefont {G.}~\bibnamefont
  {T{\'e}llez}}\ and\ \bibinfo {author} {\bibfnamefont {E.}~\bibnamefont
  {Trizac}},\ }\bibfield  {title} {\bibinfo {title} {Screening like charges in
  one-dimensional {C}oulomb systems: Exact results},\ }\href
  {https://doi.org/10.1103/PhysRevE.92.042134} {\bibfield  {journal} {\bibinfo
  {journal} {Phys. Rev. E}\ }\textbf {\bibinfo {volume} {92}},\ \bibinfo
  {pages} {042134} (\bibinfo {year} {2015})}\BibitemShut {NoStop}%
\bibitem [{\citenamefont {Dhar}\ \emph {et~al.}(2017)\citenamefont {Dhar},
  \citenamefont {Kundu}, \citenamefont {Majumdar}, \citenamefont
  {Sabhapandit},\ and\ \citenamefont {Schehr}}]{dhar2017}%
  \BibitemOpen
  \bibfield  {author} {\bibinfo {author} {\bibfnamefont {A.}~\bibnamefont
  {Dhar}}, \bibinfo {author} {\bibfnamefont {A.}~\bibnamefont {Kundu}},
  \bibinfo {author} {\bibfnamefont {S.~N.}\ \bibnamefont {Majumdar}}, \bibinfo
  {author} {\bibfnamefont {S.}~\bibnamefont {Sabhapandit}},\ and\ \bibinfo
  {author} {\bibfnamefont {G.}~\bibnamefont {Schehr}},\ }\bibfield  {title}
  {\bibinfo {title} {Exact extremal statistics in the classical 1d {C}oulomb
  gas},\ }\href {https://doi.org/10.1103/PhysRevLett.119.060601} {\bibfield
  {journal} {\bibinfo  {journal} {Phys. Rev. Lett.}\ }\textbf {\bibinfo
  {volume} {119}},\ \bibinfo {pages} {060601} (\bibinfo {year}
  {2017})}\BibitemShut {NoStop}%
\bibitem [{\citenamefont {James}(1998)}]{james1998}%
  \BibitemOpen
  \bibfield  {author} {\bibinfo {author} {\bibfnamefont {D.~F.~V.}\
  \bibnamefont {James}},\ }\bibfield  {title} {\bibinfo {title} {Quantum
  dynamics of cold trapped ions with application to quantum computation},\
  }\href {https://doi.org/10.1007/s003400050373} {\bibfield  {journal}
  {\bibinfo  {journal} {Appl. Phys. B}\ }\textbf {\bibinfo {volume} {66}},\
  \bibinfo {pages} {181} (\bibinfo {year} {1998})}\BibitemShut {NoStop}%
\bibitem [{\citenamefont {Abanov}\ \emph {et~al.}(2011)\citenamefont {Abanov},
  \citenamefont {Gromov},\ and\ \citenamefont {Kulkarni}}]{Abanov2011}%
  \BibitemOpen
  \bibfield  {author} {\bibinfo {author} {\bibfnamefont {A.~G.}\ \bibnamefont
  {Abanov}}, \bibinfo {author} {\bibfnamefont {A.}~\bibnamefont {Gromov}},\
  and\ \bibinfo {author} {\bibfnamefont {M.}~\bibnamefont {Kulkarni}},\
  }\bibfield  {title} {\bibinfo {title} {Soliton solutions of a calogero model
  in a harmonic potential},\ }\href
  {https://doi.org/10.1088/1751-8113/44/29/295203} {\bibfield  {journal}
  {\bibinfo  {journal} {J. Phys. A: Math. Theor.}\ }\textbf {\bibinfo {volume}
  {44}},\ \bibinfo {pages} {295203} (\bibinfo {year} {2011})}\BibitemShut
  {NoStop}%
\bibitem [{\citenamefont {Lahaye}\ \emph {et~al.}(2009)\citenamefont {Lahaye},
  \citenamefont {Menotti}, \citenamefont {Santos}, \citenamefont {Lewenstein},\
  and\ \citenamefont {Pfau}}]{lahaye2009}%
  \BibitemOpen
  \bibfield  {author} {\bibinfo {author} {\bibfnamefont {T.}~\bibnamefont
  {Lahaye}}, \bibinfo {author} {\bibfnamefont {C.}~\bibnamefont {Menotti}},
  \bibinfo {author} {\bibfnamefont {L.}~\bibnamefont {Santos}}, \bibinfo
  {author} {\bibfnamefont {M.}~\bibnamefont {Lewenstein}},\ and\ \bibinfo
  {author} {\bibfnamefont {T.}~\bibnamefont {Pfau}},\ }\bibfield  {title}
  {\bibinfo {title} {The physics of dipolar bosonic quantum gases},\ }\href
  {https://doi.org/10.1088/0034-4885/72/12/126401} {\bibfield  {journal}
  {\bibinfo  {journal} {Rep. Prog. Phys.}\ }\textbf {\bibinfo {volume} {72}},\
  \bibinfo {pages} {126401} (\bibinfo {year} {2009})}\BibitemShut {NoStop}%
\bibitem [{\citenamefont {Chomaz}\ \emph {et~al.}(2023)\citenamefont {Chomaz},
  \citenamefont {Ferrier-Barbut}, \citenamefont {Ferlaino}, \citenamefont
  {Laburthe-Tolra}, \citenamefont {Lev},\ and\ \citenamefont
  {Pfau}}]{chomaz2023}%
  \BibitemOpen
  \bibfield  {author} {\bibinfo {author} {\bibfnamefont {L.}~\bibnamefont
  {Chomaz}}, \bibinfo {author} {\bibfnamefont {I.}~\bibnamefont
  {Ferrier-Barbut}}, \bibinfo {author} {\bibfnamefont {F.}~\bibnamefont
  {Ferlaino}}, \bibinfo {author} {\bibfnamefont {B.}~\bibnamefont
  {Laburthe-Tolra}}, \bibinfo {author} {\bibfnamefont {B.~L.}\ \bibnamefont
  {Lev}},\ and\ \bibinfo {author} {\bibfnamefont {T.}~\bibnamefont {Pfau}},\
  }\bibfield  {title} {\bibinfo {title} {Dipolar physics: a review of
  experiments with magnetic quantum gases},\ }\href
  {https://doi.org/10.1088/1361-6633/aca814} {\bibfield  {journal} {\bibinfo
  {journal} {Rep. Prog. Phys.}\ }\textbf {\bibinfo {volume} {86}},\ \bibinfo
  {pages} {026401} (\bibinfo {year} {2023})}\BibitemShut {NoStop}%
\bibitem [{\citenamefont {Brown}\ and\ \citenamefont
  {Carrington}(2003)}]{brown_CUP_2003}%
  \BibitemOpen
  \bibfield  {author} {\bibinfo {author} {\bibfnamefont {J.~M.}\ \bibnamefont
  {Brown}}\ and\ \bibinfo {author} {\bibfnamefont {A.}~\bibnamefont
  {Carrington}},\ }\href@noop {} {\emph {\bibinfo {title} {Rotational
  spectroscopy of diatomic molecules}}}\ (\bibinfo  {publisher} {Cambridge
  University Press},\ \bibinfo {year} {2003})\BibitemShut {NoStop}%
\bibitem [{\citenamefont {Agarwal}\ \emph {et~al.}(2019)\citenamefont
  {Agarwal}, \citenamefont {Dhar}, \citenamefont {Kulkarni}, \citenamefont
  {Kundu}, \citenamefont {Majumdar}, \citenamefont {Mukamel},\ and\
  \citenamefont {Schehr}}]{agarwal2019}%
  \BibitemOpen
  \bibfield  {author} {\bibinfo {author} {\bibfnamefont {S.}~\bibnamefont
  {Agarwal}}, \bibinfo {author} {\bibfnamefont {A.}~\bibnamefont {Dhar}},
  \bibinfo {author} {\bibfnamefont {M.}~\bibnamefont {Kulkarni}}, \bibinfo
  {author} {\bibfnamefont {A.}~\bibnamefont {Kundu}}, \bibinfo {author}
  {\bibfnamefont {S.~N.}\ \bibnamefont {Majumdar}}, \bibinfo {author}
  {\bibfnamefont {D.}~\bibnamefont {Mukamel}},\ and\ \bibinfo {author}
  {\bibfnamefont {G.}~\bibnamefont {Schehr}},\ }\bibfield  {title} {\bibinfo
  {title} {Harmonically confined particles with long-range repulsive
  interactions},\ }\href {https://doi.org/10.1103/PhysRevLett.123.100603}
  {\bibfield  {journal} {\bibinfo  {journal} {Phys. Rev. Lett.}\ }\textbf
  {\bibinfo {volume} {123}},\ \bibinfo {pages} {100603} (\bibinfo {year}
  {2019})}\BibitemShut {NoStop}%
\bibitem [{\citenamefont {Kethepalli}\ \emph {et~al.}(2021)\citenamefont
  {Kethepalli}, \citenamefont {Kulkarni}, \citenamefont {Kundu}, \citenamefont
  {Majumdar}, \citenamefont {Mukamel},\ and\ \citenamefont
  {Schehr}}]{kethepalli2021}%
  \BibitemOpen
  \bibfield  {author} {\bibinfo {author} {\bibfnamefont {J.}~\bibnamefont
  {Kethepalli}}, \bibinfo {author} {\bibfnamefont {M.}~\bibnamefont
  {Kulkarni}}, \bibinfo {author} {\bibfnamefont {A.}~\bibnamefont {Kundu}},
  \bibinfo {author} {\bibfnamefont {S.~N.}\ \bibnamefont {Majumdar}}, \bibinfo
  {author} {\bibfnamefont {D.}~\bibnamefont {Mukamel}},\ and\ \bibinfo {author}
  {\bibfnamefont {G.}~\bibnamefont {Schehr}},\ }\bibfield  {title} {\bibinfo
  {title} {Harmonically confined long-ranged interacting gas in the presence of
  a hard wall},\ }\href {https://doi.org/10.1088/1742-5468/ac2896} {\bibfield
  {journal} {\bibinfo  {journal} {J. Stat. Mech.}\ ,\ \bibinfo {pages}
  {103209}} (\bibinfo {year} {2021})}\BibitemShut {NoStop}%
\bibitem [{\citenamefont {Kethepalli}\ \emph {et~al.}(2022)\citenamefont
  {Kethepalli}, \citenamefont {Kulkarni}, \citenamefont {Kundu}, \citenamefont
  {Majumdar}, \citenamefont {Mukamel},\ and\ \citenamefont
  {Schehr}}]{kethepalli2022}%
  \BibitemOpen
  \bibfield  {author} {\bibinfo {author} {\bibfnamefont {J.}~\bibnamefont
  {Kethepalli}}, \bibinfo {author} {\bibfnamefont {M.}~\bibnamefont
  {Kulkarni}}, \bibinfo {author} {\bibfnamefont {A.}~\bibnamefont {Kundu}},
  \bibinfo {author} {\bibfnamefont {S.~N.}\ \bibnamefont {Majumdar}}, \bibinfo
  {author} {\bibfnamefont {D.}~\bibnamefont {Mukamel}},\ and\ \bibinfo {author}
  {\bibfnamefont {G.}~\bibnamefont {Schehr}},\ }\bibfield  {title} {\bibinfo
  {title} {Edge fluctuations and third-order phase transition in harmonically
  confined long-range systems},\ }\href
  {https://doi.org/https://doi.org/10.1088/1742-5468/ac52b} {\bibfield
  {journal} {\bibinfo  {journal} {J. Stat. Mech.}\ ,\ \bibinfo {pages}
  {033203}} (\bibinfo {year} {2022})}\BibitemShut {NoStop}%
\bibitem [{\citenamefont {Dandekar}\ \emph {et~al.}(2023)\citenamefont
  {Dandekar}, \citenamefont {Krapivsky},\ and\ \citenamefont
  {Mallick}}]{dandekar2023}%
  \BibitemOpen
  \bibfield  {author} {\bibinfo {author} {\bibfnamefont {R.}~\bibnamefont
  {Dandekar}}, \bibinfo {author} {\bibfnamefont {P.~L.}\ \bibnamefont
  {Krapivsky}},\ and\ \bibinfo {author} {\bibfnamefont {K.}~\bibnamefont
  {Mallick}},\ }\bibfield  {title} {\bibinfo {title} {Dynamical fluctuations in
  the {R}iesz gas},\ }\href {https://doi.org/10.1103/PhysRevE.107.044129}
  {\bibfield  {journal} {\bibinfo  {journal} {Phys. Rev. E}\ }\textbf {\bibinfo
  {volume} {107}},\ \bibinfo {pages} {044129} (\bibinfo {year}
  {2023})}\BibitemShut {NoStop}%
\bibitem [{\citenamefont {Mukherjee}\ \emph {et~al.}(2026)\citenamefont
  {Mukherjee}, \citenamefont {Dhar},\ and\ \citenamefont
  {Kulkarni}}]{mukherjee2026}%
  \BibitemOpen
  \bibfield  {author} {\bibinfo {author} {\bibfnamefont {I.}~\bibnamefont
  {Mukherjee}}, \bibinfo {author} {\bibfnamefont {A.}~\bibnamefont {Dhar}},\
  and\ \bibinfo {author} {\bibfnamefont {M.}~\bibnamefont {Kulkarni}},\
  }\href@noop {} {\bibinfo {title} {Confined {R}iesz gas: From {N}ewtonian
  dynamics to emergent hydrodynamics}} (\bibinfo {year} {2026}),\ \Eprint
  {https://arxiv.org/abs/2610.06626} {arXiv:2610.06626 [cond-mat.stat-mech]}
  \BibitemShut {NoStop}%
\bibitem [{\citenamefont {{B. K. S.}}\ \emph {et~al.}(2021)\citenamefont {{B.
  K. S.}}, \citenamefont {Huse},\ and\ \citenamefont {Kulkarni}}]{bks2021}%
  \BibitemOpen
  \bibfield  {author} {\bibinfo {author} {\bibnamefont {{B. K. S.}}}, \bibinfo
  {author} {\bibfnamefont {D.~A.}\ \bibnamefont {Huse}},\ and\ \bibinfo
  {author} {\bibfnamefont {M.}~\bibnamefont {Kulkarni}},\ }\bibfield  {title}
  {\bibinfo {title} {Spatiotemporal spread of perturbations in power-law models
  at low temperatures: Exact results for classical out-of-time-order
  correlators},\ }\href {https://doi.org/10.1103/PhysRevE.104.044117}
  {\bibfield  {journal} {\bibinfo  {journal} {Phys. Rev. E}\ }\textbf {\bibinfo
  {volume} {104}},\ \bibinfo {pages} {044117} (\bibinfo {year}
  {2021})}\BibitemShut {NoStop}%
\bibitem [{\citenamefont {Engelberg}\ \emph {et~al.}(2001)\citenamefont
  {Engelberg}, \citenamefont {Liu},\ and\ \citenamefont
  {Tadmor}}]{engelberg2001}%
  \BibitemOpen
  \bibfield  {author} {\bibinfo {author} {\bibfnamefont {S.}~\bibnamefont
  {Engelberg}}, \bibinfo {author} {\bibfnamefont {H.}~\bibnamefont {Liu}},\
  and\ \bibinfo {author} {\bibfnamefont {E.}~\bibnamefont {Tadmor}},\
  }\bibfield  {title} {\bibinfo {title} {Critical thresholds in
  {E}uler--{P}oisson equations},\ }\href@noop {} {\bibfield  {journal}
  {\bibinfo  {journal} {Indiana Univ. Math. J.}\ }\textbf {\bibinfo {volume}
  {50}},\ \bibinfo {pages} {109} (\bibinfo {year} {2001})},\ \Eprint
  {https://arxiv.org/abs/math/0112014} {arXiv:math/0112014} \BibitemShut
  {NoStop}%
\bibitem [{\citenamefont {Henning}\ \emph {et~al.}(2008)\citenamefont
  {Henning}, \citenamefont {Fujioka}, \citenamefont {Ludwig}, \citenamefont
  {Piel}, \citenamefont {Melzer},\ and\ \citenamefont {Bonitz}}]{henning2008}%
  \BibitemOpen
  \bibfield  {author} {\bibinfo {author} {\bibfnamefont {C.}~\bibnamefont
  {Henning}}, \bibinfo {author} {\bibfnamefont {K.}~\bibnamefont {Fujioka}},
  \bibinfo {author} {\bibfnamefont {P.}~\bibnamefont {Ludwig}}, \bibinfo
  {author} {\bibfnamefont {A.}~\bibnamefont {Piel}}, \bibinfo {author}
  {\bibfnamefont {A.}~\bibnamefont {Melzer}},\ and\ \bibinfo {author}
  {\bibfnamefont {M.}~\bibnamefont {Bonitz}},\ }\bibfield  {title} {\bibinfo
  {title} {Existence and vanishing of the breathing mode in strongly correlated
  finite systems},\ }\href {https://doi.org/10.1103/PhysRevLett.101.045002}
  {\bibfield  {journal} {\bibinfo  {journal} {Phys. Rev. Lett.}\ }\textbf
  {\bibinfo {volume} {101}},\ \bibinfo {pages} {045002} (\bibinfo {year}
  {2008})}\BibitemShut {NoStop}%
\bibitem [{\citenamefont {Kohn}(1961)}]{kohn1961}%
  \BibitemOpen
  \bibfield  {author} {\bibinfo {author} {\bibfnamefont {W.}~\bibnamefont
  {Kohn}},\ }\bibfield  {title} {\bibinfo {title} {Cyclotron resonance and de
  {H}aas--van {A}lphen oscillations of an interacting electron gas},\ }\href
  {https://doi.org/10.1103/PhysRev.123.1242} {\bibfield  {journal} {\bibinfo
  {journal} {Phys. Rev.}\ }\textbf {\bibinfo {volume} {123}},\ \bibinfo {pages}
  {1242} (\bibinfo {year} {1961})}\BibitemShut {NoStop}%
\bibitem [{\citenamefont {Dobson}(1994)}]{dobson1994}%
  \BibitemOpen
  \bibfield  {author} {\bibinfo {author} {\bibfnamefont {J.~F.}\ \bibnamefont
  {Dobson}},\ }\bibfield  {title} {\bibinfo {title} {Harmonic-potential
  theorem: Implications for approximate many-body theories},\ }\href
  {https://doi.org/10.1103/PhysRevLett.73.2244} {\bibfield  {journal} {\bibinfo
   {journal} {Phys. Rev. Lett.}\ }\textbf {\bibinfo {volume} {73}},\ \bibinfo
  {pages} {2244} (\bibinfo {year} {1994})}\BibitemShut {NoStop}%
\bibitem [{\citenamefont {Pinney}(1950)}]{pinney1950}%
  \BibitemOpen
  \bibfield  {author} {\bibinfo {author} {\bibfnamefont {E.}~\bibnamefont
  {Pinney}},\ }\bibfield  {title} {\bibinfo {title} {The nonlinear differential
  equation $y''+p(x)y+cy^{-3}=0$},\ }\href
  {https://doi.org/10.1090/S0002-9939-1950-0037979-4} {\bibfield  {journal}
  {\bibinfo  {journal} {Proc. Amer. Math. Soc.}\ }\textbf {\bibinfo {volume}
  {1}},\ \bibinfo {pages} {681} (\bibinfo {year} {1950})}\BibitemShut {NoStop}%
\bibitem [{\citenamefont {Mesquita}\ \emph {et~al.}(2026)\citenamefont
  {Mesquita}, \citenamefont {Kulkarni}, \citenamefont {Majumdar},\ and\
  \citenamefont {Sabhapandit}}]{GPE_reset}%
  \BibitemOpen
  \bibfield  {author} {\bibinfo {author} {\bibfnamefont {N.}~\bibnamefont
  {Mesquita}}, \bibinfo {author} {\bibfnamefont {M.}~\bibnamefont {Kulkarni}},
  \bibinfo {author} {\bibfnamefont {S.~N.}\ \bibnamefont {Majumdar}},\ and\
  \bibinfo {author} {\bibfnamefont {S.}~\bibnamefont {Sabhapandit}},\
  }\href@noop {} {\bibinfo {title} {Weakly interacting bose-einstein condensate
  with stochastic resetting}} (\bibinfo {year} {2026}),\ \Eprint
  {https://arxiv.org/abs/2609.37110} {arXiv:2609.37110 [cond-mat.stat-mech]}
  \BibitemShut {NoStop}%
\bibitem [{\citenamefont {Nye}(1999)}]{nye1999}%
  \BibitemOpen
  \bibfield  {author} {\bibinfo {author} {\bibfnamefont {J.~F.}\ \bibnamefont
  {Nye}},\ }\href@noop {} {\emph {\bibinfo {title} {Natural Focusing and Fine
  Structure of Light: Caustics and Wave Dislocations}}}\ (\bibinfo  {publisher}
  {Institute of Physics Publishing},\ \bibinfo {address} {Bristol},\ \bibinfo
  {year} {1999})\BibitemShut {NoStop}%
\bibitem [{\citenamefont {Berry}(1981)}]{berry1981}%
  \BibitemOpen
  \bibfield  {author} {\bibinfo {author} {\bibfnamefont {M.~V.}\ \bibnamefont
  {Berry}},\ }\bibfield  {title} {\bibinfo {title} {Singularities in waves and
  rays},\ }in\ \href@noop {} {\emph {\bibinfo {booktitle} {Physics of Defects,
  Les Houches Session XXXV, 1980}}},\ Vol.~\bibinfo {volume} {35},\ \bibinfo
  {editor} {edited by\ \bibinfo {editor} {\bibfnamefont {R.}~\bibnamefont
  {Balian}}, \bibinfo {editor} {\bibfnamefont {M.}~\bibnamefont {Kl{\'e}man}},\
  and\ \bibinfo {editor} {\bibfnamefont {J.-P.}\ \bibnamefont {Poirier}}}\
  (\bibinfo  {publisher} {North-Holland},\ \bibinfo {address} {Amsterdam},\
  \bibinfo {year} {1981})\ pp.\ \bibinfo {pages} {453--543}\BibitemShut
  {NoStop}%
\bibitem [{\citenamefont {Berry}\ and\ \citenamefont
  {Upstill}(1980)}]{BERRY1980257}%
  \BibitemOpen
  \bibfield  {author} {\bibinfo {author} {\bibfnamefont {M.}~\bibnamefont
  {Berry}}\ and\ \bibinfo {author} {\bibfnamefont {C.}~\bibnamefont
  {Upstill}},\ }\bibfield  {title} {\bibinfo {title} {{IV Catastrophe Optics}:
  Morphologies of caustics and their diffraction patterns},\ }in\ \href
  {https://doi.org/https://doi.org/10.1016/S0079-6638(08)70215-4} {\emph
  {\bibinfo {booktitle} {Progress in Optics}}},\ Vol.~\bibinfo {volume} {18},\
  \bibinfo {editor} {edited by\ \bibinfo {editor} {\bibfnamefont
  {E.}~\bibnamefont {Wolf}}}\ (\bibinfo  {publisher} {Elsevier},\ \bibinfo
  {year} {1980})\ pp.\ \bibinfo {pages} {257--346}\BibitemShut {NoStop}%
\bibitem [{\citenamefont {Chalker}\ and\ \citenamefont
  {Shapiro}(2009)}]{Chalker2009}%
  \BibitemOpen
  \bibfield  {author} {\bibinfo {author} {\bibfnamefont {J.~T.}\ \bibnamefont
  {Chalker}}\ and\ \bibinfo {author} {\bibfnamefont {B.}~\bibnamefont
  {Shapiro}},\ }\bibfield  {title} {\bibinfo {title} {Caustic formation in
  expanding condensates of cold atoms},\ }\href
  {https://doi.org/10.1103/PhysRevA.80.013603} {\bibfield  {journal} {\bibinfo
  {journal} {Phys. Rev. A}\ }\textbf {\bibinfo {volume} {80}},\ \bibinfo
  {pages} {013603} (\bibinfo {year} {2009})}\BibitemShut {NoStop}%
\bibitem [{\citenamefont {Mumford}\ \emph {et~al.}(2017)\citenamefont
  {Mumford}, \citenamefont {Kirkby},\ and\ \citenamefont
  {O’Dell}}]{mumford2017}%
  \BibitemOpen
  \bibfield  {author} {\bibinfo {author} {\bibfnamefont {J.}~\bibnamefont
  {Mumford}}, \bibinfo {author} {\bibfnamefont {W.}~\bibnamefont {Kirkby}},\
  and\ \bibinfo {author} {\bibfnamefont {D.~H.~J.}\ \bibnamefont {O’Dell}},\
  }\bibfield  {title} {\bibinfo {title} {Catastrophes in non-equilibrium
  many-particle wave functions: universality and critical scaling},\ }\href
  {https://doi.org/10.1088/1361-6455/aa56af} {\bibfield  {journal} {\bibinfo
  {journal} {J. Phys. B: At. Mol. Opt. Phys.}\ }\textbf {\bibinfo {volume}
  {50}},\ \bibinfo {pages} {044005} (\bibinfo {year} {2017})}\BibitemShut
  {NoStop}%
\bibitem [{\citenamefont {O'Dell}(2012)}]{ODell2012}%
  \BibitemOpen
  \bibfield  {author} {\bibinfo {author} {\bibfnamefont {D.~H.~J.}\
  \bibnamefont {O'Dell}},\ }\bibfield  {title} {\bibinfo {title} {Quantum
  catastrophes and ergodicity in the dynamics of bosonic josephson junctions},\
  }\href {https://doi.org/10.1103/PhysRevLett.109.150406} {\bibfield  {journal}
  {\bibinfo  {journal} {Phys. Rev. Lett.}\ }\textbf {\bibinfo {volume} {109}},\
  \bibinfo {pages} {150406} (\bibinfo {year} {2012})}\BibitemShut {NoStop}%
\bibitem [{\citenamefont {Agarwal}\ \emph {et~al.}(2023)\citenamefont
  {Agarwal}, \citenamefont {Kulkarni},\ and\ \citenamefont
  {O'Dell}}]{agarwal2023}%
  \BibitemOpen
  \bibfield  {author} {\bibinfo {author} {\bibfnamefont {A.}~\bibnamefont
  {Agarwal}}, \bibinfo {author} {\bibfnamefont {M.}~\bibnamefont {Kulkarni}},\
  and\ \bibinfo {author} {\bibfnamefont {D.~H.~J.}\ \bibnamefont {O'Dell}},\
  }\bibfield  {title} {\bibinfo {title} {Caustics in the sine-{G}ordon model
  from quenches in coupled one-dimensional {B}ose gases},\ }\href
  {https://doi.org/10.1103/PhysRevA.108.013312} {\bibfield  {journal} {\bibinfo
   {journal} {Phys. Rev. A}\ }\textbf {\bibinfo {volume} {108}},\ \bibinfo
  {pages} {013312} (\bibinfo {year} {2023})}\BibitemShut {NoStop}%
\bibitem [{\citenamefont {Kirkby}\ \emph {et~al.}(2022)\citenamefont {Kirkby},
  \citenamefont {Yee}, \citenamefont {Shi},\ and\ \citenamefont
  {O'Dell}}]{kirkby2022}%
  \BibitemOpen
  \bibfield  {author} {\bibinfo {author} {\bibfnamefont {W.}~\bibnamefont
  {Kirkby}}, \bibinfo {author} {\bibfnamefont {Y.}~\bibnamefont {Yee}},
  \bibinfo {author} {\bibfnamefont {K.}~\bibnamefont {Shi}},\ and\ \bibinfo
  {author} {\bibfnamefont {D.~H.~J.}\ \bibnamefont {O'Dell}},\ }\bibfield
  {title} {\bibinfo {title} {Caustics in quantum many-body dynamics},\ }\href
  {https://doi.org/10.1103/PhysRevResearch.4.013105} {\bibfield  {journal}
  {\bibinfo  {journal} {Phys. Rev. Res.}\ }\textbf {\bibinfo {volume} {4}},\
  \bibinfo {pages} {013105} (\bibinfo {year} {2022})}\BibitemShut {NoStop}%
\bibitem [{SM()}]{SM}%
  \BibitemOpen
  \href@noop {} {\bibinfo  {journal} {See Supplementary Material}\
  }\BibitemShut {NoStop}%
\bibitem [{\citenamefont {Sch{\"u}tzhold}\ \emph {et~al.}(2007)\citenamefont
  {Sch{\"u}tzhold}, \citenamefont {Uhlmann}, \citenamefont {Petersen},
  \citenamefont {Schmitz}, \citenamefont {Friedenauer},\ and\ \citenamefont
  {Sch{\"a}tz}}]{ralf_PRL}%
  \BibitemOpen
\bibfield  {journal} {  }\bibfield  {author} {\bibinfo {author} {\bibfnamefont
  {R.}~\bibnamefont {Sch{\"u}tzhold}}, \bibinfo {author} {\bibfnamefont
  {M.}~\bibnamefont {Uhlmann}}, \bibinfo {author} {\bibfnamefont
  {L.}~\bibnamefont {Petersen}}, \bibinfo {author} {\bibfnamefont
  {H.}~\bibnamefont {Schmitz}}, \bibinfo {author} {\bibfnamefont
  {A.}~\bibnamefont {Friedenauer}},\ and\ \bibinfo {author} {\bibfnamefont
  {T.}~\bibnamefont {Sch{\"a}tz}},\ }\bibfield  {title} {\bibinfo {title}
  {Analogue of cosmological particle creation in an ion trap},\ }\href
  {https://doi.org/10.1103/PhysRevLett.99.201301} {\bibfield  {journal}
  {\bibinfo  {journal} {Phys. Rev. Lett.}\ }\textbf {\bibinfo {volume} {99}},\
  \bibinfo {pages} {201301} (\bibinfo {year} {2007})}\BibitemShut {NoStop}%
\bibitem [{\citenamefont {Simmons}(2016)}]{simmons2016}%
  \BibitemOpen
  \bibfield  {author} {\bibinfo {author} {\bibfnamefont {G.~F.}\ \bibnamefont
  {Simmons}},\ }\href@noop {} {\emph {\bibinfo {title} {Differential Equations
  with Applications and Historical Notes}}},\ \bibinfo {edition} {3rd}\ ed.,\
  Textbooks in Mathematics\ (\bibinfo  {publisher} {CRC Press},\ \bibinfo
  {address} {Boca Raton, FL},\ \bibinfo {year} {2016})\BibitemShut {NoStop}%
\bibitem [{\citenamefont {Coddington}\ and\ \citenamefont
  {Levinson}(1955)}]{coddington1955}%
  \BibitemOpen
  \bibfield  {author} {\bibinfo {author} {\bibfnamefont {E.~A.}\ \bibnamefont
  {Coddington}}\ and\ \bibinfo {author} {\bibfnamefont {N.}~\bibnamefont
  {Levinson}},\ }\href@noop {} {\emph {\bibinfo {title} {Theory of Ordinary
  Differential Equations}}}\ (\bibinfo  {publisher} {McGraw-Hill},\ \bibinfo
  {address} {New York},\ \bibinfo {year} {1955})\BibitemShut {NoStop}%
\bibitem [{\citenamefont {Teschl}(2012)}]{teschl2012}%
  \BibitemOpen
  \bibfield  {author} {\bibinfo {author} {\bibfnamefont {G.}~\bibnamefont
  {Teschl}},\ }\href {https://doi.org/10.1090/gsm/140} {\emph {\bibinfo {title}
  {Ordinary Differential Equations and Dynamical Systems}}},\ \bibinfo {series}
  {Graduate Studies in Mathematics}, Vol.\ \bibinfo {volume} {140}\ (\bibinfo
  {publisher} {American Mathematical Society},\ \bibinfo {address} {Providence,
  RI},\ \bibinfo {year} {2012})\BibitemShut {NoStop}%
\bibitem [{\citenamefont {Palmero}\ \emph {et~al.}(2015)\citenamefont
  {Palmero}, \citenamefont {Mart{\'\i}nez-Garaot}, \citenamefont {Alonso},
  \citenamefont {Home},\ and\ \citenamefont {Muga}}]{palmero2015}%
  \BibitemOpen
  \bibfield  {author} {\bibinfo {author} {\bibfnamefont {M.}~\bibnamefont
  {Palmero}}, \bibinfo {author} {\bibfnamefont {S.}~\bibnamefont
  {Mart{\'\i}nez-Garaot}}, \bibinfo {author} {\bibfnamefont {J.}~\bibnamefont
  {Alonso}}, \bibinfo {author} {\bibfnamefont {J.~P.}\ \bibnamefont {Home}},\
  and\ \bibinfo {author} {\bibfnamefont {J.~G.}\ \bibnamefont {Muga}},\
  }\bibfield  {title} {\bibinfo {title} {Fast expansions and compressions of
  trapped-ion chains},\ }\href {https://doi.org/10.1103/PhysRevA.91.053411}
  {\bibfield  {journal} {\bibinfo  {journal} {Phys. Rev. A}\ }\textbf {\bibinfo
  {volume} {91}},\ \bibinfo {pages} {053411} (\bibinfo {year}
  {2015})}\BibitemShut {NoStop}%
\bibitem [{\citenamefont {Gu{\'e}ry-Odelin}\ \emph {et~al.}(2019)\citenamefont
  {Gu{\'e}ry-Odelin}, \citenamefont {Ruschhaupt}, \citenamefont {Kiely},
  \citenamefont {Torrontegui}, \citenamefont {Mart{\'\i}nez-Garaot},\ and\
  \citenamefont {Muga}}]{guery2019}%
  \BibitemOpen
  \bibfield  {author} {\bibinfo {author} {\bibfnamefont {D.}~\bibnamefont
  {Gu{\'e}ry-Odelin}}, \bibinfo {author} {\bibfnamefont {A.}~\bibnamefont
  {Ruschhaupt}}, \bibinfo {author} {\bibfnamefont {A.}~\bibnamefont {Kiely}},
  \bibinfo {author} {\bibfnamefont {E.}~\bibnamefont {Torrontegui}}, \bibinfo
  {author} {\bibfnamefont {S.}~\bibnamefont {Mart{\'\i}nez-Garaot}},\ and\
  \bibinfo {author} {\bibfnamefont {J.~G.}\ \bibnamefont {Muga}},\ }\bibfield
  {title} {\bibinfo {title} {Shortcuts to adiabaticity: Concepts, methods, and
  applications},\ }\href {https://doi.org/10.1103/RevModPhys.91.045001}
  {\bibfield  {journal} {\bibinfo  {journal} {Rev. Mod. Phys.}\ }\textbf
  {\bibinfo {volume} {91}},\ \bibinfo {pages} {045001} (\bibinfo {year}
  {2019})}\BibitemShut {NoStop}%
\bibitem [{\citenamefont {Olivetti}\ \emph {et~al.}(2009)\citenamefont
  {Olivetti}, \citenamefont {Barr{\'e}}, \citenamefont {Marcos}, \citenamefont
  {Bouchet},\ and\ \citenamefont {Kaiser}}]{OlivettiPRL}%
  \BibitemOpen
  \bibfield  {author} {\bibinfo {author} {\bibfnamefont {A.}~\bibnamefont
  {Olivetti}}, \bibinfo {author} {\bibfnamefont {J.}~\bibnamefont {Barr{\'e}}},
  \bibinfo {author} {\bibfnamefont {B.}~\bibnamefont {Marcos}}, \bibinfo
  {author} {\bibfnamefont {F.}~\bibnamefont {Bouchet}},\ and\ \bibinfo {author}
  {\bibfnamefont {R.}~\bibnamefont {Kaiser}},\ }\bibfield  {title} {\bibinfo
  {title} {Breathing mode for systems of interacting particles},\ }\href
  {https://doi.org/10.1103/PhysRevLett.103.224301} {\bibfield  {journal}
  {\bibinfo  {journal} {Phys. Rev. Lett.}\ }\textbf {\bibinfo {volume} {103}},\
  \bibinfo {pages} {224301} (\bibinfo {year} {2009})}\BibitemShut {NoStop}%
\bibitem [{\citenamefont {Abraham}\ \emph {et~al.}(2014)\citenamefont
  {Abraham}, \citenamefont {Bonitz}, \citenamefont {McDonald}, \citenamefont
  {Orlando},\ and\ \citenamefont {Brabec}}]{abraham2014}%
  \BibitemOpen
  \bibfield  {author} {\bibinfo {author} {\bibfnamefont {J.~W.}\ \bibnamefont
  {Abraham}}, \bibinfo {author} {\bibfnamefont {M.}~\bibnamefont {Bonitz}},
  \bibinfo {author} {\bibfnamefont {C.}~\bibnamefont {McDonald}}, \bibinfo
  {author} {\bibfnamefont {G.}~\bibnamefont {Orlando}},\ and\ \bibinfo {author}
  {\bibfnamefont {T.}~\bibnamefont {Brabec}},\ }\bibfield  {title} {\bibinfo
  {title} {Quantum breathing mode of trapped systems in one and two
  dimensions},\ }\href {https://doi.org/10.1088/1367-2630/16/1/013001}
  {\bibfield  {journal} {\bibinfo  {journal} {New J. Phys.}\ }\textbf {\bibinfo
  {volume} {16}},\ \bibinfo {pages} {013001} (\bibinfo {year}
  {2014})}\BibitemShut {NoStop}%
\bibitem [{\citenamefont {Abraham}\ and\ \citenamefont
  {Bonitz}(2014)}]{abraham2014review}%
  \BibitemOpen
  \bibfield  {author} {\bibinfo {author} {\bibfnamefont {J.~W.}\ \bibnamefont
  {Abraham}}\ and\ \bibinfo {author} {\bibfnamefont {M.}~\bibnamefont
  {Bonitz}},\ }\bibfield  {title} {\bibinfo {title} {Quantum breathing mode of
  trapped particles: From nanoplasmas to ultracold gases},\ }\href
  {https://doi.org/10.1002/ctpp.201300066} {\bibfield  {journal} {\bibinfo
  {journal} {Contrib. Plasma Phys.}\ }\textbf {\bibinfo {volume} {54}},\
  \bibinfo {pages} {27} (\bibinfo {year} {2014})}\BibitemShut {NoStop}%
\bibitem [{\citenamefont {Pyka}\ \emph {et~al.}(2013)\citenamefont {Pyka},
  \citenamefont {Keller}, \citenamefont {Partner}, \citenamefont {Nigmatullin},
  \citenamefont {Burgermeister}, \citenamefont {Meier}, \citenamefont
  {Kuhlmann}, \citenamefont {Retzker}, \citenamefont {Plenio}, \citenamefont
  {Zurek}, \citenamefont {del Campo},\ and\ \citenamefont
  {Mehlst{\"a}ubler}}]{pyka2013}%
  \BibitemOpen
  \bibfield  {author} {\bibinfo {author} {\bibfnamefont {K.}~\bibnamefont
  {Pyka}}, \bibinfo {author} {\bibfnamefont {J.}~\bibnamefont {Keller}},
  \bibinfo {author} {\bibfnamefont {H.~L.}\ \bibnamefont {Partner}}, \bibinfo
  {author} {\bibfnamefont {R.}~\bibnamefont {Nigmatullin}}, \bibinfo {author}
  {\bibfnamefont {T.}~\bibnamefont {Burgermeister}}, \bibinfo {author}
  {\bibfnamefont {D.~M.}\ \bibnamefont {Meier}}, \bibinfo {author}
  {\bibfnamefont {K.}~\bibnamefont {Kuhlmann}}, \bibinfo {author}
  {\bibfnamefont {A.}~\bibnamefont {Retzker}}, \bibinfo {author} {\bibfnamefont
  {M.~B.}\ \bibnamefont {Plenio}}, \bibinfo {author} {\bibfnamefont {W.~H.}\
  \bibnamefont {Zurek}}, \bibinfo {author} {\bibfnamefont {A.}~\bibnamefont
  {del Campo}},\ and\ \bibinfo {author} {\bibfnamefont {T.~E.}\ \bibnamefont
  {Mehlst{\"a}ubler}},\ }\bibfield  {title} {\bibinfo {title} {Topological
  defect formation and spontaneous symmetry breaking in ion {C}oulomb
  crystals},\ }\href {https://doi.org/10.1038/ncomms3291} {\bibfield  {journal}
  {\bibinfo  {journal} {Nat. Commun.}\ }\textbf {\bibinfo {volume} {4}},\
  \bibinfo {pages} {2291} (\bibinfo {year} {2013})}\BibitemShut {NoStop}%
\bibitem [{\citenamefont {Ulm}\ \emph {et~al.}(2013)\citenamefont {Ulm},
  \citenamefont {Ro{\ss}nagel}, \citenamefont {Jacob}, \citenamefont
  {Deg{\"u}nther}, \citenamefont {Dawkins}, \citenamefont {Poschinger},
  \citenamefont {Nigmatullin}, \citenamefont {Retzker}, \citenamefont {Plenio},
  \citenamefont {Schmidt-Kaler},\ and\ \citenamefont {Singer}}]{ulm2013}%
  \BibitemOpen
  \bibfield  {author} {\bibinfo {author} {\bibfnamefont {S.}~\bibnamefont
  {Ulm}}, \bibinfo {author} {\bibfnamefont {J.}~\bibnamefont {Ro{\ss}nagel}},
  \bibinfo {author} {\bibfnamefont {G.}~\bibnamefont {Jacob}}, \bibinfo
  {author} {\bibfnamefont {C.}~\bibnamefont {Deg{\"u}nther}}, \bibinfo {author}
  {\bibfnamefont {S.~T.}\ \bibnamefont {Dawkins}}, \bibinfo {author}
  {\bibfnamefont {U.~G.}\ \bibnamefont {Poschinger}}, \bibinfo {author}
  {\bibfnamefont {R.}~\bibnamefont {Nigmatullin}}, \bibinfo {author}
  {\bibfnamefont {A.}~\bibnamefont {Retzker}}, \bibinfo {author} {\bibfnamefont
  {M.~B.}\ \bibnamefont {Plenio}}, \bibinfo {author} {\bibfnamefont
  {F.}~\bibnamefont {Schmidt-Kaler}},\ and\ \bibinfo {author} {\bibfnamefont
  {K.}~\bibnamefont {Singer}},\ }\bibfield  {title} {\bibinfo {title}
  {Observation of the {K}ibble--{Z}urek scaling law for defect formation in ion
  crystals},\ }\href {https://doi.org/10.1038/ncomms3290} {\bibfield  {journal}
  {\bibinfo  {journal} {Nat. Commun.}\ }\textbf {\bibinfo {volume} {4}},\
  \bibinfo {pages} {2290} (\bibinfo {year} {2013})}\BibitemShut {NoStop}%
\bibitem [{\citenamefont {Moritz}\ \emph {et~al.}(2003)\citenamefont {Moritz},
  \citenamefont {St{\"o}ferle}, \citenamefont {K{\"o}hl},\ and\ \citenamefont
  {Esslinger}}]{moritz2003}%
  \BibitemOpen
  \bibfield  {author} {\bibinfo {author} {\bibfnamefont {H.}~\bibnamefont
  {Moritz}}, \bibinfo {author} {\bibfnamefont {T.}~\bibnamefont
  {St{\"o}ferle}}, \bibinfo {author} {\bibfnamefont {M.}~\bibnamefont
  {K{\"o}hl}},\ and\ \bibinfo {author} {\bibfnamefont {T.}~\bibnamefont
  {Esslinger}},\ }\bibfield  {title} {\bibinfo {title} {Exciting collective
  oscillations in a trapped {1D} gas},\ }\href
  {https://doi.org/10.1103/PhysRevLett.91.250402} {\bibfield  {journal}
  {\bibinfo  {journal} {Phys. Rev. Lett.}\ }\textbf {\bibinfo {volume} {91}},\
  \bibinfo {pages} {250402} (\bibinfo {year} {2003})}\BibitemShut {NoStop}%
\bibitem [{\citenamefont {Fang}\ \emph {et~al.}(2014)\citenamefont {Fang},
  \citenamefont {Carleo}, \citenamefont {Johnson},\ and\ \citenamefont
  {Bouchoule}}]{fang2014}%
  \BibitemOpen
  \bibfield  {author} {\bibinfo {author} {\bibfnamefont {B.}~\bibnamefont
  {Fang}}, \bibinfo {author} {\bibfnamefont {G.}~\bibnamefont {Carleo}},
  \bibinfo {author} {\bibfnamefont {A.}~\bibnamefont {Johnson}},\ and\ \bibinfo
  {author} {\bibfnamefont {I.}~\bibnamefont {Bouchoule}},\ }\bibfield  {title}
  {\bibinfo {title} {Quench-induced breathing mode of one-dimensional {B}ose
  gases},\ }\href {https://doi.org/10.1103/PhysRevLett.113.035301} {\bibfield
  {journal} {\bibinfo  {journal} {Phys. Rev. Lett.}\ }\textbf {\bibinfo
  {volume} {113}},\ \bibinfo {pages} {035301} (\bibinfo {year}
  {2014})}\BibitemShut {NoStop}%
\bibitem [{\citenamefont {Chafa{\"\i}}\ \emph {et~al.}(2022)\citenamefont
  {Chafa{\"\i}}, \citenamefont {Saff},\ and\ \citenamefont
  {Womersley}}]{CHAFAI2022126367}%
  \BibitemOpen
  \bibfield  {author} {\bibinfo {author} {\bibfnamefont {D.}~\bibnamefont
  {Chafa{\"\i}}}, \bibinfo {author} {\bibfnamefont {E.~B.}\ \bibnamefont
  {Saff}},\ and\ \bibinfo {author} {\bibfnamefont {R.~S.}\ \bibnamefont
  {Womersley}},\ }\bibfield  {title} {\bibinfo {title} {On the solution of a
  riesz equilibrium problem and integral identities for special functions},\
  }\href {https://doi.org/https://doi.org/10.1016/j.jmaa.2022.126367}
  {\bibfield  {journal} {\bibinfo  {journal} {J. Math. Anal. Appl.}\ }\textbf
  {\bibinfo {volume} {515}},\ \bibinfo {pages} {126367} (\bibinfo {year}
  {2022})}\BibitemShut {NoStop}%
\bibitem [{\citenamefont {Byun}\ \emph {et~al.}(2026)\citenamefont {Byun},
  \citenamefont {Forrester}, \citenamefont {Majumdar},\ and\ \citenamefont
  {Schehr}}]{BYUN2026130813}%
  \BibitemOpen
  \bibfield  {author} {\bibinfo {author} {\bibfnamefont {S.-S.}\ \bibnamefont
  {Byun}}, \bibinfo {author} {\bibfnamefont {P.~J.}\ \bibnamefont {Forrester}},
  \bibinfo {author} {\bibfnamefont {S.~N.}\ \bibnamefont {Majumdar}},\ and\
  \bibinfo {author} {\bibfnamefont {G.}~\bibnamefont {Schehr}},\ }\bibfield
  {title} {\bibinfo {title} {Equilibrium measures for higher dimensional
  rotationally symmetric riesz gases},\ }\href
  {https://doi.org/https://doi.org/10.1016/j.jmaa.2026.130813} {\bibfield
  {journal} {\bibinfo  {journal} {J. Math. Anal. Appl.}\ }\textbf {\bibinfo
  {volume} {563}},\ \bibinfo {pages} {130813} (\bibinfo {year}
  {2026})}\BibitemShut {NoStop}%
\end{thebibliography}%


\begin{thebibliography}{15}%
\makeatletter
\providecommand \@ifxundefined [1]{%
 \@ifx{#1\undefined}
}%
\providecommand \@ifnum [1]{%
 \ifnum #1\expandafter \@firstoftwo
 \else \expandafter \@secondoftwo
 \fi
}%
\providecommand \@ifx [1]{%
 \ifx #1\expandafter \@firstoftwo
 \else \expandafter \@secondoftwo
 \fi
}%
\providecommand \natexlab [1]{#1}%
\providecommand \enquote  [1]{``#1''}%
\providecommand \bibnamefont  [1]{#1}%
\providecommand \bibfnamefont [1]{#1}%
\providecommand \citenamefont [1]{#1}%
\providecommand \href@noop [0]{\@secondoftwo}%
\providecommand \href [0]{\begingroup \@sanitize@url \@href}%
\providecommand \@href[1]{\@@startlink{#1}\@@href}%
\providecommand \@@href[1]{\endgroup#1\@@endlink}%
\providecommand \@sanitize@url [0]{\catcode `\\12\catcode `\$12\catcode
  `\&12\catcode `\#12\catcode `\^12\catcode `\_12\catcode `\%12\relax}%
\providecommand \@@startlink[1]{}%
\providecommand \@@endlink[0]{}%
\providecommand \url  [0]{\begingroup\@sanitize@url \@url }%
\providecommand \@url [1]{\endgroup\@href {#1}{\urlprefix }}%
\providecommand \urlprefix  [0]{URL }%
\providecommand \Eprint [0]{\href }%
\providecommand \doibase [0]{https://doi.org/}%
\providecommand \selectlanguage [0]{\@gobble}%
\providecommand \bibinfo  [0]{\@secondoftwo}%
\providecommand \bibfield  [0]{\@secondoftwo}%
\providecommand \translation [1]{[#1]}%
\providecommand \BibitemOpen [0]{}%
\providecommand \bibitemStop [0]{}%
\providecommand \bibitemNoStop [0]{.\EOS\space}%
\providecommand \EOS [0]{\spacefactor3000\relax}%
\providecommand \BibitemShut  [1]{\csname bibitem#1\endcsname}%
\let\auto@bib@innerbib\@empty
\bibitem [{\citenamefont {Riesz}(1938)}]{riesz1938integrales}%
  \BibitemOpen
  \bibfield  {author} {\bibinfo {author} {\bibfnamefont {M.}~\bibnamefont
  {Riesz}},\ }\bibfield  {title} {\bibinfo {title} {Int{\'e}grales de
  riemann-liouville et potentiels},\ }\href
  {https://www.acta.bibl.u-szeged.hu/13487/1/math_009_001-042.pdf} {\bibfield
  {journal} {\bibinfo  {journal} {Acta Sci. Math. Szeged}\ }\textbf {\bibinfo
  {volume} {9}},\ \bibinfo {pages} {1} (\bibinfo {year} {1938})}\BibitemShut
  {NoStop}%
\bibitem [{\citenamefont {S.}\ \emph {et~al.}(2021)\citenamefont {S.},
  \citenamefont {Huse},\ and\ \citenamefont {Kulkarni}}]{bks21}%
  \BibitemOpen
  \bibfield  {author} {\bibinfo {author} {\bibfnamefont {B.~K.}\ \bibnamefont
  {S.}}, \bibinfo {author} {\bibfnamefont {D.~A.}\ \bibnamefont {Huse}},\ and\
  \bibinfo {author} {\bibfnamefont {M.}~\bibnamefont {Kulkarni}},\ }\bibfield
  {title} {\bibinfo {title} {Spatiotemporal spread of perturbations in
  power-law models at low temperatures: Exact results for classical
  out-of-time-order correlators},\ }\href
  {https://doi.org/10.1103/PhysRevE.104.044117} {\bibfield  {journal} {\bibinfo
   {journal} {Phys. Rev. E}\ }\textbf {\bibinfo {volume} {104}},\ \bibinfo
  {pages} {044117} (\bibinfo {year} {2021})}\BibitemShut {NoStop}%
\bibitem [{\citenamefont {Henning}\ \emph {et~al.}(2008)\citenamefont
  {Henning}, \citenamefont {Fujioka}, \citenamefont {Ludwig}, \citenamefont
  {Piel}, \citenamefont {Melzer},\ and\ \citenamefont {Bonitz}}]{henning2008}%
  \BibitemOpen
  \bibfield  {author} {\bibinfo {author} {\bibfnamefont {C.}~\bibnamefont
  {Henning}}, \bibinfo {author} {\bibfnamefont {K.}~\bibnamefont {Fujioka}},
  \bibinfo {author} {\bibfnamefont {P.}~\bibnamefont {Ludwig}}, \bibinfo
  {author} {\bibfnamefont {A.}~\bibnamefont {Piel}}, \bibinfo {author}
  {\bibfnamefont {A.}~\bibnamefont {Melzer}},\ and\ \bibinfo {author}
  {\bibfnamefont {M.}~\bibnamefont {Bonitz}},\ }\bibfield  {title} {\bibinfo
  {title} {Existence and vanishing of the breathing mode in strongly correlated
  finite systems},\ }\href {https://doi.org/10.1103/PhysRevLett.101.045002}
  {\bibfield  {journal} {\bibinfo  {journal} {Phys. Rev. Lett.}\ }\textbf
  {\bibinfo {volume} {101}},\ \bibinfo {pages} {045002} (\bibinfo {year}
  {2008})}\BibitemShut {NoStop}%
\bibitem [{\citenamefont {{Calogero}}(1971)}]{Calogero1971}%
  \BibitemOpen
  \bibfield  {author} {\bibinfo {author} {\bibfnamefont {F.}~\bibnamefont
  {{Calogero}}},\ }\bibfield  {title} {\bibinfo {title} {Solution of the
  one-dimensional $n$-body problems with quadratic and/or inversely quadratic
  pair potentials},\ }\href {https://doi.org/10.1063/1.1665604} {\bibfield
  {journal} {\bibinfo  {journal} {J. Math. Phys.}\ }\textbf {\bibinfo {volume}
  {12}},\ \bibinfo {pages} {419} (\bibinfo {year} {1971})}\BibitemShut
  {NoStop}%
\bibitem [{\citenamefont {Abanov}\ \emph {et~al.}(2011)\citenamefont {Abanov},
  \citenamefont {Gromov},\ and\ \citenamefont {Kulkarni}}]{Abanov2011}%
  \BibitemOpen
  \bibfield  {author} {\bibinfo {author} {\bibfnamefont {A.~G.}\ \bibnamefont
  {Abanov}}, \bibinfo {author} {\bibfnamefont {A.}~\bibnamefont {Gromov}},\
  and\ \bibinfo {author} {\bibfnamefont {M.}~\bibnamefont {Kulkarni}},\
  }\bibfield  {title} {\bibinfo {title} {Soliton solutions of a calogero model
  in a harmonic potential},\ }\href
  {https://doi.org/10.1088/1751-8113/44/29/295203} {\bibfield  {journal}
  {\bibinfo  {journal} {J. Phys. A: Math. Theor.}\ }\textbf {\bibinfo {volume}
  {44}},\ \bibinfo {pages} {295203} (\bibinfo {year} {2011})}\BibitemShut
  {NoStop}%
\bibitem [{\citenamefont {Pinney}(1950)}]{Pinney1950}%
  \BibitemOpen
  \bibfield  {author} {\bibinfo {author} {\bibfnamefont {E.}~\bibnamefont
  {Pinney}},\ }\bibfield  {title} {\bibinfo {title} {The nonlinear differential
  equation $y''+p(x)y+cy^{-3}=0$},\ }\href
  {https://doi.org/10.1090/S0002-9939-1950-0037979-4} {\bibfield  {journal}
  {\bibinfo  {journal} {Proc. Amer. Math. Soc.}\ }\textbf {\bibinfo {volume}
  {1}},\ \bibinfo {pages} {681} (\bibinfo {year} {1950})}\BibitemShut {NoStop}%
\bibitem [{\citenamefont {Choquard}\ \emph {et~al.}(1981)\citenamefont
  {Choquard}, \citenamefont {Kunz}, \citenamefont {Martin},\ and\ \citenamefont
  {Navet}}]{ocp1}%
  \BibitemOpen
  \bibfield  {author} {\bibinfo {author} {\bibfnamefont {P.}~\bibnamefont
  {Choquard}}, \bibinfo {author} {\bibfnamefont {H.}~\bibnamefont {Kunz}},
  \bibinfo {author} {\bibfnamefont {P.~A.}\ \bibnamefont {Martin}},\ and\
  \bibinfo {author} {\bibfnamefont {M.}~\bibnamefont {Navet}},\ }\bibfield
  {title} {\bibinfo {title} {One-dimensional {C}oulomb systems},\ }in\
  \href@noop {} {\emph {\bibinfo {booktitle} {Physics in One Dimension}}},\
  \bibinfo {series} {Springer Series in Solid-State Sciences}, Vol.~\bibinfo
  {volume} {23},\ \bibinfo {editor} {edited by\ \bibinfo {editor}
  {\bibfnamefont {J.}~\bibnamefont {Bernasconi}}\ and\ \bibinfo {editor}
  {\bibfnamefont {T.}~\bibnamefont {Schneider}}}\ (\bibinfo  {publisher}
  {Springer},\ \bibinfo {address} {Berlin, Heidelberg},\ \bibinfo {year}
  {1981})\ pp.\ \bibinfo {pages} {335--350}\BibitemShut {NoStop}%
\bibitem [{\citenamefont {Lenard}(1961)}]{ocp2}%
  \BibitemOpen
  \bibfield  {author} {\bibinfo {author} {\bibfnamefont {A.}~\bibnamefont
  {Lenard}},\ }\bibfield  {title} {\bibinfo {title} {Exact statistical
  mechanics of a one-dimensional system with {C}oulomb forces},\ }\href
  {https://doi.org/10.1063/1.1703757} {\bibfield  {journal} {\bibinfo
  {journal} {J. Math. Phys.}\ }\textbf {\bibinfo {volume} {2}},\ \bibinfo
  {pages} {682} (\bibinfo {year} {1961})}\BibitemShut {NoStop}%
\bibitem [{\citenamefont {Prager}(1962)}]{ocp3}%
  \BibitemOpen
  \bibfield  {author} {\bibinfo {author} {\bibfnamefont {S.}~\bibnamefont
  {Prager}},\ }\bibfield  {title} {\bibinfo {title} {The one-dimensional
  plasma},\ }in\ \href {https://doi.org/10.1002/9780470143506.ch5} {\emph
  {\bibinfo {booktitle} {Advances in Chemical Physics}}},\ Vol.~\bibinfo
  {volume} {4},\ \bibinfo {editor} {edited by\ \bibinfo {editor} {\bibfnamefont
  {I.}~\bibnamefont {Prigogine}}}\ (\bibinfo  {publisher} {John Wiley \&
  Sons},\ \bibinfo {address} {Hoboken, NJ},\ \bibinfo {year} {1962})\ p.\
  \bibinfo {pages} {201}\BibitemShut {NoStop}%
\bibitem [{\citenamefont {Baxter}(1963)}]{ocp4}%
  \BibitemOpen
  \bibfield  {author} {\bibinfo {author} {\bibfnamefont {R.~J.}\ \bibnamefont
  {Baxter}},\ }\bibfield  {title} {\bibinfo {title} {Statistical mechanics of a
  one-dimensional {C}oulomb system with a uniform charge background},\ }\href
  {https://doi.org/10.1017/S0305004100003790} {\bibfield  {journal} {\bibinfo
  {journal} {Proc. Cambridge Philos. Soc.}\ }\textbf {\bibinfo {volume} {59}},\
  \bibinfo {pages} {779} (\bibinfo {year} {1963})}\BibitemShut {NoStop}%
\bibitem [{\citenamefont {T{\'e}llez}\ and\ \citenamefont
  {Trizac}(2015)}]{ocp6}%
  \BibitemOpen
  \bibfield  {author} {\bibinfo {author} {\bibfnamefont {G.}~\bibnamefont
  {T{\'e}llez}}\ and\ \bibinfo {author} {\bibfnamefont {E.}~\bibnamefont
  {Trizac}},\ }\bibfield  {title} {\bibinfo {title} {Screening like charges in
  one-dimensional {C}oulomb systems: Exact results},\ }\href
  {https://doi.org/10.1103/PhysRevE.92.042134} {\bibfield  {journal} {\bibinfo
  {journal} {Phys. Rev. E}\ }\textbf {\bibinfo {volume} {92}},\ \bibinfo
  {pages} {042134} (\bibinfo {year} {2015})}\BibitemShut {NoStop}%
\bibitem [{\citenamefont {Dhar}\ \emph {et~al.}(2017)\citenamefont {Dhar},
  \citenamefont {Kundu}, \citenamefont {Majumdar}, \citenamefont
  {Sabhapandit},\ and\ \citenamefont {Schehr}}]{dhar2017}%
  \BibitemOpen
  \bibfield  {author} {\bibinfo {author} {\bibfnamefont {A.}~\bibnamefont
  {Dhar}}, \bibinfo {author} {\bibfnamefont {A.}~\bibnamefont {Kundu}},
  \bibinfo {author} {\bibfnamefont {S.~N.}\ \bibnamefont {Majumdar}}, \bibinfo
  {author} {\bibfnamefont {S.}~\bibnamefont {Sabhapandit}},\ and\ \bibinfo
  {author} {\bibfnamefont {G.}~\bibnamefont {Schehr}},\ }\bibfield  {title}
  {\bibinfo {title} {Exact extremal statistics in the classical 1d {C}oulomb
  gas},\ }\href {https://doi.org/10.1103/PhysRevLett.119.060601} {\bibfield
  {journal} {\bibinfo  {journal} {Phys. Rev. Lett.}\ }\textbf {\bibinfo
  {volume} {119}},\ \bibinfo {pages} {060601} (\bibinfo {year}
  {2017})}\BibitemShut {NoStop}%
\bibitem [{\citenamefont {Mukherjee}\ \emph {et~al.}(2026)\citenamefont
  {Mukherjee}, \citenamefont {Dhar},\ and\ \citenamefont
  {Kulkarni}}]{mukherjee2026}%
  \BibitemOpen
  \bibfield  {author} {\bibinfo {author} {\bibfnamefont {I.}~\bibnamefont
  {Mukherjee}}, \bibinfo {author} {\bibfnamefont {A.}~\bibnamefont {Dhar}},\
  and\ \bibinfo {author} {\bibfnamefont {M.}~\bibnamefont {Kulkarni}},\
  }\href@noop {} {\bibinfo {title} {Confined {R}iesz gas: From {N}ewtonian
  dynamics to emergent hydrodynamics}} (\bibinfo {year} {2026}),\ \Eprint
  {https://arxiv.org/abs/2610.06626} {arXiv:2610.06626 [cond-mat.stat-mech]}
  \BibitemShut {NoStop}%
\bibitem [{\citenamefont {Agarwal}\ \emph {et~al.}(2019)\citenamefont
  {Agarwal}, \citenamefont {Dhar}, \citenamefont {Kulkarni}, \citenamefont
  {Kundu}, \citenamefont {Majumdar}, \citenamefont {Mukamel},\ and\
  \citenamefont {Schehr}}]{Riesz_avg_density_SA}%
  \BibitemOpen
  \bibfield  {author} {\bibinfo {author} {\bibfnamefont {S.}~\bibnamefont
  {Agarwal}}, \bibinfo {author} {\bibfnamefont {A.}~\bibnamefont {Dhar}},
  \bibinfo {author} {\bibfnamefont {M.}~\bibnamefont {Kulkarni}}, \bibinfo
  {author} {\bibfnamefont {A.}~\bibnamefont {Kundu}}, \bibinfo {author}
  {\bibfnamefont {S.~N.}\ \bibnamefont {Majumdar}}, \bibinfo {author}
  {\bibfnamefont {D.}~\bibnamefont {Mukamel}},\ and\ \bibinfo {author}
  {\bibfnamefont {G.}~\bibnamefont {Schehr}},\ }\bibfield  {title} {\bibinfo
  {title} {Harmonically confined particles with long-range repulsive
  interactions},\ }\href {https://doi.org/10.1103/PhysRevLett.123.100603}
  {\bibfield  {journal} {\bibinfo  {journal} {Phys. Rev. Lett.}\ }\textbf
  {\bibinfo {volume} {123}},\ \bibinfo {pages} {100603} (\bibinfo {year}
  {2019})}\BibitemShut {NoStop}%
\bibitem [{\citenamefont {Sonin}(1954)}]{Sonin1954}%
  \BibitemOpen
  \bibfield  {author} {\bibinfo {author} {\bibfnamefont {N.~{\relax Ya}.}\
  \bibnamefont {Sonin}},\ }\href@noop {} {\emph {\bibinfo {title} {Studies of
  Cylinder Functions and Special Polynomials}}}\ (\bibinfo  {publisher}
  {Gostekhizdat},\ \bibinfo {address} {Moscow},\ \bibinfo {year} {1954})\
  \bibinfo {note} {(in Russian)}\BibitemShut {NoStop}%
\end{thebibliography}%

\end{document}


\title{\underline{Supplemental Material}\\ Nonlinear Breathing and Caustic Collapse in Quenched Long-Range Interacting Gases}

\author{Manas Kulkarni}
\email{manas.kulkarni@icts.res.in }
\affiliation{International Centre for Theoretical Sciences, Tata Institute of Fundamental Research, Bangalore 560089, India}

\author{Sanjib Sabhapandit}
\email{sanjib@rri.res.in}
\affiliation{Raman Research Institute, Bangalore 560080, India}

\date{\today}

\begin{abstract}
    Here, we give details of some of the results presented in the main text in a self contained manner. 
\end{abstract}

\maketitle
\tableofcontents

\section{The microscopic model}

We consider $N$ classical particles in 1D with pairwise interaction confined by an external harmonic trap. This is the so called Riesz gas \cite{riesz1938integrales,bks21} given by,
\begin{subequations}\label{Hamiltonian}
\begin{align}
H =& \sum_{i=1}^{N} \frac{p_{i}^{2}}{2 m} +U(\{x_j\}, t)\\
U(\{x_j\}, t)=&\sum_{i=1}^{N}\left[\frac{m\, \omega^2(t)}{2}\bigl[x_i-a(t)\bigr]^2+\frac{J\,\text{sgn}(k)}{2}\sum_{j\neq i}\frac{1}{|x_i-x_j|^k}\right]\,.
\end{align}
\end{subequations}
Here $x_{i}$ is the position of the $i^{th}$ particle, $p_{i}$ is the corresponding conjugate momentum, $m$ is the mass of each particle, $J$ is the interaction strength, $\omega(t)$ is a time-dependent frequency of the external trap, and $a(t)$ is a time-dependent shift of the location of the trap. Therefore, the Hamilton's equations of motion, $\dot{x}_i =\partial_{p_i} H$ and $\dot{p}_i =-\partial_{x_i} H$,  become
\noindent \begin{equation}\label{HamiltonEquations}
\dot{x}_{i} = \frac{p_{i}}{m}\quad \text{and}\quad
\dot{p}_{i}= -m\, \omega^2 (t)\, [x_{i}-a(t)]+Jk\,\text{sgn}(k)\sum_{j\neq i}\frac{ \mathrm{sgn}(x_{i}-x_{j})}{\left|x_{i}-x_{j}\right|^{k+1}}\,, 
\end{equation} 
respectively.

\subsection{Trap quench and exact dilation dynamics}\label{sec:dilation}

\noindent{\it Quench protocol.---}
We prepare the system in the zero-temperature equilibrium state of the trap with frequency $\omega(t=0)=\omega_0$, shift $a(0)=0$, zero momentum $p_i(0)=0$, and $\dot{p}_i(0)=0$. Therefore, from Eq.~\eqref{HamiltonEquations}, the initial positions $\{x_i(0)\equiv x_i^0\}$ satisfy,
\begin{equation}\label{eq:equilibrium}
m\omega_0^2\, x_i^0 = F_i\big(\{x_j^0\}\big)\quad\text{with}\quad
F_i\big(\{x_j\}\big)\equiv Jk\,\text{sgn}(k)\sum_{j\neq i}\frac{\mathrm{sgn}(x_i-x_j)}{|x_i-x_j|^{k+1}}\,.
\end{equation}
At $t=0^+$, the initial trap frequency $\omega_0$ is suddenly changed to a time dependent frequency, $\omega_0\to\omega(t)$, and the location of the trap $a(t)$ is made time dependent.  Therefore, the particles evolve according to Eq.~\eqref{HamiltonEquations} starting with the initial conditions $\omega(0)=\omega_0$, $a(0)=0$, $p_i(0)=0$, $\dot{p}_i(0)=0$, and the initial positions $\{x_i(0)\equiv x_i^0\}$ are chosen from the zero temperature equilibrium configuration such that they satisfy Eq.~\eqref{eq:equilibrium}. \\

\noindent{\it Exact dilation.---}
Equation~\eqref{HamiltonEquations} suggests a change of variable 
\begin{equation}
 x_i(t) = \tilde{x}_i(t) + X_{CM}(t)\quad \text{and} \quad p_i(t) = \tilde{p}_i(t)+  P(t)   \, .
\end{equation}
Substituting these in Eq.~\eqref{HamiltonEquations} and setting $P(t) = m\, \dot{X}_{CM}(t)$,  we get the equation of motion for $X_{CM}(t)$ as 
\begin{equation}
\label{eq:Xt}
    \ddot{X}_{CM}(t) = -\omega^2(t) \bigl[X_{CM}(t) -a(t)\bigr],\quad\text{with}\quad X_{CM}(0)=\dot{X}_{CM}(0)=0\, .
\end{equation}
On the other hand, $\{\tilde{x}_i, \tilde{p}_i)\}$ satisfy the Hamilton's equations of motion
\noindent \begin{equation}\label{HamiltonEquations-tilde}
\dot{\tilde{x}}_{i} = \frac{\tilde p_{i}}{m}\quad \text{and}\quad
\dot{\tilde p}_{i}= -m\, \omega^2 (t)\, \tilde x_{i} +Jk\,\text{sgn}(k)\sum_{j\neq i}\frac{ \mathrm{sgn}(\tilde x_{i}-\tilde x_{j})}{\left|\tilde x_{i}-\tilde x_{j}\right|^{k+1}}\,, 
\end{equation} 
with the initial conditions $\tilde{p}_i(0) =0$ and $\tilde{x}_i(0) = x_i^0$, where $x_i^0$ satisfies the equilibrium initial condition $F_i$ defined in Eq.~\eqref{eq:equilibrium}.

The interaction force in Eq.~\eqref{HamiltonEquations-tilde} [also defined in Eq.~\eqref{eq:equilibrium}] is a homogeneous function of degree $-(k+1)$ of the positions, i.e., 
\begin{equation}\label{eq:homogeneity}
F_i(\{\Lambda x_j\}) = \mathrm{sgn}(\Lambda)\,|\Lambda|^{-(k+1)}\,F_i(\{x_j\}) \quad \text{for any real } \Lambda\neq 0.
\end{equation}
This suggests a uniform dilation (or contraction) of the initial configuration,
\begin{equation}\label{eq:dilation}
\tilde x_i(t)=b(t)\,x_i^0,\quad\text{and}\quad  \tilde p_i(t)=m\,\dot b(t)\,x_i^0,\quad \text{with}\quad b(0)=1\quad\text{and}\quad \dot b(0)=0 .
\end{equation}
Substituting Eq.~\eqref{eq:dilation} in Eq.~\eqref{HamiltonEquations-tilde} yields a equation for $b(t)$ as
\begin{equation}
\ddot b(t)
= -\omega^2(t)\, b + \frac{\mathrm{sgn}(b)\, \omega_0^2}{|b|^{k+1}} \, ,
\quad\text{with}\quad  b(0)=1\quad\text{and}\quad \dot b(0)=0 .
\label{eq:b_eq}
\end{equation}
where we have used the first relation in Eq.~\eqref{eq:equilibrium}. Therefore, all the $N$-equations in Eq.~\eqref{HamiltonEquations-tilde} reduce to a single equation in Eq.~\eqref{eq:b_eq} via Eq.~\eqref{eq:dilation}.

It is useful to write Eq.~\eqref{eq:b_eq} in dimensionless time $\omega_0 t \to t$ and dimensionless frequency $\alpha(t) =\omega(t)/\omega_0$, which yields
\begin{equation}
\ddot b(t)
= -\alpha^2(t)\, b + \frac{\mathrm{sgn}(b)}{|b|^{k+1}} \, ,
\quad\text{with}\quad  b(0)=1\quad\text{and}\quad \dot b(0)=0 .
\label{eq:b_eq2}
\end{equation}

In general, it is difficult to write an explicit solution of $X_{CM}(t)$ and $b(t)$ for arbitrary $\omega(t)$ and $a(t)$. Therefore, for illustration, we consider the specific case, where $\alpha(t)=\alpha$ and $a(t)=a$ for $t>0$ are constants. In this scenario, the solution of Eq.~\eqref{eq:Xt}, in dimensionless time, is given by
\begin{equation}
\label{eq:Xsol}
    X_{CM}(t) = a (1-\cos \alpha t)\, .
\end{equation}
On the other hand, Eq.~\eqref{eq:b_eq} can be written as 
\begin{equation}
    \ddot{b}(t) = - V'(b)\quad\text{where}\quad V(b) = \frac{1}{2}\alpha^2 b^2 + \frac{1}{k \,|b|^{k}},
    \label{eq:bw1}
\end{equation}
is the effective potential. It is easy to see that the initial acceleration from Eq.~\eqref{eq:b_eq2} is given by
\begin{equation}
\label{eq:bacc0}
    \ddot{b}(0)=1-\alpha^2\, .
\end{equation}
Multiplying both sides of Eq.~\eqref{eq:bw1} by $\dot{b}(t)$ and integrating over time using the initial conditions $b(0)=1$ and $\dot{b}(0)=0$, we get
\begin{equation}
    \dot{b}^2(t) = \Phi(b) \quad\text{where} \quad \Phi(b)=2\bigl[V(1)-V(b) \bigr]=\alpha^2(1-b^2) + \frac{2}{k} \left(1-\frac{1}{|b|^k}\right) \, . 
    \label{eq:bphi}
\end{equation}

The two important parameters are the dimensionless frequency $\alpha$ and power-law exponent $k$. For $\alpha<1$ the gas experiences an expansion whereas for $\alpha>1$, it undergoes contraction. 
It is worth noting that for every $k>-2$, the effective potential $V(b)$ has a minimum on the positive side at
$b_\mathrm{min}=\alpha^{-2/(k+2)}$, where
$V''(b_\mathrm{min})=\alpha^2+(k+1)\,b_\mathrm{min}^{-(k+2)}=(k+2)\,\alpha^2>0$.
Small oscillations of $b$ about $b_\mathrm{min}$ therefore have frequency
$\sqrt{k+2}\,\alpha$ (i.e.\ $\sqrt{k+2}\,\omega$ in physical units) which
reduces to the universal breathing frequency $\sqrt{k+2}\,\omega_0$
of Ref.~\cite{henning2008} for a weak quench, $\alpha\to1$.

Before proceeding further, let us first analyze the potential $V(b)$ in Eq.~\eqref{eq:bw1} in the $(\alpha,k)$ plane.

\subsection{Structure of the potential $V(b)$ and the solution of $b(t)$ in the $(\alpha, k)$ plane}
\label{subsec:bt}

\subsubsection{$k>0$}

\begin{figure}
    \centering
    \includegraphics[width=0.45\textwidth]{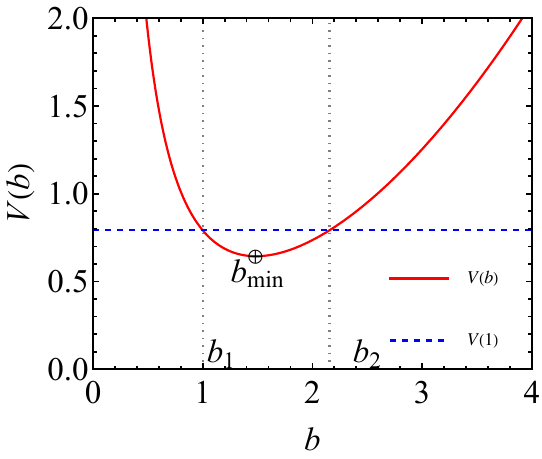}~~~~~\includegraphics[width=0.425\textwidth]{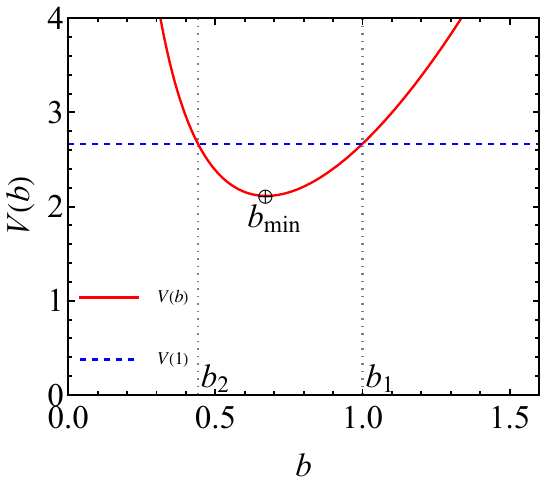}
    \caption{The red solid lines show the potential $V(b)$ in Eq.~\eqref{eq:bw1} for $\alpha = 0.5$ (left) and $\alpha=2$ (right). We take $k=1.5$. The function $b(t)$, starting with $b(0)=b_1=1$ remains confined between the turning points $b_1$ and $b_2$. The blue dashed lines represent the total energy, which is also the value of the potential $V(1)$ at $t=0$. The minimum of the potential $b_\mathrm{min}$ is shown by the plus-circle. }
    \label{fig:s1}
\end{figure}

For $k>0$, the potential $V(b)$ diverges as $b\to 0$ and $b\to \infty$, with a minimum at [see Fig.~\ref{fig:s1}] 
\begin{equation}
\label{eq:bmin}
    b_\mathrm{min} = \alpha^{-\frac{2}{k+2}}\, .
\end{equation}
Therefore, starting from $b=1$ at $t=0$, the solution $b(t)$ of Eq.~\eqref{eq:bw1} remains positive for all $t>0$. Moreover, $b(t)$ is confined within the potential with the two turning points given by $b_1=1$ and $b_2$, which is the unique positive non-trivial root ($b_2\ne 1$) of the equation $\Phi(b_2)=0$, i.e., 
\begin{equation}
    \alpha^2(1-b_2^2) + \frac{2}{k} \left(1-\frac{1}{b_2^k}\right) =0\, .
    \label{eq:b2}
\end{equation}
For $\alpha <1$, from Eq.~\eqref{eq:bmin}, $b_\mathrm{min} >1$. Therefore, $b_2 > b_\mathrm{min} >1$ [see Fig.~\ref{fig:s1} (left)]. It is also consistent with the initial acceleration in Eq.~\eqref{eq:bacc0}, $\ddot{b}(0)>0$.
On the other hand,  $0< b_\mathrm{min}<1$  and $\ddot{b}(0)<0$ for $\alpha >1$. Therefore, $0< b_2 < b_\mathrm{min} <1$ [see Fig.~\ref{fig:s1} (right)].

It is evident that $b(t)$ undergoes a periodic motion, whose time period $T$ can be determined by integrating Eq.~\eqref{eq:bphi} as
\begin{equation}
    T = 2 \left|\int_{b_1}^{b_2} \frac{db}{\sqrt{\Phi(b)}} \right|\, ,
    \label{eq:T}
\end{equation}
where we recall that $b_1=1$ and the information about $\alpha$ and $k$ are encoded in $\Phi(b)$ through Eq.~\eqref{eq:bphi}.

{\it Special case: $k=2$.---} The solution of $b(t)$ in Eq.~\eqref{eq:b_eq2} for the special case $k=2$, known as the Calogero-Moser model~\cite{Calogero1971,Abanov2011}, can be worked out explicitly. In this case, Eq.~\eqref{eq:b_eq2} reduces to the Ermakov--Pinney equation~\cite{Pinney1950},
\begin{equation}
\ddot b(t)=-\alpha^2(t)\, b+\frac{1}{b^3}\,.
\label{eq:k2-beq}
\end{equation}
For $\alpha(t)=\alpha$ constant, the function $\Phi(b)$ in Eq.~\eqref{eq:bphi} factorizes as
\begin{equation}
\Phi(b)=\frac{(1-b^2)\,(\alpha^2b^2-1)}{b^2}\,.
\label{eq:k2-phi}
\end{equation}
Therefore, the turning points are $b_1=1$ and $b_2=1/\alpha$, implying that the dynamics of $b(t)$ is confined on the positive side.  The minimum of the potential in Eq.~\eqref{eq:bmin} is at  $b_\mathrm{min}=\alpha^{-1/2}$. Moreover, $u=b^2$ obeys the linear equation $\ddot u=-4\alpha^2u+2(1+\alpha^2)$, which follows from Eqs.~\eqref{eq:k2-beq} and \eqref{eq:bphi}. Hence,
\begin{equation}
b(t)=\frac{1}{\sqrt2\,\alpha}\sqrt{(1+\alpha^2)+ (\alpha^2-1)\cos2\alpha t}\,,
\label{eq:k2-bsol}
\end{equation}
which oscillates between $b_1=1$ and $b_2=1/\alpha$ with the period $T=\pi/\alpha$. Figure.~\eqref{fig:b_exact} (left) shows the time evolution of $b(t)$ for both $\alpha<1$ (expansion) and $\alpha >1$ (compression).

\begin{figure}
    \centering
    \includegraphics[width=0.45\textwidth]{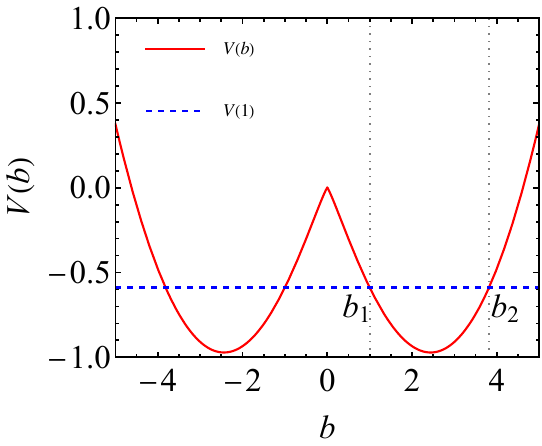}~~~~~\includegraphics[width=0.45\textwidth]{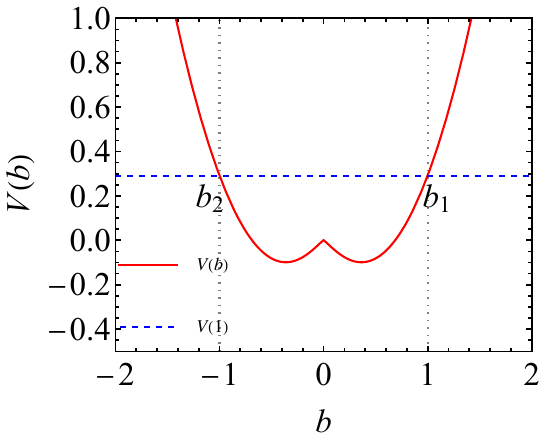}
    \caption{The red solid lines show the potential $V(b)$ in Eq.~\eqref{eq:Vb2} for $\alpha = 0.7$ (left) and $\alpha=1.5$ (right). We take $k=-1.2$ in both cases which gives $\alpha^*(k) = 1.29$ from Eq.~\eqref{eq:alphastar}. For the case $\alpha = 0.7$ (left), which satisfies  $0<\alpha<1<\alpha^*$,  the function $b(t)$, starting with $b(0)=b_1=1$ remains confined between the turning points (shown with vertical dotted lines) $b_1=1$ and $b_2>0$  given by the non-trivial root of Eq.~\eqref{eq:b2}. For the case $\alpha = 1.5$ (right), which satisfies $\alpha>\alpha^*$, the function $b(t)$, starting with $b(0)=b_1=1$ remains confined between the turning points $b_1=1$ and $b_2=-1$. The blue dashed lines represent the total energy, which is also the value of the potential $V(1)$ at $t=0$. The case $1<\alpha<\alpha^*(k)$ is not shown as it is similar to the left figure with only difference being $0<b_2<b_1$.}
    \label{fig:s2}
\end{figure}

\subsubsection{$-2<k<0$}

In this case, from Eq.~\eqref{eq:bw1}, we have
\begin{equation}
    V(b) = \frac{1}{2}\alpha^2 b^2 - \frac{|b|^{|k|}}{|k|}\, .
    \label{eq:Vb2}
\end{equation}
Therefore, 
$V(0)=0$, whereas $V(b)$ diverges as $b\to \pm\infty$. Therefore, depending on the value of $V(1)=\alpha^2/2-1/|k|$, starting with $b(0)=1$ at $t=0$, the dynamics of $b(t)$ is either confined on the positive side between the two turning points $b_1=1$ and $b_2>0$ [Fig.~\ref{fig:s2}~(left)], or it oscillates between the turning points $b_1=1$ and $b_2=-1$ [Fig.~\ref{fig:s2}~(right)]. More precisely, if $V(1) <0$, the dynamics of $b(t)$ is confined on the positive side between $b_1=1$ and $b_2>0$ given by the non-trivial root of Eq.~\eqref{eq:b2}. The condition $V(1)<0$ yields $\alpha < \alpha^*(k)$, where
\begin{equation}
    \alpha^*(k)=\sqrt\frac{2}{|k|}\, .
    \label{eq:alphastar}
\end{equation}
Since $k$ is bounded by $-2$ from below ($k>-2$), $\alpha^*(k)$ is bounded from below by unity, i.e., $\alpha^*(k)\to 1^+$ as $k\to -2^+$. Evidently, as $\alpha^*\to \infty$ as $k\to 0^-$. Therefore, $\alpha^*(k) \in (1, \infty)$.
For $\alpha <\alpha^*(k)$, the analysis is similar to that of $k>0$ case discussed above including the expression of the time period given in Eq.~\eqref{eq:T}. The turning point $b_2$ is given by the non-trivial root of Eq.~\eqref{eq:b2}: (1) $b_2>b_\mathrm{min} >1$ for $0<\alpha <1$ [Fig.~\ref{fig:s2} (left)] and (2) $0<b_2<b_\mathrm{min}<1$ for $1<\alpha<\alpha^*(k)$ [not shown in Fig.~\ref{fig:s2}].

\begin{figure}
    \centering
    \includegraphics[width=0.48\linewidth]{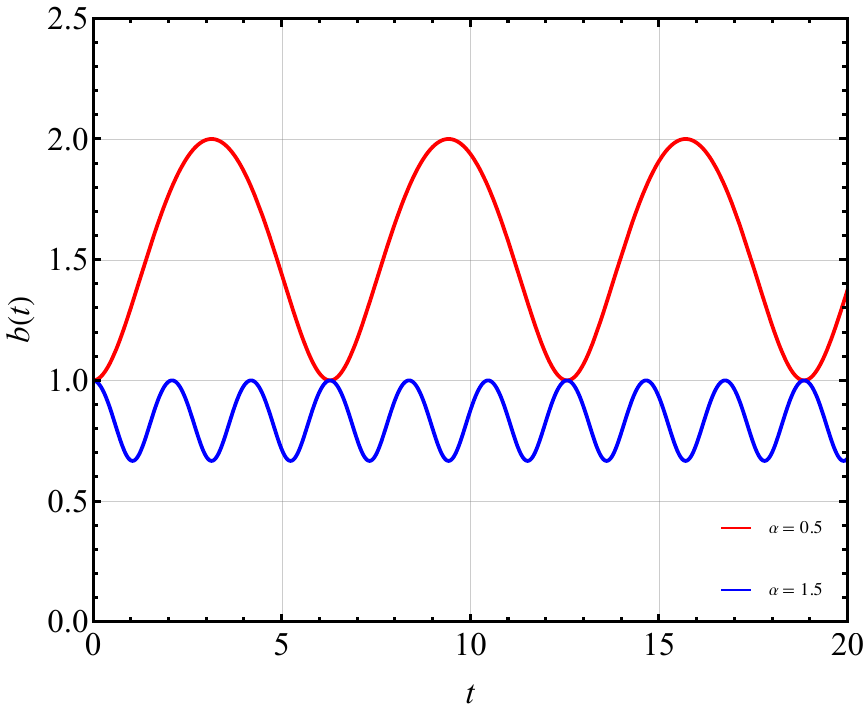}
    \includegraphics[width=0.48\linewidth]{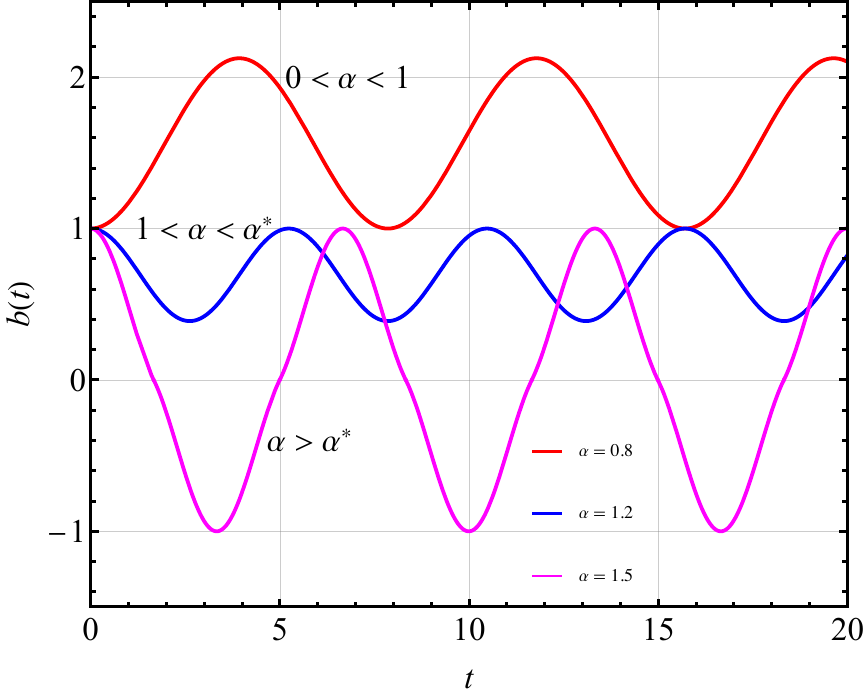}
    \caption{Time evolution of  $b(t)$ for the two exactly solvable cases.
Left: the Calogero--Moser model ($k=2$) given by Eq.~\eqref{eq:k2-bsol}.
Here $b(t)$ remains positive and oscillates between $b_1=1$ and $b_2=1/\alpha$ with period $T=\pi/\alpha$.
Right: the one-dimensional one-component plasma ($k=-1$).
Here for $\alpha <\alpha^* =\sqrt{2}$, $b(t)$ is given by Eq.~\eqref{eq:km1-bsol}. For $\alpha>\alpha^*$, $b(t)$ is given by Eq.~\eqref{eq:km1-b-2}.
}
\label{fig:b_exact}
\end{figure}

For $\alpha >\alpha^*(k)$, the initial energy $V(1)$ is greater than the potential barrier $V(0)=0$ at the origin. Since the potential $V(b)$ is symmetric about the origin ($b=0$), starting with $b_1=1$, the motion continues to the negative side till the other turning point at $b_2=-1$ [Fig.~\ref{fig:s2} (right)]. Thus, $b(t)$ undergoes a periodic motion between the two turning points $\pm 1$, with the time period given by 
\begin{equation}
    T= 2 \int_{-1}^{1} \frac{db}{\sqrt{\Phi(b)}} \, .
    \label{eq:T2}
\end{equation}

Since $b(t)$ crosses the origin, the case $\alpha>\alpha^*(k)$ corresponds to the scenario in which all particles $\tilde{x}_i(t)$ [recall Eq.~\eqref{eq:dilation}] cross the origin at the same time, causing the onset of a perfect caustic.

Exactly at the  threshold $\alpha=\alpha^*(k)$ [recall Fig.~2 of the main text], starting with $b(0)=1$, Eq.~\eqref{eq:bphi} for $b>0$ reduces to
\begin{equation}
  -  \frac{db}{\sqrt{b^{|k|}-b^2}} =\alpha^* dt\, .
\end{equation}
Integrating this equation gives 
\begin{equation}
    \frac{2}{2+k}\, \sin^{-1} \left(b^{(2+k)/2}\right) = - \alpha^* t +C\, ,
    \label{eq:b-alpha*}
\end{equation}
where $C$ is a constant to be determined from the initial condition. Indeed, using $b(0)=1$ gives $C=\pi/(2+k)$. Substituting $C$ in Eq.~\eqref{eq:b-alpha*}, after simplification we get
\begin{equation}
b(t)=\Big[\cos\Big(\frac{2+k}{2}\,\alpha^*t\Big)\Big]^{2/(2+k)}\, .
\label{eq:separatrix}
\end{equation}
Starting with $b(0)=1$, according to Eq.~\eqref{eq:separatrix}, $b(t)$ decreases monotonically till the argument of cosine function hits the value $\pi/2$, which occurs at time $t=t_c$, where
\begin{equation}
    t_c =  \frac{\pi}{(2+k)\, \alpha^*(k)}\, .
\end{equation}
Therefore, at $t=t_c$, from Eq.~\eqref{eq:separatrix} we get $b(t_c)=0$. Since $b(t)$ in Eq.~\eqref{eq:separatrix} is derived at the threshold $\alpha=\alpha^*$, it separates the solutions of $b(t)$ that crosses the origin for $\alpha>\alpha^*$ from the ones that remain on the positive side of $b$ for $\alpha<\alpha^*$. 
Therefore, we denote this special separatrix in Eq.~\eqref{eq:separatrix} by $b^*(t)$ in the main text.

\medskip
{\it Special case $k=-1$.---} The special case $k=-1$ is the well-known one dimensional one component plasma~\cite{ocp1, ocp2, ocp3, ocp4, ocp6, dhar2017}. In this case,  Eq.~\eqref{eq:bw1} [see also Eq.~\eqref{eq:Vb2}] becomes 
\begin{equation}
\label{eq:km1-beq}
  V(b)=\frac{\alpha^2b^2}{2}-|b|\,, \quad\text{and}\quad
  \ddot b(t)=-\alpha^2 b+\mathrm{sgn}(b)\, .
\end{equation}
Moreover, from Eq.~\eqref{eq:alphastar}, we have the threshold $\alpha^*=\sqrt{2}$.

For $\alpha\le\sqrt{2}$, $b(t)$ is confined on the positive side, and therefore, there is no caustic.
The solution of Eq.~\eqref{eq:km1-beq} is
\begin{equation}
b(t)=\frac{1}{\alpha^2}+\Big(1-\frac{1}{\alpha^2}\Big)\cos\alpha t\,,
\label{eq:km1-bsol}
\end{equation}
which oscillates between $b_1=1$ and $b_2=2/\alpha^2-1$ [the root of Eq.~\eqref{eq:b2}] with the time period $T=2\pi/\alpha$. At the threshold $\alpha=\sqrt{2}$, $b(t)=\cos^2(\alpha t/2)$ touches zero with zero velocity, signaling the onset of the caustic. Figure~\eqref{fig:b_exact}~(right) shows that $b(t)$ is confined on the positive side for the $0<\alpha<1$ (red) and $1<\alpha<\alpha^*$ (blue).

For $\alpha>\sqrt{2}$, 
Eq.~\eqref{eq:km1-bsol} holds until $b(t)$ reaches zero at
\begin{equation}
t_c=\frac{1}{\alpha}\cos^{-1}\!\Big(\frac{-1}{\alpha^2-1}\Big)\,,
\label{eq:km1-tc}
\end{equation}
when all particles pass simultaneously through origin with $\dot b(t_c)=-\sqrt{\alpha^2-2}$. By the symmetry of $V(b)$, the motion after the crossing the origin, mirrors the motion before it. Therefore, the time period $T=4t_c$ [which is Eq.~\eqref{eq:T2} in closed form] and we have
\begin{equation}
    b(t) =\begin{cases}
    \displaystyle
        b_+(t-4nt_c) & \text{for}\quad (4n-1)\,t_c < t<(4n+1)\,t_c \\[3mm]
        \displaystyle
        -  b_+(t-(4n+2)\,t_c) & \text{for}\quad (4n+1)\,t_c < t<(4n+3)t_c
    \end{cases}
    \label{eq:km1-b-2}
\end{equation}
where $n=\left\lfloor \frac{t+t_c}{4t_c}\right\rfloor$ and 
\begin{equation}
b_+(t)=\frac{1}{\alpha^2}+\Big(1-\frac{1}{\alpha^2}\Big)\cos\alpha t\,.
\label{eq:km1-bsol+}
\end{equation}
Figure.~\eqref{fig:b_exact}~(right) shows that $b(t)$ spans on both positive and negative sides for $\alpha>\alpha^*$ (magenta).

\section{Collective equations of motion}
\label{sec:collective}

For large $N$, an alternative approach to studying the behaviour of this system is to investigate the collective field theory. In this continuum description, the individual positions and momenta $\{x_i, p_i=mv_i\}$ of the particles are replaced by a macroscopic density profile $\rho(x)$, normalized to unity, $\int_{-\infty}^\infty\rho(x)\, dx=1$, and a velocity field $v(x)$. The description applies when the local inter-particle spacing, $1/[N\rho(x)]$, is much smaller than the length scale over which $\rho(x)$ varies, and when the velocities of nearby particles differ only slightly. Our aim is to express the energy $H$ in Eq.~\eqref{Hamiltonian}, which is given in terms of the microscopic configuration $\{x_i, p_i\}$, as a functional of $\rho(x)$ and $v(x)$. This energy functional is the central object of the continuum description. The equilibrium density is the profile that minimizes it with $v=0$, among all profiles satisfying the normalization $\int_{-\infty}^\infty\rho\,dx=1$. The functional derivative of the energy gives the force that drives the evolution of $\rho$ and $v$ (Sec.~\ref{sec:hydro}). Obtaining both from the same functional ensures that the statics and the dynamics are consistent with each other. Since $\mathcal{E}$ contains no entropy, its minimizer is, strictly speaking, the equilibrium density at zero temperature. At a finite temperature $T$ (with the Boltzmann constant set to unity), one should instead minimize the free energy $\mathcal{E}-TS$, where $S=-N\int_{-\infty}^{\infty}\rho\ln\rho\,dx$ is the entropy of the density profile. The entropy grows with $N$ essentially linearly, up to a logarithmic factor, whereas the energy grows faster, since the energy per particle increases as a power of $N$. Therefore, at any temperature that does not grow with $N$, the entropy is negligible for large $N$, and the minimizer of $\mathcal{E}$ gives the equilibrium density to leading order.

All the continuum approximations discussed below assume coarse-graining over an interval $(x, x+dx)$, which is short compared with the scale $L$ over which $\rho$ and $v$ vary, yet contain many particles,
\begin{equation}
    n_{dx}(x) \approx N\rho(x)\, dx \gg 1 ,
    \label{eq:ndx}
\end{equation}
where $n_{dx}(x)$ denotes the number of particles in $(x,x+dx)$.

Let us first consider the confining potential term, which is of the form
\begin{equation}
  \sum_{i=1}^N f(x_i) \quad\text{with}\quad  f(x) = \frac{1}{2} m \omega^2\, x^2,
  \label{eq:sumfi}
\end{equation}
where we have set dynamics of the trap center $a(t)=0$ for simplicity. The sum over the particle index $i$ can equivalently be computed by dividing space into disjoint intervals of length $dx$, summing within each interval $x_i\in (x, x+dx)$, and then summing over all the disjoint intervals $(x, x+dx)$, i.e.,
\begin{equation}
   \sum_{i=1}^N f(x_i)=\sum_{(x, x+dx)}\ \sum_{x_i\in (x, x+dx)} f(x_i)\, .
   \label{eq:sumxi}
\end{equation}
Since the variation of $f(x_i)$ within an interval $x_i\in(x,x+dx)$ is very small, we can approximate $f(x_i)\approx f(x)$ for every particle in the interval. Therefore, the inner sum in Eq.~\eqref{eq:sumxi} becomes 
becomes
\begin{equation}
\sum_{x_i\in (x, x+dx)} f(x_i) \approx   f(x)\, n_{dx}(x)\approx f(x)\, N\rho(x)\, dx\, ,
\label{eq:innerS}
\end{equation}
where we recall $n_{dx}(x)$ from Eq.~\eqref{eq:ndx}.
Finally, substituting Eq.~\eqref{eq:innerS} in Eq.~\eqref{eq:sumxi}, we get
\begin{equation}
   \sum_{i=1}^N f(x_i) \approx \sum_{(x, x+dx)} f(x)\, N\rho(x)\, dx = N \int_{-\infty}^\infty f(x)\, \rho(x)\, dx .
   \label{eq:intxi}
\end{equation}

An alternative way to arrive at Eq.~\eqref{eq:intxi} is to start with the empirical density (normalized to unity)
\begin{equation}
    \hat{\rho}(x)=\frac{1}{N}\sum_{i=1}^N\delta(x-x_i)\, ,
    \label{eq:rhohat}
\end{equation}
and the identity
\begin{equation}
    f(x_i) = \int_{-\infty}^\infty f(x)\, \delta(x-x_i)\, dx.
    \label{eq:iden1}
\end{equation}
Summing Eq.~\eqref{eq:iden1} over $i$ and using Eq.~\eqref{eq:rhohat}, we get
\begin{equation}
 \sum_{i=1}^N f(x_i) = N\int_{-\infty}^\infty f(x)\, \hat{\rho}(x)\, dx \, .
 \label{eq:intx}
\end{equation}
Equation~\eqref{eq:intx} is exact for all $N$. For large $N$, the empirical density $\hat{\rho}(x)$ may be replaced by the macroscopic density profile $\rho(x)$ inside the integral in Eq.~\eqref{eq:intx}. This does not mean that $\hat{\rho}(x)\to\rho(x)$ pointwise, since $\hat{\rho}(x)$ is a sum of delta functions. Rather, for any $f(x)$ that varies slowly on the scale of the inter-particle spacing,
\begin{equation}
    \int_{-\infty}^\infty f(x)\, \hat{\rho}(x)\, dx \;\longrightarrow\; \int_{-\infty}^\infty f(x)\, \rho(x)\, dx ,
\end{equation}
which gives Eq.~\eqref{eq:intxi}.

The kinetic energy is treated in the same way. In the continuum description, the velocity of each particle is given by the local velocity field, $v_i\approx v(x_i)$. The kinetic energy is then a one-body sum of the form~\eqref{eq:sumfi} with $f(x)=\frac{1}{2}m v^2(x)$, and Eq.~\eqref{eq:intxi} applies directly. Here we assume that all particles near $x$ move with the same velocity $v(x)$. A local spread of velocities about $v(x)$, with variance $\sigma^2(x)$ adds a thermal kinetic energy $\frac{m}{2}N\int_{-\infty}^\infty\rho(x)\,\sigma^2(x)\,dx$, which for $\sigma^2(x) \propto T(x)$ gives rise to the thermal pressure $\rho(x) T(x)$ in the Euler equation for the velocity field~\cite{mukherjee2026}. In our zero temperature (or restricted to low temperature) description, we do not consider this contribution.  Therefore, the noninteracting part of the Hamiltonian in Eq.~\eqref{Hamiltonian} becomes
\begin{equation}
    H_0 (\{x_i, v_i\})=\sum_{i=1}^N \left[\frac{1}{2}m v_i^2 + \frac{1}{2} m \omega^2 x_i^2 \right] \approx \mathcal{E}_0 [\rho(x), v(x)] ,
    \label{eq:H0}
\end{equation}
where
\begin{equation}
    \mathcal{E}_0 [\rho(x), v(x)]  = \frac{m}{2}\, N\int_{-\infty}^\infty \rho(x)\, v^2(x)\, dx + \frac{m\omega^2}{2}\, N\int_{-\infty}^\infty x^2\, \rho(x)\, dx.
    \label{eq:mcH0}
\end{equation}

\begin{figure}[t]
    \centering
    \includegraphics[width=.8\linewidth]{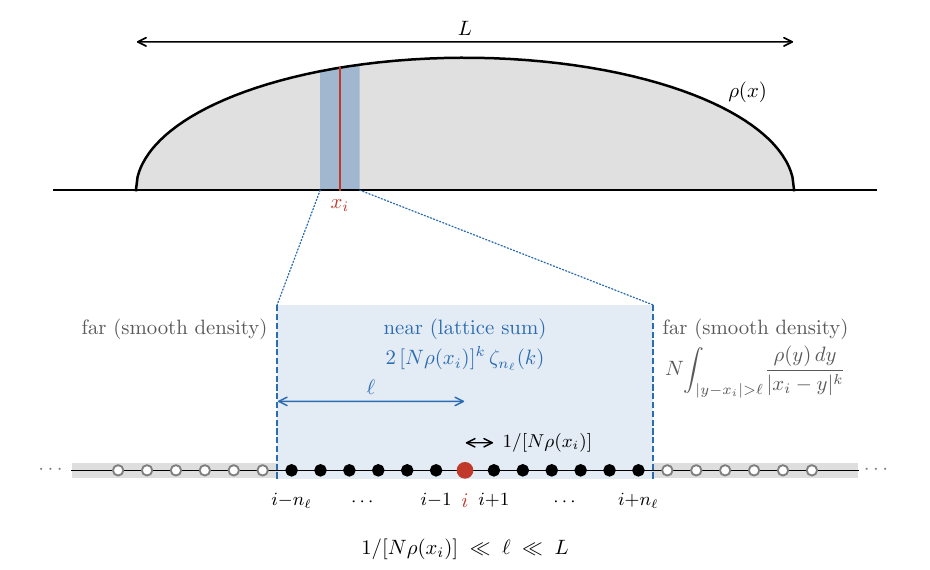}
    \caption{Schematic of the split of the pair interaction into near and far parts in Eq.~\eqref{eq:sum-brk}. The upper panel shows the macroscopic density $\rho(x)$, which varies on the scale $L$, with a window of half-width $\ell$ around the position $x_i$ of the $i$-th particle. The lower panel zooms into this window. Within distance $\ell$ of $x_i$ (shaded in blue), the particles are locally equally spaced at $1/[N\rho(x_i)]$, and their contribution is the lattice sum $2[N\rho(x_i)]^k\zeta_{n_\ell}(k)$ [Eq.~\eqref{eq:nearS1}]. Beyond $\ell$, the particles are replaced by the smooth density $\rho(y)$, and their contribution becomes an integral [Eq.~\eqref{eq:xixj2}]. Both approximations rely on the separation of scales $1/[N\rho(x_i)]\ll\ell\ll L$ [Eq.~\eqref{eq:scale-sep}].}
    \label{fig:near-far}
\end{figure}

Let us now consider the interaction term in Eq.~\eqref{Hamiltonian},
\begin{equation}
    V_\mathrm{int}(\{x_i\}) =  \frac{J\,\text{sgn}(k)}{2}\sum_{i=1}^N\sum_{\substack{j=1\\ j\neq i}}^{N}\frac{1}{|x_i-x_j|^k}\, .
    \label{eq:Vint}
\end{equation}
Replacing both sums by integrals, as in the one-body case gives $N^2\iint \rho(x)\rho(y)\,|x-y|^{-k}\,dx\,dy$. However, this is justified only for $k<1$, where the integral is convergent. For $k\geq 1$ the kernel $|x-y|^{-k}$ is not integrable at $y=x$, and the sum is dominated by close neighbours, whose separations are set by the microscopic arrangement of the particles rather than by the smooth density. We therefore treat near and far neighbours separately. For each particle $i$, we split the sum over $j$ at a mesoscopic distance $\ell$, which is much larger than the local spacing $1/[N\rho(x_i)]$ but much smaller than the scale $L$ over which $\rho$ varies (see Fig.~\ref{fig:near-far})
\begin{eqnarray}
\frac{1}{N\rho(x_i)}\ll\ell\ll L\, .
\label{eq:scale-sep}
\end{eqnarray}
Because of this separation of the scales, it is convenient to express
\begin{equation}
    \sum_{i=1}^N\sum_{\substack{j=1\\ j\neq i}}^{N} \frac{1}{|x_i-x_j|^k}
    = \sum_{i=1}^N \sum_{0<|x_j-x_i|<\ell} \frac{1}{|x_i-x_j|^k}
    + \sum_{i=1}^N \sum_{|x_j-x_i|>\ell} \frac{1}{|x_i-x_j|^k}\, .
    \label{eq:sum-brk}
\end{equation}

In the second (far) sum on the right-hand side of Eq.~\eqref{eq:sum-brk}, the separations are much larger than the local spacing and the summand varies smoothly, so both sums can be converted into integrals exactly as in the one-body case, giving
\begin{equation}
    \sum_{i=1}^N \sum_{|x_j-x_i|>\ell} \frac{1}{|x_i-x_j|^k}
    \approx N^2 \int_{-\infty}^\infty dx\, \rho(x) \int_{|y-x|>\ell} dy\, \frac{\rho(y)}{|x-y|^k}\, .
    \label{eq:xixj2}
\end{equation}

In the first (near) sum on the right-hand side of Eq.~\eqref{eq:sum-brk}, the density is essentially constant, equal to $\rho(x_i)$, over the range $|x_j-x_i|<\ell$. We further assume that on this scale the particles are locally equally spaced (locally crystalline order), as in the ground state. Labelling the particles in order of position, the $m$-th neighbour on either side of particle $i$, $j=i\pm m$, then lies at distance
\begin{equation}
    |x_i-x_{i\pm m}| \approx \frac{m}{N\rho(x_i)}, \quad\text{so that}\quad
    \frac{1}{|x_i-x_{i\pm m}|^k} \approx \frac{[N\rho(x_i)]^k}{m^k},\quad\text{with}\quad m=1,2,\ldots,n_\ell\, ,
    \label{eq:lattice}
\end{equation}
where $n_\ell = N\rho(x_i)\,\ell \gg 1$ is the number of neighbours (for a given $x_i$) on each side within distance $\ell$. Summing over $m$, and over the neighbours on both sides, gives
\begin{equation}
    \sum_{0<|x_j-x_i|<\ell} \frac{1}{|x_i-x_j|^k} \approx 2\,[N\rho(x_i)]^k\, \zeta_{n_\ell}(k),
    \quad\text{where}\quad
    \zeta_{n_\ell}(k) =\sum_{m=1}^{n_\ell} \frac{1}{m^k}
    \label{eq:nearS1}
\end{equation}
is the truncated Riemann zeta function. The right-hand side of Eq.~\eqref{eq:nearS1} is a smooth function of $x_i$, and summing it over $i$ using Eq.~\eqref{eq:intxi} gives
\begin{equation}
    \sum_{i=1}^N \sum_{0<|x_j-x_i|<\ell} \frac{1}{|x_i-x_j|^k}
    \approx 2 N^{k+1}\int_{-\infty}^\infty dx\, [\rho(x)]^{k+1}\, \zeta_{N\rho(x)\ell}(k)\, .
    \label{eq:firstsum}
\end{equation}

Separately, Eqs.~\eqref{eq:xixj2} and \eqref{eq:firstsum} depend on the arbitrary length $\ell$, but their sum cannot, since the left-hand side of Eq.~\eqref{eq:sum-brk} does not. The cancellation is made explicit by the large-$n$ behaviour of the truncated zeta function,
\begin{equation}
    \zeta_n(k) = \zeta(k) + \frac{n^{1-k}}{1-k} + O(n^{-k}) \qquad (k\neq 1),
    \label{eq:zeta-asym}
\end{equation}
which follows from the Euler--Maclaurin formula. Here $\zeta(k)$ is the Riemann zeta function, analytically continued to $k<1$. The term $n^{1-k}/(1-k)$ in Eq.~\eqref{eq:zeta-asym} is what one obtains by replacing the sum over $m$ by an integral, $\sum_m m^{-k}\to\int dm\, m^{-k}$, i.e., by treating the near neighbours as a continuum. With $n=N\rho(x)\ell$, its contribution to Eq.~\eqref{eq:firstsum} is
\begin{equation}
    2N^{k+1}\int_{-\infty}^\infty dx\,[\rho(x)]^{k+1}\,\frac{[N\rho(x)\,\ell]^{1-k}}{1-k}
    = \frac{2\ell^{1-k}}{1-k}\,N^2\int_{-\infty}^\infty [\rho(x)]^2\,dx\, ,
    \label{eq:ell-term}
\end{equation}
which is of order $N^2$, the same order as the far part~\eqref{eq:xixj2}. As we show below, this term cancels the $\ell$-dependence of the far part, both for $k<1$ and for $k>1$. The remaining term, $\zeta(k)$, is independent of $\ell$ and accounts for the discreteness of the particles. To proceed further, we consider the two cases separately.

\subsection{$-2<k<1$}
For $k<1$, the kernel $|x-y|^{-k}$ is integrable at $y=x$. The far part~\eqref{eq:xixj2} is therefore the full double integral minus the contribution of the strip $|y-x|<\ell$. Since $\rho(y)\approx\rho(x)$ within the strip, we get
\begin{equation}
    N^2 \int_{-\infty}^\infty dx\, \rho(x) \int_{|y-x|>\ell} dy\, \frac{\rho(y)}{|x-y|^k}
    \approx N^2 \int_{-\infty}^\infty dx \int_{-\infty}^\infty dy\, \frac{\rho(x)\rho(y)}{|x-y|^k}
    - \frac{2\ell^{1-k}}{1-k}\, N^2 \int_{-\infty}^\infty [\rho(x)]^2\, dx\, .
    \label{eq:fark-1}
\end{equation}
Using Eq.~\eqref{eq:zeta-asym} with $n=N\rho(x)\ell$, the near part~\eqref{eq:firstsum} becomes
\begin{equation}
    2 N^{k+1}\int_{-\infty}^\infty dx\, [\rho(x)]^{k+1}\, \zeta_{N\rho(x)\ell}(k)
    \approx \frac{2\ell^{1-k}}{1-k}\, N^2 \int_{-\infty}^\infty [\rho(x)]^2\, dx
    + 2\zeta(k)\, N^{k+1} \int_{-\infty}^\infty [\rho(x)]^{k+1}\, dx\, .
    \label{eq:neark-1}
\end{equation}
The $\ell$-dependent terms cancel when adding Eqs~\eqref{eq:fark-1} and \eqref{eq:neark-1}, and Eq.~\eqref{eq:sum-brk} becomes
\begin{equation}
    \sum_{i=1}^N\sum_{\substack{j=1\\ j\neq i}}^{N} \frac{1}{|x_i-x_j|^k}
    \approx N^2 \int_{-\infty}^\infty dx \int_{-\infty}^\infty dy\, \frac{\rho(x)\rho(y)}{|x-y|^k}
    + 2\zeta(k)\, N^{k+1} \int_{-\infty}^\infty [\rho(x)]^{k+1}\, dx\, .
    \label{eq:k-1}
\end{equation}
Both integrals in Eq.~\eqref{eq:k-1} have the same dimension $L^{-k}$, where $L$ is the characteristic length scale over which the density $\rho(x)$ varies. Therefore, the second term in Eq.~\eqref{eq:k-1} is smaller than the first by a relative factor of order $1/N^{1-k}$, which vanishes in the large $N$ limit for $k<1$. To leading order, the interaction term~\eqref{eq:Vint} is therefore given by the naive continuum limit,
\begin{equation}
    V_\mathrm{int}(\{x_i\}) \approx \mathcal{E}_\mathrm{int}[\rho(x)] \quad\text{with}\quad
    \mathcal{E}_\mathrm{int}[\rho(x)] =  \frac{J\,\text{sgn}(k)}{2}\, N^2 \int_{-\infty}^\infty dx \int_{-\infty}^\infty dy \,\frac{\rho(x)\rho(y)}{|x-y|^k}\, .
    \label{eq:Vint1}
\end{equation}

\subsection{$k>1$}
For $k>1$, the kernel is not integrable at $y=x$, and the far part~\eqref{eq:xixj2} diverges as $\ell\to0$. To isolate the divergence, we write $\rho(y)=\rho(x)+[\rho(y)-\rho(x)]$ in the inner integral. Since $\int_{|y-x|>\ell}|x-y|^{-k}\,dy = 2\ell^{1-k}/(k-1)$, this gives
\begin{equation}
    \int_{|y-x|>\ell} dy\, \frac{\rho(y)}{|x-y|^k}
    = \frac{2\ell^{1-k}}{k-1}\, \rho(x) + \int_{|y-x|>\ell} dy\, \frac{\rho(y)-\rho(x)}{|x-y|^k}\, .
    \label{eq:isolate-div}
\end{equation}
The first term carries the divergence. The numerator in the second term vanishes at $y=x$, which softens the singularity, and for $1<k<3$ the integral has a finite limit as $\ell\to0$ (for $2\leq k<3$ owing to the cancellation of the term linear in $y-x$, which is odd). For $k\geq3$, higher-order terms in the Taylor expansion of $\rho(y)$ about $x$ produce further divergent terms. These involve derivatives of $\rho$, and therefore cannot be canceled by the uniform-lattice  [Eq.~\eqref{eq:lattice}] result in Eq.~\eqref{eq:zeta-asym}. Since the left-hand side of Eq.~\eqref{eq:sum-brk} does not depend on $\ell$, they must cancel against corrections to the near sum from the variation of the spacing within the window, which we have neglected. After this cancellation, only terms of lower order in $N$ remain. Multiplying by $N^2\rho(x)$ and integrating over $x$, the far part~\eqref{eq:xixj2} becomes
\begin{equation}
    N^2 \int_{-\infty}^\infty dx\, \rho(x) \int_{|y-x|>\ell} dy\, \frac{\rho(y)}{|x-y|^k}
    = \frac{2\ell^{1-k}}{k-1}\, N^2 \int_{-\infty}^\infty [\rho(x)]^2\, dx + N^2 I_k \, ,
    \label{eq:far-large-k}
\end{equation}
where
\begin{equation}
    I_k = \int_{-\infty}^\infty dx\, \rho(x) \int_{|x-y>\ell} dy\, \frac{\rho(y)-\rho(x)}{|x-y|^k}
\end{equation}
and $\ell$-dependence in $I_k$ cancels with additional terms arising from considering the variation of the interparticle density in the lattice-sum. In the near part, the series defining $\zeta_n(k)$ now converges, and Eq.~\eqref{eq:zeta-asym} gives $\zeta_n(k)=\zeta(k)-n^{1-k}/(k-1)+\cdots$. With $n=N\rho(x)\ell$, the near part~\eqref{eq:firstsum} becomes
\begin{equation}
    2 N^{k+1}\int_{-\infty}^\infty dx\, [\rho(x)]^{k+1}\, \zeta_{N\rho(x)\ell}(k)
    \approx 2\zeta(k)\, N^{k+1} \int_{-\infty}^\infty [\rho(x)]^{k+1}\, dx
    - \frac{2\ell^{1-k}}{k-1}\, N^2 \int_{-\infty}^\infty [\rho(x)]^2\, dx\, .
    \label{eq:near-large-k}
\end{equation}
The truncation of the near sum thus produces the same $\ell^{1-k}$ term as the divergence of the far part, with the opposite sign. This is to be expected, since the near sum leaves out precisely the neighbours just beyond $\ell$, and their contribution, computed as an integral, is the divergent part of the far integral. Adding Eqs.~\eqref{eq:far-large-k} and \eqref{eq:near-large-k}, the $\ell$-dependent terms cancel, and
\begin{equation}
    \sum_{i=1}^N\sum_{\substack{j=1\\ j\neq i}}^{N} \frac{1}{|x_i-x_j|^k}
    \approx 2\zeta(k)\, N^{k+1} \int_{-\infty}^\infty [\rho(x)]^{k+1}\, dx + N^2 I_k\, .
\end{equation}
The first term, which comes from the near neighbours, dominates the second by a factor of order $N^{k-1}$. Thus, for $k>1$ the interaction energy is local in the density. Since $\text{sgn}(k)=1$, to leading order we get
\begin{equation}
    V_\mathrm{int}(\{x_i\}) \approx \mathcal{E}_\mathrm{int}[\rho(x)] \quad\text{with}\quad
    \mathcal{E}_\mathrm{int}[\rho(x)] =  J\,\zeta(k)\, N^{k+1} \int_{-\infty}^\infty  [\rho(x)]^{k+1}\,  dx \, .
    \label{eq:Vint2}
\end{equation}

\section{Hydrodynamic equations}
\label{sec:hydro}

The energy functional 
\begin{equation}
  \mathcal{E}[\rho,v]=\mathcal{E}_0[\rho,v]+\mathcal{E}_\mathrm{int}[\rho]\, ,
  \label{eq:energy-func}
\end{equation}
given by Eqs.~\eqref{eq:mcH0}, \eqref{eq:Vint1} and \eqref{eq:Vint2}, determines the time evolution of the density $\rho(x,t)$ and the velocity field $v(x,t)$. The continuity equation follows from the conservation of particles alone. The energy functional enters through the Euler equation, where it supplies the force on a particle in terms of $\rho$, instead of the positions of all the other particles. For $k>1$ this is a considerable simplification. On a locally uniform lattice the forces from the neighbours on the two sides cancel, and the net force arises only from the density gradient. Computing it directly from the microscopic would require expanding the lattice sums to first order in the gradient. In contrast, $\mathcal{E}_\mathrm{int}$ required only the uniform lattice sum $\zeta(k)$, and its functional derivative gives the force at once.

The most direct way to obtain the evolution equations is to start from Newton's equations for the particles and take the continuum limit in Lagrangian coordinates, in which each particle is followed by its label.
We label the particles in order of position. For the smooth flows considered here, nearby particles move together and this order is preserved in time. We assign to the $i$-th particle the label $s=i/N$, which becomes a continuous variable $s\in(0,1)$ for large $N$. The positions $x_i(t)$ are then described by a smooth function $X(s,t)$, with 
\begin{equation}
    x_i(t)=X(i/N,t). 
    \label{eq:xiXs}
\end{equation}
Evidently, the number of particles whose  labels lie in $(s,s+ds)$ is $n_{ds}= Nds$. On the other hand, the positions on these particles lie in the spatial interval $(X, X+\partial_s X\, ds)$, and therefore, the number of particles within this spatial interval is alternatively given by $n_{dX}=N\rho(X,t)\,\partial_s X\, ds$ particles. Since, these are the same particles, setting $n_{ds}=n_{dX}$ we get
\begin{equation}
    \rho\big(X(s,t),t\big)\, \partial_s X(s,t) = 1 .
    \label{eq:rho-label}
\end{equation}
The velocity field is the velocity of the particle at that position, i.e., 
\begin{equation}
    v\big(X(s,t),t\big) = \partial_t X(s,t) .
    \label{eq:v-label}
\end{equation}

\subsection{Continuity equation}

Differentiating Eq.~\eqref{eq:rho-label} with respect to $t$ at fixed $s$ gives
\begin{equation}
    \big(\partial_t\rho + v\,\partial_x\rho\big)\, \partial_s X + \rho\, \partial_s \partial_t X = 0 .
    \label{eq:E1}
\end{equation}
From Eq.~\eqref{eq:v-label}, $\partial_s\partial_t X = \partial_s v(X,t) = \partial_x v\, \partial_s X$. Substituting this in the last term of Eq.~\eqref{eq:E1}, then dividing the resulting equation by $\partial_s X$ and combining the two spatial derivative terms, we obtain the continuity equation
\begin{equation}
    \partial_t \rho + \partial_x(\rho\, v) = 0 .
    \label{eq:continuity}
\end{equation}
This equation expresses the conservation of the number of particles and does not rely on any other details. Indeed, the empirical density in Eq.~\eqref{eq:rhohat} satisfies $\partial_t\hat\rho + \partial_x\big[\frac{1}{N}\sum_{i} v_i\,\delta(x-x_i)\big]=0$ exactly.

\subsection{Euler equation}

The dynamics enters through Newton's equations
\begin{equation}
    m\,\ddot{x}_i = -\frac{\partial U}{\partial x_i},
    \qquad U(\{x_i\}) = \sum_{i=1}^N \frac{1}{2}m\omega^2 x_i^2 + V_\mathrm{int}(\{x_i\}),
    \label{eq:newton}
\end{equation}
where $U$ is the potential energy. In the continuum limit, $U(\{x_i\})\approx\mathcal{U}[\rho]$, with
\begin{equation}
    \mathcal{U}[\rho] = \frac{m\omega^2}{2}\, N\int_{-\infty}^\infty x^2\, \rho(x)\, dx + \mathcal{E}_\mathrm{int}[\rho] .
    \label{eq:U-rho}
\end{equation}
The kinetic part of $\mathcal{E}_0$ does not contribute to the force. It gives the inertial term on the left-hand side of Eq.~\eqref{eq:newton}.

To express $\partial U/\partial x_i$ in terms of the functional derivative $\delta\mathcal{U}/\delta\rho$, consider a small displacement of the particles, $X(s)\to X(s)+\delta X(s)$. Varying Eq.~\eqref{eq:rho-label} at fixed $s$, in the same way as in the derivation of Eq.~\eqref{eq:continuity}, the corresponding change of the density is
\begin{equation}
    \delta\rho = -\partial_x\big(\rho\, \delta X\big) .
    \label{eq:del-rho}
\end{equation}
The change in the potential energy is therefore
\begin{equation}
    \delta\mathcal{U} = \int_{-\infty}^\infty dx\, \frac{\delta\mathcal{U}}{\delta\rho(x)}\, \delta\rho(x)
    = \int_{-\infty}^\infty dx\, \rho(x)\, \delta X\, \partial_x \frac{\delta\mathcal{U}}{\delta\rho(x)} 
    = \int_0^1 ds\, \delta X(s)\, \partial_x \frac{\delta\mathcal{U}}{\delta\rho(x)}\bigg|_{x=X(s)} ,
    \label{eq:dU1}
\end{equation}
where we have first used Eq.~\eqref{eq:del-rho}, then integrated by parts, and finally used $\rho\, dx = ds$ from Eq.~\eqref{eq:rho-label}. On the other hand, with $\delta x_i = \delta X(i/N)$ and the sum over $i$ replaced by $N\int_0^1 ds$,
\begin{equation}
    \delta U = \sum_{i=1}^N \frac{\partial U}{\partial x_i}\, \delta x_i \approx N \int_0^1 ds\, \frac{\partial U}{\partial x_i}\, \delta X(s) .
     \label{eq:dU2}
\end{equation}
Comparing Eq.~\eqref{eq:dU1} and \eqref{eq:dU2} gives
\begin{equation}
    \frac{\partial U}{\partial x_i} \approx \frac{1}{N}\, \partial_x \frac{\delta\mathcal{U}}{\delta\rho(x)}\bigg|_{x=x_i} .
\end{equation}
Finally, differentiating Eq.~\eqref{eq:v-label} with respect to $t$ gives the acceleration of a particle as the material derivative of the velocity field, 
\begin{equation}
    \partial_t^2 X = \partial_t v + v\,\partial_x v. 
\end{equation}
Newton's equations~\eqref{eq:newton} then become
\begin{equation}
    \partial_t v + v\, \partial_x v = -\omega^2 x - \frac{1}{mN}\, \partial_x \frac{\delta\mathcal{E}_\mathrm{int}}{\delta\rho(x)} .
    \label{eq:euler}
\end{equation}

Equations~\eqref{eq:continuity} and \eqref{eq:euler} form a closed set for $\rho(x,t)$ and $v(x,t)$. Setting $v=0$ and integrating Eq.~\eqref{eq:euler} once, we get
\begin{equation}
    \frac{1}{2} m\omega^2 x^2 + \frac{1}{N}\, \frac{\delta\mathcal{E}_\mathrm{int}}{\delta\rho(x)} = \mu ,
    \label{eq:stationary}
\end{equation}
where $\mu$ is a constant. This is the condition for the minimum of $\mathcal{U}[\rho]$, given in Eq.~\eqref{eq:U-rho}, under the constraint $\int_{-\infty}^\infty\rho\, dx = 1$, with $\mu$ being the Lagrange multiplier. Hence, the equilibrium density is a stationary solution of the hydrodynamic equations, as it should be.

We now evaluate the interaction term in Eq.~\eqref{eq:euler} in the two regimes.

\subsubsection{$-2<k<1$}

Using $\mathcal{E}_\mathrm{int}$ from Eq.~\eqref{eq:Vint1},
\begin{equation}
    \frac{1}{N}\, \frac{\delta\mathcal{E}_\mathrm{int}}{\delta\rho(x)} = J\, \text{sgn}(k)\, N \int_{-\infty}^\infty dy\, \frac{\rho(y)}{|x-y|^k} ,
\end{equation}
where the factor $1/2$ in Eq.~\eqref{eq:Vint1} is cancelled by the two equal contributions from varying $\rho(x)$ and $\rho(y)$. Using $\partial_x |x-y|^{-k} = -k\, \text{sgn}(x-y)\, |x-y|^{-k-1}$ and $k\, \text{sgn}(k) = |k|$, Eq.~\eqref{eq:euler} becomes
\begin{equation}
    \partial_t v + v\, \partial_x v = -\omega^2 x + \frac{J|k|N}{m}\, \fint_{-\infty}^\infty dy\, \frac{\text{sgn}(x-y)\, \rho(y)}{|x-y|^{k+1}} .
    \label{eq:euler-small-k}
\end{equation}
The force is nonlocal, i.e., it depends on the density everywhere and not only near $x$. For $0<k<1$ the integral exists only as a Cauchy principal value, denoted by $\fint$. Since the kernel $\text{sgn}(x-y)\,|x-y|^{-k-1}$ is odd about $y=x$, the contributions from the two sides cancel, and the principal value is finite for $k<1$. In the limit $k\to 0$ with $J|k|$ held fixed, the force becomes proportional to $\fint dy\, \rho(y)/(x-y)$, which is the Hilbert transform of the density familiar from the log-gas.

The stationary condition~\eqref{eq:stationary} gives the equilibrium density as the solution of the integral equation 
 \begin{eqnarray}
     \frac{1}{2}m\, \omega^2 x^2 + J\, \text{sgn}(k)\, N \int_{-\infty}^\infty dy\, \frac{\rho(y)}{|x-y|^k} =\mu\, . 
     \label{eq:inteq}
 \end{eqnarray}
Inverting this equation yields~\cite{Riesz_avg_density_SA,Sonin1954}, 
\begin{equation}
    \rho(x) = \frac{1}{B}\, \left[1-\left(\frac{x}{x_0}\right)^2\right]^{\frac{k+1}{2}} \quad\text{for}\quad |x|<x_0, \quad\text{where}\quad x_0 = \sqrt{\frac{2k\,\mu}{m\,\omega^2}}\quad\text{and}\quad B= \frac{J\,\text{sgn}(k)\,N\,\Gamma\big(\frac{1-k}{2}\big)\,\Gamma\big(\frac{k+3}{2}\big)\,x_0^{1-k}}{\mu}\, .
\end{equation}
The density $\rho(x)=0$ outside the support $(-x_0,x_0)$ and the Lagrange multiplier $\mu$ is fixed by the normalization of the density to unity, i.e., $\int_{-\infty}^\infty\rho(x)\, dx=1$.
\subsubsection{$k>1$}

From Eq.~\eqref{eq:Vint2},
\begin{equation}
    \frac{1}{N}\, \frac{\delta\mathcal{E}_\mathrm{int}}{\delta\rho(x)} = J\, \zeta(k)\, (k+1)\, N^k\, [\rho(x)]^k ,
\end{equation}
and Eq.~\eqref{eq:euler} becomes
\begin{equation}
    \partial_t v + v\, \partial_x v = -\omega^2 x - \frac{J\, \zeta(k)\, (k+1)\, N^k}{m}\, \partial_x \rho^k .
    \label{eq:euler-large-k}
\end{equation}
The force is now local. Equation~\eqref{eq:euler-large-k} can be written as 
\begin{equation}
  \partial_t v + v\,\partial_x v = -\omega^2 x - \frac{\partial_x P}{m\, n}  
\end{equation}
with the number density and the pressure, respectively
\begin{equation}
 n = N\rho\quad\text{and}\quad    P = k\, J\, \zeta(k)\, n^{k+1} .
    \label{eq:pressure}
\end{equation}
Therefore, for $k>1$ the system behaves as a polytropic fluid with the adiabatic index $\gamma = k+1$. For example, the Calogero case $k=2$ corresponds to $\gamma=3$. The stationary condition~\eqref{eq:stationary} gives the equilibrium density
\begin{equation}
    \rho(x) = \frac{1}{B}\, \left[1-\left(\frac{x}{x_0}\right)^2\right]^{1/k} \quad\text{for}\quad |x|<x_0, \quad\text{where}\quad x_0 = \sqrt{\frac{2\mu}{m\,\omega^2}}\quad\text{and}\quad B= \left[\frac{J\, \zeta(k)\, (k+1)\, N^k}{\mu}\right]^{1/k}\, .
\end{equation}
The density $\rho(x)=0$ outside the support $(-x_0,x_0)$ and $\mu$ is fixed by the normalization $\int_{-\infty}^\infty\rho(x)\, dx=1$.

\subsection{Derivation from the Hamiltonian formalism}

The hydrodynamic equations \eqref{eq:continuity} and \eqref{eq:euler} can also be obtained directly from the energy functional, once the canonical variables of the continuum description are identified. For the particles, the canonical variables are the pairs $(x_i,p_i)$. Hamilton's equations,
\begin{equation}
    \dot{x}_i=\frac{\partial H}{\partial p_i} \quad\text{and}\quad 
    \dot{p}_i=-\frac{\partial H}{\partial x_i}\, ,
\end{equation} 
follow from requiring the phase-space action
\begin{equation}
    S = \int dt\, \bigg[\sum_{i=1}^N p_i\,\dot{x}_i - H(\{x_i,p_i\})\bigg]
    \label{eq:action-particles}
\end{equation}
to be stationary under independent variations of $x_i(t)$ and $p_i(t)$. The first term in Eq.~\eqref{eq:action-particles} determines which variables are canonically conjugate. The variable whose time derivative appears ($x_i$) is a coordinate, and the variable multiplying that time derivative ($p_i$) is its conjugate momentum. To find the canonical variables of the continuum description, we therefore express this term in terms of fields.
 
We first introduce a velocity potential field $\varphi(x,t)$. In one dimension any velocity field can be written as a derivative,
\begin{equation}
    m\, v(x,t) = \partial_x\varphi(x,t), \qquad \varphi(x,t) = m\int^x v(x',t)\, dx' ,
    \label{eq:phi-def}
\end{equation}
where the lower limit of the integral is arbitrary. Changing it adds to $\varphi$ a function of $t$ only, which does not change $v$. In the continuum description, all particles near $x$ move with the velocity $v(x)$, so the momentum of the $i$-th particle is $p_i = m\, v(x_i) = \partial_x\varphi(x_i)$. Using the identity~\eqref{eq:iden1}, the first term in Eq.~\eqref{eq:action-particles} becomes
\begin{equation}
    \sum_{i=1}^N p_i\,\dot{x}_i = \sum_{i=1}^N \partial_x\varphi(x_i)\, \dot{x}_i
    = \int_{-\infty}^\infty dx\, \partial_x\varphi(x)\, \hat{j}(x),
    \label{eq:pxdot}
\end{equation}
where
\begin{equation}
    \hat{j}(x) = \sum_{i=1}^N \dot{x}_i\, \delta(x-x_i)
\end{equation}
is the empirical particle current, i.e., the number of particles crossing the point $x$ per unit time. The current is related to the empirical number density
\begin{equation}
    \hat{n}(x)=\sum_{i=1}^N\delta(x-x_i) = N\hat{\rho}(x)
\end{equation}
[see Eq.~\eqref{eq:rhohat}] by the continuity equation. Indeed, differentiating $\hat{n}(x)$ with respect to time gives
\begin{equation}
    \partial_t \hat{n}(x) = -\sum_{i=1}^N \dot{x}_i\, \delta'(x-x_i) = -\partial_x \hat{j}(x) .
    \label{eq:continuity-n}
\end{equation}
Integrating Eq.~\eqref{eq:pxdot} by parts and using Eq.~\eqref{eq:continuity-n}, we get
\begin{equation}
    \sum_{i=1}^N p_i\,\dot{x}_i = -\int_{-\infty}^\infty dx\, \varphi\, \partial_x \hat{j} = \int_{-\infty}^\infty dx\, \varphi\, \partial_t \hat{n} .
    \label{eq:symplectic}
\end{equation}
Similarly, the kinetic energy is 
\begin{equation}
   \sum_{i=1}^N \frac{p_i^2}{2m} = \frac{1}{2m}\int_{-\infty}^\infty dx\, \hat{n}\, (\partial_x\varphi)^2\, .
   \label{eq:ken}
\end{equation}

We now take the continuum limit. As for the one-body sums in Sec.~\ref{sec:collective}, the empirical density $\hat{n}(x)$ appears in these expressions only inside integrals against the smooth field $\varphi$, so it can be replaced by the macroscopic number density $n(x)=N\rho(x)$. Therefore, the right hand sides of Eqs.~\eqref{eq:symplectic} and Eq.~\eqref{eq:ken} become $\int dx\,\varphi \partial_t n$ and $\int dx\, n(x)\, (\partial_x\varphi)^2/(2m)$ respectively. The microscopic Hamiltonian in Eq.~\eqref{eq:action-particles} becomes an energy functional $\mathcal{E}$.
The action in Eq.~\eqref{eq:action-particles} thus becomes
\begin{equation}
    S = \int dt\, \bigg[\int_{-\infty}^\infty dx\, \varphi\, \partial_t n - \mathcal{E}[n,\varphi]\bigg] .
    \label{eq:action-field}
\end{equation}
 
Comparing Eqs.~\eqref{eq:action-particles} and \eqref{eq:action-field}, the discrete index $i$ is replaced by the continuous position $x$, and the sum over $i$ by the integral over $x$. The density $n(x)$ plays the role of the coordinate $x_i$, and $\varphi(x)$ plays the role of its conjugate momentum $p_i$. Accordingly, the Poisson bracket $\{x_i,p_j\}=\delta_{ij}$ is replaced by
\begin{equation}
    \{n(x),\varphi(y)\} = \delta(x-y) .
\end{equation}
The equations of motion follow by requiring the action~\eqref{eq:action-field} to be stationary under independent variations of $n(x)$ and $\varphi(x)$, exactly as for the particles. Varying $\varphi$ gives $\partial_t n - \delta\mathcal{E}/\delta\varphi = 0$. Varying $n$, and integrating the first term in Eq.~\eqref{eq:action-field} by parts in time, gives $-\partial_t\varphi - \delta\mathcal{E}/\delta n = 0$. Finally, since $n=N\rho$, we have $\delta\mathcal{E}/\delta n = N^{-1}\,\delta\mathcal{E}/\delta\rho$, and Hamilton's equations read
\begin{equation}
    \partial_t \rho = \frac{1}{N}\, \frac{\delta\mathcal{E}}{\delta\varphi(x)} ,
    \qquad
    \partial_t \varphi = -\frac{1}{N}\, \frac{\delta\mathcal{E}}{\delta\rho(x)} ,
    \label{eq:hamilton}
\end{equation}
with $\mathcal{E}=\mathcal{E}[\rho,\varphi]$ expressed in terms of $\rho$ and $\varphi$.
 
We now check that Eqs.~\eqref{eq:hamilton} reproduce the hydrodynamic equations \eqref{eq:continuity} and \eqref{eq:euler}. Only the kinetic energy depends on $\varphi$. Varying $(N/2m)\int\rho\,(\partial_x\varphi)^2\, dx$ with respect to $\varphi$ and integrating by parts gives
\begin{equation}
    \frac{\delta\mathcal{E}}{\delta\varphi(x)} = -\frac{N}{m}\, \partial_x\big(\rho\, \partial_x\varphi\big) = -N\, \partial_x(\rho\, v) ,
\end{equation}
and the first of Eqs.~\eqref{eq:hamilton} reproduces the continuity equation~\eqref{eq:continuity}. For the second, we need
\begin{equation}
    \frac{\delta\mathcal{E}}{\delta\rho(x)} = N\left[\frac{(\partial_x\varphi)^2}{2m} + \frac{1}{2} m\omega^2 x^2\right] + \frac{\delta\mathcal{E}_\mathrm{int}}{\delta\rho(x)} ,
\end{equation}
with $(\partial_x\varphi)^2/(2m) = \frac{1}{2}mv^2$. The second of Eqs.~\eqref{eq:hamilton} then gives
\begin{equation}
    \partial_t \varphi = -\frac{1}{2} m v^2 - \frac{1}{2} m\omega^2 x^2 - \frac{1}{N}\, \frac{\delta\mathcal{E}_\mathrm{int}}{\delta\rho(x)} .
\end{equation}
Taking a derivative with respect to $x$ and dividing by $m$, i.e.,  $m^{-1}\,\partial_x$, on both sides, and using $\partial_x\varphi = mv$ and $\partial_x\big(\frac{1}{2}mv^2\big) = m\, v\,\partial_x v$, reproduces the Euler equation~\eqref{eq:euler}.

\section{Self-similar solutions to the continuity and Euler equations}
\label{sec:selfsimilar}

We now use the hydrodynamic equations~\eqref{eq:continuity} and \eqref{eq:euler} to describe the quench protocol of Sec.~\ref{sec:dilation}, with the trap centered, $a(t)=0$. The system is initially at rest, $v(x,0)=0$, in the equilibrium state of a trap with frequency $\omega_0$. At $t=0^+$ the trap frequency is changed to $\omega(t)$, so that for $t>0$ the Euler equation~\eqref{eq:euler} applies with $\omega\to\omega(t)$.

The initial density $\rho(x,0)$ is the equilibrium profile in the trap with frequency $\omega_0$, i.e., it satisfies the stationary condition~\eqref{eq:stationary} with $\omega\to\omega_0$. Its explicit form is not needed in what follows. We only use the fact that the density profile (normalized to unity) is even in $x$  (since the trap is symmetric) and supported on a finite interval $x\in (-x_0,x_0)$. We write it as
\begin{equation}
    \rho(x,0) = \frac{1}{x_0}\,R\Big(\frac{x}{x_0}\Big),
    \label{eq:ss-initial}
\end{equation}
where the scaled profile $R(z)$ vanishes for $|z|>1$, satisfies $R(-z)=R(z)$, and is normalized, $\int_{-\infty}^\infty R(z)\,dz=1$.

We make the ansatz that during the subsequent evolution starting from $\rho(x,0)$, the density $\rho(x,t)$ at any time $t>0$ is given by the self-similar scaling form
\begin{equation}
    \rho(x,t) = \frac{1}{x_0\,|\lambda(t)|}\; R\bigg(\frac{x}{x_0\,|\lambda(t)|}\bigg),
    \qquad\text{with}\qquad \lambda(0)=1 .
    \label{eq:ss-ansatz}
\end{equation}

\subsection{Velocity field from the continuity equation}

Differentiating Eq.~\eqref{eq:ss-ansatz} with respect to $t$ at fixed $x$, we get
\begin{equation}
    \partial_t\rho
    = -\frac{\mathrm{sgn}(\lambda)\, \dot{\lambda}}{x_0\, \lambda^2}\,\frac{d}{dz}\big[z\,R(z)\big] \quad\text{where}\quad z=\frac{x}{x_0|\lambda (t)|}\, .
\end{equation}
The continuity equation~\eqref{eq:continuity}, $\partial_x(\rho v)=-\partial_t\rho$, then becomes
\begin{equation}
    \partial_z(\rho v) = \frac{\dot{\lambda}}{ \lambda}\,\frac{d}{dz}\big[z\,R(z)\big] .
\end{equation}
Integrating over $z$ yields
\begin{equation}
    \rho v = \frac{\dot{\lambda}}{\lambda}\,\big[z\,R(z)\big] + C(t) ,
    \label{eq:rv}
\end{equation}
where $C(t)$ is a function independent of $z$.
Outside the support, both $\rho v$ and $zR(z)$ vanish, which gives $C(t)=0$. Substituting $\rho(x,t)$ from  Eq.~\eqref{eq:ss-ansatz} in Eq.~\eqref{eq:rv} finally gives 
the velocity field inside the support as
\begin{equation}
    v(x,t) = \frac{\dot\lambda(t)}{\lambda(t)}\; x \, .
    \label{eq:ss-velocity}
\end{equation}
The velocity field is linear in $x$, and the initial condition $v(x,0)=0$ requires $\dot\lambda(0)=0$.

\subsection{Equation for $\lambda(t)$ from the Euler equation}

We recall the Euler equation~\eqref{eq:euler} 
\begin{equation}
    \partial_t v + v\,\partial_x v = -\omega^2(t)\,x - \frac{1}{m}\,\partial_x V_\mathrm{eff}(x,t) \, \quad\text{where}\quad V_\mathrm{eff}(x,t) = \frac{1}{N}\,\frac{\delta\mathcal{E}_\mathrm{int}}{\delta\rho(x)}
    \label{eq:ss-euler}
\end{equation}
is the effective interaction potential created by the density at time $t$.

With the velocity field~\eqref{eq:ss-velocity}, we have $\partial_t v=(\ddot\lambda/\lambda-\dot\lambda^2/\lambda^2)\,x$ and $v\,\partial_x v=(\dot\lambda^2/\lambda^2)\,x$. The left-hand side of the Euler equation~\eqref{eq:ss-euler} is therefore
\begin{equation}
    \partial_t v + v\,\partial_x v = \frac{\ddot\lambda}{\lambda}\,x .
    \label{eq:ss-lhs}
\end{equation}

Using $\mathcal{E}_\mathrm{int}$ from Eq.~\eqref{eq:Vint1}  for $-2<k<1$, and Eq.~\eqref{eq:Vint2} for $k>1$, it is easy to see that $V_\mathrm{eff}(x,t)$ obeys the scaling form 
\begin{equation}
    V_\mathrm{eff}(x,t) = |\lambda|^{-k}\; V_\mathrm{eff}^0\Big(\frac{x}{|\lambda|}\Big) \, ,
    \label{eq:ss-Vmf}
\end{equation}
where $V_\mathrm{eff}^0(x)=  V_\mathrm{eff}(x,0)$ is the initial mean-field interaction potential. The stationary condition~\eqref{eq:stationary} at $t=0$, with $\omega\to\omega_0$, gives
\begin{equation}
    V_\mathrm{eff}^0(x) = \mu - \frac{1}{2}\,m\,\omega_0^2\,x^2 \qquad\text{for}\quad |x|<x_0 .
    \label{eq:ss-eq0}
\end{equation}

Substituting Eq.~\eqref{eq:ss-eq0} with $x\to x/|\lambda|$ in Eq.~\eqref{eq:ss-Vmf}, we find
\begin{equation}
    V_\mathrm{eff}(x,t) = \frac{\mu}{|\lambda|^{k}} - \frac{1}{2}\,m\,\frac{\omega_0^2}{|\lambda|^{k+2}}\,x^2
    \qquad\text{for}\quad |x|<x_0|\lambda| .
    \label{eq:ss-Vmf-t}
\end{equation}
The constant term in Eq.~\eqref{eq:ss-Vmf-t} does not contribute to the force, and
\begin{equation}
    \partial_x V_\mathrm{eff}(x,t) = -\frac{m\,\omega_0^2}{|\lambda|^{k+2}}\,x
    \qquad\text{for}\quad |x|<x_0|\lambda| .
    \label{eq:ss-force}
\end{equation}

Substituting Eqs.~\eqref{eq:ss-lhs} and \eqref{eq:ss-force} in the Euler equation~\eqref{eq:ss-euler}, after simplification, we get
\begin{equation}
    \ddot\lambda(t) = -\omega^2(t)\,\lambda + \frac{\text{sgn}(\lambda)\,\omega_0^2}{|\lambda|^{k+1}} ,
    \qquad\text{with}\qquad \lambda(0)=1\quad\text{and}\quad \dot\lambda(0)=0 .
    \label{eq:ss-lambda}
\end{equation}
In the dimensionless time $\omega_0 t\to t$ and with $\alpha(t)=\omega(t)/\omega_0$, this becomes
\begin{equation}
    \ddot\lambda(t) = -\alpha^2(t)\,\lambda + \frac{\text{sgn}(\lambda)}{|\lambda|^{k+1}} ,
    \qquad\text{with}\qquad \lambda(0)=1\quad\text{and}\quad \dot\lambda(0)=0 .
    \label{eq:ss-lambda2}
\end{equation}

We remark that neither the explicit form of $\mathcal{E}_\mathrm{int}$ nor that of $R(z)$ entered the derivation. Only the homogeneity of the interaction energy [Eq.~\eqref{eq:ss-Vmf}] and the equilibrium condition~\eqref{eq:ss-eq0} were used. Therefore, Eq.~\eqref{eq:ss-lambda} holds in the same form for $-2<k<1$, where the interaction is nonlocal, and for $k>1$, where it is local.

Equation~\eqref{eq:ss-lambda} and its initial conditions are identical to Eq.~\eqref{eq:b_eq} for the microscopic scale factor $b(t)$, including the factor $\text{sgn}(\lambda)$. Hence $\lambda(t)=b(t)$, and the self-similar solution is the continuum counterpart of the exact dilation~\eqref{eq:dilation}. Indeed, with $a(t)=0$ the particles move as $x_i(t)=b(t)\,x_i^0$, and the empirical density~\eqref{eq:rhohat} is
\begin{equation}
    \hat\rho(x,t) = \frac{1}{N}\sum_{i=1}^N \delta\big(x-b\,x_i^0\big)
    = \frac{1}{|b|}\,\hat\rho\Big(\frac{x}{|b|},0\Big)
    \label{eq:ss-empirical}
\end{equation}
for any $N$, where we have used $\delta(x-b\,x_i^0)=|b|^{-1}\delta(x/b-x_i^0)$ and the symmetry of the equilibrium configuration $\{x_i^0\}$ under $x\to-x$. Equation~\eqref{eq:ss-empirical} has the form of the ansatz~\eqref{eq:ss-ansatz}, including the absolute value, and $\dot x_i=(\dot b/b)\,x_i$ corresponds to the velocity field~\eqref{eq:ss-velocity}. Since $\lambda(t)$ and $b(t)$ obey the same equation with the same initial conditions, the results for $b(t)$ obtained in Sec.~\ref{subsec:bt} apply directly to $\lambda(t)$.

\bibliography{references}